\documentclass[twocolumn,prb,showpacs,noshowkeys,floatfix]{revtex4}

\usepackage{amssymb}
\usepackage{amsmath}
\usepackage{graphicx}
\usepackage{verbatim}
\usepackage{mathrsfs}
\usepackage{amsfonts}
\usepackage{comment}
\usepackage{color}
\usepackage{hyperref}

\DeclareMathOperator{\sgn}{sgn}
\newcommand{\ud}{\mathrm{d}}

\newcommand*{\dt}{\!\!\ud t\,}
\newcommand*{\dk}{\!\!\ud k\,}
\newcommand*{\dv}{\!\!\ud v\,}
\newcommand*{\bv}[1]{\boldsymbol{#1}}
\definecolor{red}{rgb}{1,0,0}

\begin{document}

\title{Intrinsic dissipation in nanomechanical resonators due to
  phonon tunneling}

\author{I.~Wilson-Rae}
\email{Ignacio.Wilson-Rae@ph.tum.de}
\affiliation{Departamento de F\'{\i}sica Te\'{o}rica de la Materia
Condensada, Universidad Aut\'{o}noma de Madrid, 28049 Madrid, Spain}
\affiliation{Technische Universit\"{a}t M\"{u}nchen, D-85748 Garching,
  Germany.}

\date{\today}

\begin{abstract}
  State of the art nanomechanical resonators present quality factors
  $Q \sim 10^3 - 10^5$, which are much lower than those that can be
  naively extrapolated from the behavior of micromechanical
  resonators. We analyze the dissipation mechanism that arises in
  nanomechanical beam-structures due to the tunneling of mesoscopic
  phonons between the beam and its supports (known as clamping
  losses). We derive the environmental force spectral density that
  determines the quantum Brownian motion of a given resonance. Our
  treatment is valid for low frequencies and provides the leading
  contribution in the aspect ratio. This yields fundamental limits for
  the $Q$-values which are described by simple scaling laws and are
  relevant for state of the art experimental structures. In this
  context, for resonant frequencies in the $0.1-1$GHz range, while
  this dissipation mechanism can limit flexural resonators it is found
  to be negligible for torsional ones. In the case of structureless 3D
  supports the corresponding environmental spectral densities are
  Ohmic for flexural resonators and super-Ohmic for torsional ones,
  while for 2D slab supports they yield $1/f$ noise. Furthermore
  analogous results are established for the case of suspended
  semiconducting single-walled carbon nanotubes. Finally, we provide a
  general expression for the spectral density that allows to extend
  our treatment to other geometries and illustrate its use by applying
  it to a microtoroid. Our analysis is relevant for applications in
  high precision measurements and for the prospects of probing quantum
  effects in a macroscopic mechanical degree of freedom.
\end{abstract}



\maketitle

\section{Introduction}\label{s:intro}

Nanomechanical resonators offer a host of novel applications
\cite{Craighead00,Roukes01,Ekinci05} in high precision measurements
\cite{Milburn94,Bocko96} and may provide a new arena for probing
fundamental aspects of quantum physics \cite{Schwab05}. A prominent
example of such novel applications is scaling magnetic resonance force
microscopy \cite{Sidles95} (MRFM) down to the level of single-spin
detection.  This degree of sensitivity would allow, for example,
three-dimensional imaging of individual biomolecules with atomic-scale
resolution. In these microscopes a magnetic particle \cite{thomas96}
mounted on a cantilever interacts with the nuclear or electron spins
in the sample via the magnetic dipole force. Hence as in many other
applications the relevant mechanical resonator has the beam geometry
on which we will mainly focus in this paper \cite{Cleland}.  Beyond
microscopy, an interesting issue is whether in the nanoscale regime
such a device could be used to probe the quantum state of a single
spin \cite{Gassmann04}.

A single spin $1/2$ coupled to a harmonic oscillator, namely the
Jaynes-Cummings model, constitutes a fundamental system in quantum
optics \cite{Walls}. Physical realizations of this system have proved
invaluable in probing and understanding quantum phenomena
\cite{Haroche01,Leibfried03}. These investigations have ranged from
studying the quantum-classical interface to proof-of-principle
demonstrations of the basic building blocks of quantum information
processing. These developments were enabled by the advent of high-$Q$
resonators for optical and microwave photons and of conservative
harmonic traps for atoms. In these systems the relevant harmonic
oscillator is furnished, respectively, by a single normal mode of the
electromagnetic field or the atomic motion \cite{Stenholm86}. On the
other hand state of the art semiconductor nanostructures can support
mechanical resonances with quality factors $Q\sim 10^3-10^5$ and
frequencies approaching the GHz regime \cite{Ekinci05}. For some of
these structures measurements of mechanical displacements with a
sensitivity approaching the quantum limit have been achieved by
exploiting capacitive coupling to a single electron transistor
\cite{Schwab04,Knobel03}. These developments suggest the possibility
of realizing a \emph{quantum phononics} realm in complete analogy to
quantum optics \cite{Schwab05,Weig04}. In this case the relevant
harmonic oscillator would be furnished instead by a mechanical
resonance and the role of the pseudospin could be played by a
capacitively coupled Cooper pair box \cite{Armour02,Martin04} or an
excitonic transition of an embedded self-assembled quantum dot
\cite{WilsonRae04}. Along these lines one could envisage observing
quantum jumps due to the discrete nature of phonons and realizing
quantum state engineering of non-classical states of motion
\cite{Hopkins03
-Clerk08}. In addition to
semiconductor planar heterostructure realizations there are other
promising possibilities like suspended single walled carbon nanotubes
\cite{Babic03,Sazonova04,Tian04} (SWNT), nanowires \cite{Borgstrom05},
and single crystal diamond beams with embedded nitrogen vacancy color
centers \cite{Olivero05Greentree06}. Another alternative for
furnishing the non-linearity needed to induce non-classical behavior
in mesoscopic mechanical oscillators are optomechanical schemes in
which the resonator couples via radiation pressure to an optical
cavity \cite{Mancini98,Kippenberg07}.  This venue has recently
witnessed significant experimental progress
\cite{Metzger04
-Schliesser07}
towards achieving \emph{ground state cooling} which is highly
desirable to enable quantum effects
\cite{WilsonRae07-
xue07}.  Finally, for resonators with
sufficiently small effective masses and high bending rigidity, yet
another alternative that has been considered is to use the Euler
instability \cite{Carr01
-savel'ev07}.

The realization of all of the aforementioned applications of
nanomechanical systems hinges on understanding and controlling the
\emph{intrinsic dissipation} and noise mechanisms \cite{Cleland02}
that limit their coherent dynamics.  For centimeter-scale
semiconductor micromechanical systems $Q\sim 10^8$ have been measured
at low temperatures \cite{Roukes01}. On the other hand when these
devices are shrunk to the nanometer-scale these values decrease
dramatically to $Q\sim 10^3-10^5$. Early work suggested that the
increase in surface-to-volume ratio combined with surface effects
might be invoked as a plausible explanation of this phenomenon
\cite{Roukes01}.  Subsequently it was realized that elastic wave
radiation into the supports --- the so-called \emph{clamping losses}
\cite{Ekinci05} --- could play an important role leading to
non-trivial scaling laws for the $Q$-values with the aspect ratio that
are intimately related to the low frequency behavior of the
corresponding transmission coefficients (cf.~Sec.~\ref{s:singlejunc}
and Refs.~[\onlinecite{Cross01,phdthesis,Photiadis04,Chang05}]).

If we consider the environment of the mechanical resonator responsible
for its dissipation, the $Q$-value is determined by the pole of the
ensuing modified propagator for the resonator's normal coordinate that
corresponds to its resonant frequency $\omega_R$. If one adopts a
Caldeira-Leggett model \cite{CaldeiraLeggett83,Weiss} --- i.e.~the
environment is assumed to consist of a thermal ensemble of harmonic
oscillators --- with a linear coupling to the environment, this
propagator can be obtained exactly from the \emph{environmental force
  spectral density} $I(\omega)$ and the $Q$-value is found to be
temperature independent.  Thus, the quantum Brownian motion of the
normal coordinate $\hat X_R$ associated with a given resonance is
characterized by the following generalized equation of motion
\begin{equation}\label{motion}
\ddot{\hat{X}}_R(t) + \int_0^t\!\!\ud
t'\,\gamma(t-t')\dot{\hat{X}}_R(t')+\omega_R^2\hat{X}_R(t)=\hat{\xi}(t) 
\end{equation}
and is completely determined by the function
$I(\omega)=\omega\int_{-\infty}^{\infty}\!\!\ud t\,\gamma(t)e^{i\omega
  t}/2\omega_R$ where $\gamma(t)$ is the symmetric dissipation-kernel
\cite{Hu92} [cf.~Eq.~(\ref{eq:I}) and Sec.~\ref{II}] and
$\hat{\xi}(t)$ corresponds to the environmental noise. When the
standard \emph{Markov approximation} is warranted
$\gamma(t)\sim\delta(t)$ and the $Q$-value is determined by
\begin{equation}\label{eq:Q1}
\frac{1}{Q}=\frac{I(\omega_R)}{\omega_R}\,.
\end{equation}
Within this approximation the value $I(\omega_R)$ is the only relevant
information about the spectral density, Eq.~(\ref{motion}) can be
interpreted as a quantum Langevin equation, and we have the standard
relation
\begin{equation}\label{eq:Q2}
\frac{1}{Q}=\frac{\langle \dot E \rangle}{\omega_R\langle E\rangle}\,,
\end{equation}
where $E$ is the total energy stored in the resonator's degree of
freedom. This is normally valid for sufficiently high $Q$ and low
temperature if $I(\omega)$ is smooth enough, or for high temperatures
if $I(\omega)$ scales linearly with frequency \cite{Ford65,Hu92,Weiss}
(classical Langevin). However in many instances it is desirable to go
beyond the Markov approximation and more precise knowledge about
$I(\omega)$ is needed. Examples of these are the cases discussed above
in which the resonator interacts with a single spin $1/2$ system if
the coupling is relatively strong and cases in which the $Q$ is not
high enough or the temperature low enough. Currently there exists no
microscopic derivation of Eq.~(\ref{motion}) and of the underlying
Caldeira-Leggett Hamiltonian and corresponding spectral density
$I(\omega)$, for which customarily an ohmic dependence is assumed.

For nanometer-sized suspended monocrystalline beam structures at low
temperatures, the relevant phonon mean free path can become larger
than the beam's length. In the case of an insulating system, a
prominent consequence of the ballistic regime that results is the
quantization of thermal conductance \cite{Rego98} in units of $G_{th}=
\pi^2 k_B^2 T/3 h$, which has been demonstrated experimentally
\cite{Schwab00}. In this low temperature, low frequency regime and for
small deflections, anharmonicity becomes irrelevant for the analysis
of dissipative effects \cite{Cleland,Lifshitz00,Jiang04}.  Thus, in
this paper we analyze the ideal limit, that ensues for the low-lying
resonances of an insulating beam close to equilibrium at low enough
temperatures, in which \emph{phonon tunneling between the beam and its
  supports is the only source of thermal noise and dissipation.} To
this effect the vibrations of the whole structure are described by a
purely harmonic Hamiltonian. Its normal modes, discussed in the
following section, form a continuum. As will become clear below, this
feature is inherited from the supports that thus provide thermal
phonon reservoirs. When the systems under consideration deviate from
the aforementioned ideal scenario (e.g.~the vacuum is imperfect, the
beam's material is amorphous or surface effects are relevant) there
will be other contributions to the mechanical damping that will add
incoherently \cite{Ekinci05}.  However it should be stressed that in
all such cases our treatment is valid for the contribution to the
dissipation arising from the \emph{vibrational degrees of freedom} and
thus our results will provide an \emph{upper bound for the
  $Q$-values.}

\begin{figure}
\includegraphics[width=8.6cm]{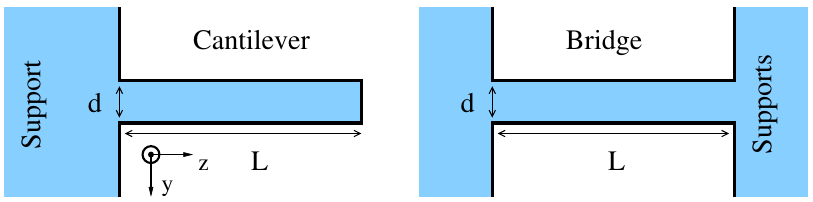}
\caption{Schematic diagram of suspended mechanical beam structures
  presenting abrupt junctions with the supports. Throughout the paper
  the $z$-axis corresponds to the beam's axis, while the $x$ and $y$
  axes are the principal axes of inertia of the beam's cross section
  at the left junction (the origin is set at the center of mass of the
  latter).
  \label{fig:beams}}
\end{figure}

Beam nanostructures for which high $Q$-values have been measured
normally involve abrupt junctions with the supports. This can be
intuitively understood in terms of impedance mismatch for the
propagation of elastic waves. We will focus mainly on this type of
structures, as depicted in Fig.~\ref{fig:beams}, and consider phonon
frequencies that correspond to $k \sim 2 \pi /L \ll 2 \pi/d$, where
$k$ is the wavevector along $z$ inside the beam, $L$ the beam's
length, and $d$ the typical dimension of the cross section
(cf.~Fig.~\ref{fig:beams} and App.~\ref{ap:beam}). It is clear that
for this beam geometry the reciprocal of the aspect ratio $d/L$
provides a natural small parameter that will underpin our analysis of
the non-trivial matching at the abrupt junctions. Given that we are
interested in phonons with wavelengths much longer than the lattice
constant, the continuum limit constitutes a good approximation. This
yields for the supports 3D isotropic elasticity
\cite{Landau,Graff,Cleland}, if one further neglects the possible
anisotropy --- or 2D ``thin plate elasticity'' for slab supports of
thickness much smaller than the phonon wavelength
\cite{Cross01,Landau}. If we consider the beam there are basically two
scenarios: ($i$) all the characteristic dimensions of the cross section
are much larger than the lattice constant (e.g.~semiconductor planar
heterostructures) or ($ii$) there are characteristic dimensions
comparable to the lattice constant (e.g.  semiconducting SWNTs and
small-radius nanowires). In case ($i$) one can start from 3D isotropic
elasticity and obtain the effective theory valid for phonon
wavelengths much larger than $d$ known as ``thin rod elasticity''
\cite{Landau,Graff,Cleland} (TRE). The latter is completely determined
by the linear mass density $\mu_b$, the mean axial moment of inertia
$\langle r^2\rangle_S$, and the extensional, torsional and bending
rigidities; and leads to the following dispersion relations for the
four low frequency branches
\begin{equation}\label{disper}
  \omega_{\beta}(k)= \tilde{c}_{\beta} k^{p_\beta}\qquad p_c=p_t=1, \
p_v=p_h=2\,.
\end{equation}
Here $\beta$ is the branch index: $c$ for compression, $t$ for
torsion, $v$ for vertical bending and $h$ for horizontal bending [the
prefactors $\tilde{c}_{\beta}$ are given in App.~\ref{ap:beam},
Eq.~(\ref{tildec})].  Though for propagation inside the beam this
effective theory will suffice, resort to the underlying 3D
``microscopic'' theory will allow for a rigorous treatment of the
matching at the junction (cf.  Sec.~\ref{s:singlejunc}) --- an
analogous procedure is feasible for a thin plate geometry
\cite{Cross01} with the ``microscopic'' theory furnished by 2D thin
plate elasticity. In case ($ii$) it is well understood that the
effective long wavelength theory is formally equivalent to TRE but the
standard ``bulk'' relations between the rigidities are no longer
warranted \cite{Segall02}. Whence heuristic considerations allow to
extend our treatment of the non-trivial matching at the abrupt
junctions to this case with suitably redefined $\tilde{c}_{\beta}$ ---
when considering this extension we will focus mainly on SWNTs.  Thus,
our treatment will amount to a derivation of the leading contribution
in the aspect ratio to the environmental spectral densities
$I(\omega)$ associated with phonon tunneling induced noise starting from
the underlying microscopic lattice Hamiltonian.

The relevance of the ensuing results is twofold: ($i$) they furnish a
very general understanding of clamping losses determining the
corresponding $Q$-values for a wide range of experimentally relevant
structures, and ($ii$) they provide an instance for which a
microscopic derivation of the quantum dissipative dynamics of a
``macroscopic'' mechanical degree of freedom can be given, which
reduces exactly to a Caldeira-Leggett model
\cite{Ford65,Ullersma1,Ullersma2}. The latter is essentially a
derivation of the basic Hamiltonian for a lossy 1D phonon-cavity and
brings together aspects of the analogous problem in quantum optics
\cite{Viviescas03} and the ballistic transport in a mesoscopic wire
\cite{Rego98}. In this respect it should be noted that the Markov
approximation, which is customarily used for the Brownian motion of
mechanical resonators, is not assumed in our derivation but instead
its range of validity emerges from the behavior of the spectral
densities $I(\omega)$ that we calculate and that allow (if necessary)
to go beyond the Markovian regime --- for these aspects the reader is
referred to the vast literature on quantum dissipation
\cite{Weiss}. Thus, the validity (for large aspect ratios) of
Eq.~(\ref{eq:Q1}), that will be used to determine the $Q$-values, is
established rather than postulated.

We will first present our main results (cf.~Subsec.~\ref{sub:results})
and subsequently give their derivation in the following Sections.  For
the sake of clarity we will mainly focus on a specific model with
maximally symmetric 3D structureless supports and the beam having the
aforementioned property ($i$) [cf.~Subsec.~\ref{sub:modes}], and
outline how the treatment can be extended (cf.~App.~\ref{ap:exten}) by
referral to this concrete realization. We will discuss both the bridge
and cantilever geometries (cf.~Fig.~\ref{fig:beams}). Finally, we will
also outline how this approach can be generalized to other geometries
and illustrate this further extension for the experimentally relevant
case of a microtoroid \cite{Kippenberg07,Anetsberger08}.

\begin{table*}
\begin{ruledtabular}
\begin{tabular}{c || c | c | c | c | c | c} 
  & \multicolumn{2}{c|}{General Relations (3D)} 
  & \multicolumn{2}{c|}{Monolithic structure ($\sigma=\frac{1}{3}$)} 
  & \multicolumn{2}{c}{Semiconducting SWNT ($\sigma_s=\frac{1}{3}$)}
  \\[2pt]  \cline{2-7}  &&&&&&\\[-10pt]
  & $I_n (\omega)$ & $Q_n$ 
  & $Q_n(L,\mathsf{w},\mathsf{t})$ & \parbox{35pt}{Typical value for
    150MHz$_{\vphantom{\displaystyle{A}}}$ } 
  & $Q_n(L,R)$ & \parbox{35pt}{Typical value for
    1GHz$_{\vphantom{\displaystyle{A}}}$} \\[5pt]   
  \hline 
  Compression & 
  $\displaystyle{\frac{\omega}{Q_n}}$ & 
  $\displaystyle{\frac{\rho_s c_t^3 L}{2 \delta \mu_b \tilde{c}_c^3
      \tilde{u}_c (\alpha) k_n} }$ & 
  $\displaystyle{\frac{0.88}{\pi\delta} \frac{L^2}{ \mathsf{w} \mathsf{t}}
    \frac{1}{n+\frac{\delta}{2}}}$
  & $3.2\times 10^4$ & 
  $\displaystyle{\frac{0.14}{\pi\delta}
    \sqrt{\frac{\sigma_G}{h \rho_s} 
      \left(\frac{E_s}{E_b}\right)^3 }^{\vphantom{\displaystyle A}}
    \frac{L^2}{h R} \frac{1}{n+\frac{\delta}{2}}}$
  & $1.5\times 10^5$ \\[15pt] 
  Torsion & 
  $\displaystyle{\frac{\omega^3}{Q_n \omega_n^2}}$ & 
  $\displaystyle{\frac{\rho_s c_t^5 L}{2 \delta \mu_b\langle
      r^2\rangle \tilde{c}_t^5
      \tilde{u}_t (\alpha,\gamma_z) k_n^3} }$ & 
  $\displaystyle{\frac{4.1}{\pi^3\delta} \frac{\mathsf{w}^2
      L^4}{\mathsf{t}^6} \frac{1}{(n+\frac{\delta}{2})^3}}$
  & $7.6\times 10^9$ &  
  $\displaystyle{\frac{2.3}{\pi^3\delta} 
    \sqrt{ \left( \frac{\sigma_G}{h \rho_s} \right)^3 \!\!
      \left(\frac{E_s}{E_b} \right)^5 }
    \frac{L^4}{h R^3} \frac{1}{(n+\frac{\delta}{2})^3}}$
  & $1.3\times 10^{11}$ \\[15pt] 
  \parbox{35pt}{Vertical bending} & 
  $\displaystyle{\frac{\omega}{Q_n}}$
  & $\displaystyle{\frac{\rho_s c_t^3 L}{4 \delta C_n \mu_b \tilde{c}_v^3
      \tilde{u}_v (\alpha) k_n^4} }$ & 
  $\displaystyle{\frac{3.9}{\pi^4 \delta C_n} \frac{L^5}{\mathsf{w}
      \mathsf{t}^4} \left( \frac{3 \pi}{2 k_n L}\right)^4 }$  
  & $9.6\times 10^5$ &
  $\displaystyle{\frac{0.043}{\pi^4 \delta C_n} 
\sqrt{\frac{\sigma_G}{h \rho_s} \left(\frac{E_s}{E_b}\right)^3 }
\frac{L^5}{h R^4} \left( \frac{3 \pi}{2 k_n L}\right)^4 }$  
  & $4.2\times 10^6$ \\[15pt] 
  \parbox{43pt}{Horizontal bending} & 
  $\displaystyle{\frac{\omega}{Q_n}}$
  & $\displaystyle{\frac{\rho_s c_t^3 L}{4 \delta C_n \mu_b \tilde{c}_h^3
      \tilde{u}_h (\alpha) k_n^4} }$ & 
  $\displaystyle{\frac{3.9}{\pi^4 \delta C_n} \frac{L^5}{\mathsf{t}
      \mathsf{w}^4} \left( \frac{3 \pi}{2 k_n L}\right)^4}$
  & $3.9\times 10^5$ & 
$\displaystyle{\frac{0.043}{\pi^4 \delta C_n} 
\sqrt{\frac{\sigma_G}{h \rho_s} \left(\frac{E_s}{E_b}\right)^3 }
\frac{L^5}{h R^4} \left( \frac{3 \pi}{2 k_n L}\right)^4 }$   
  & $4.2\times 10^6$  
\end{tabular}
\end{ruledtabular}
\caption{First and second columns: general formulas for the
  environmental force spectral densities $I_n(\omega)$ and the
  $Q$-values $Q_n$ corresponding to the different resonances of a beam
  suspended from structureless 3D support(s) --- $n=0,1,\ldots$ labels
  the harmonics for each of the four branches. These general formulas
  are specialized for two cases: (third column) a monolithic structure
  with rectangular beam cross section of thickness $\mathsf{t}$ and
  width $\mathsf{w}$, and (fifth column) a suspended semiconducting
  SWNT of radius $R$. The nanotube is modeled as a cylindrical shell
  of effective thickness $h$ and Poisson ratio\cite{Yakobson96}
  $\sigma_b=0.19$. All formulas are valid for both the cantilever
  ($\delta=1$) and bridge ($\delta=2$) geometries. The
  dispersion-relation prefactors $\tilde{c}_\beta$ are given in
  Eq.~(\ref{tildec}), $\mu_b$ is the linear mass density of the beam
  (for the nanotube $\mu_b=2 \pi R \sigma_G$, where $\sigma_G$ is the
  surface density of graphene), $\rho_s$ and $c_t$ are, respectively,
  the mass density and transverse speed of sound for the supports'
  material, and $\langle r^2\rangle\equiv I_z/S$ is the beam's mean
  axial moment of inertia.  In the case of torsion (second row), for
  the monolithic structure (third and fourth columns), we specialize
  for $\mathsf{t} \ll \mathsf{w}$ so that $\gamma_z^2\approx1$ and the
  torsional rigidity \cite{Landau} reads $C \approx E_b \mathsf{w}
  \mathsf{t}^3/6(1+\sigma_b)$ [for a cylindrical shell we have instead
  $C = E_b\pi h R^3/(1+\sigma_b)$] --- $E_b$ [$E_s$] is the Young's
  modulus for the material of the beam [support(s)].  The
  dimensionless displacements and angles $\tilde{u}_\beta$, given in
  Subsec.~\ref{sub:tcoeff} (cf.~Fig.~\ref{fig:tildeu}), take the
  following values for $\sigma_s=1/3$
  [$\alpha\equiv(1-2\sigma_s)/2(1-\sigma_s)=1/4$]: $\tilde{u}_c (1/4)
  = 0.13$, $\tilde{u}_t (1/4,\gamma_z)= 1/12\pi + 0.019\gamma_z^2 $,
  and $\tilde{u}_{v/h} (1/4) = 0.12$, where $\gamma_z\equiv
  (I_y-I_x)/I_z$ and we have defined $\tilde{u}_t
  (\alpha,\gamma_z)=\tilde{u}_t^{(A)}
  +\tilde{u}_t^{(S)}(\alpha)\gamma_z^2$. The resulting typical values
  for the $Q_n$ (fourth and sixth columns) correspond to the lowest
  lying resonances of \emph{bridge geometries} (wavevectors$^a$
  $k_{0,c/t}=\pi/L$ and $k_{0,v/h} \approx 1.51 \pi/ L$) with a
  different length for each branch ($L_\beta$) chosen so that the
  comparison is made for equal frequencies$^b$: 150MHz for the
  monolithic structure (parameters: $\mathsf{w}=100\,$nm and
  $\mathsf{t}=20\,$nm) and 1GHz for the nanotube [parameters:
  $h\!=\!0.66$\AA (cf.~Ref.~\onlinecite{Yakobson96}),
  $\sigma_G\!=\!7.7\!\times\! 10^{-7}$Kg/m$^2$, $E_b\!=\!1$TPa
  (cf.~Ref.~\onlinecite{Babic03}), and $R\!=\!1$nm; we assume Si
  supports: $\rho_s \!=\! 2.3 \!\times\! 10^{3}$Kg/m$^3$,
  $E_s\!=\!112$GPa]. \\[3pt] \footnotesize{$^a$For the cantilever
  geometry these would be instead $k_{0,c/t}=\pi/2 L$ and $k_{0,v/h}
  \approx 0.60 \pi/ L$.}\\[2pt] \footnotesize{$^b$For the monolithic
  case (nanotube): $L_c\!=\! 23\mu$ ($2.3\mu$), $L_t\!=\!4.7\mu$
  ($1.5\mu$), $L_{v}\!=\!0.98\mu$ ($0.11\mu$), $L_h\!=\!2.4\mu$
  ($0.11\mu$).}
  \label{table}}
\end{table*}

\subsection{Environmental force spectral densities and
  \texorpdfstring{$Q$}{Q}-values for each resonance}\label{sub:results} 

To study the dissipation induced by the coupling to the supports it
proves useful to introduce the concept of an ``effective environmental
density of states'' for each branch given by $\tilde
\rho_\beta(\omega)$.  These functions will be defined in
Sec.~\ref{s:singlejunc}. There we will find that for 3D supports they
bear simple relations with properties of the decoupled support ---
i.e.~subject to free boundary conditions --- that are closely related
to its density of states (DOS); namely, the displacement and angle
(twist) vacuum spectrum at the junction \footnote{More precisely, at
the origin taken as the center of mass of the beam's cross section at
the junction. We note that the anomalous frequency scaling of the
$\tilde\rho_\beta(\omega)$ [cf.~Eq.~(\ref{tautilde})] when viewed as
the DOS of a 1D fictitious environment expresses just
  the non-adiabatic nature of the junction.}.  The cornerstone of our
analysis will be furnished by the following novel pair of
\emph{fundamental relations}, each of which completely specifies the
environmental force spectral density for a given resonance to
\emph{lowest order in the reciprocal of the aspect ratio} $d/L$:
\begin{align}
I_{n,\beta}(\omega)  &= \delta C_{n,\beta}\left[\frac{\ud\omega_\beta}{\ud
k} \!(\omega_{n,\beta})\right]^2\,\frac{\tilde \rho_\beta(\omega)}{2L}\,
\frac{\omega_{n,\beta}}{\omega}\,,\label{eq:spectralrho}
\\
I_{n,\beta}(\omega)  &= \delta C_{n,\beta}\frac{\ud\omega_\beta}{\ud
k} (\omega)\,\frac{\tau_\beta(\omega)}{2L}\,
\left(\frac{\omega_{n,\beta}}{\omega}\right)^{p_\beta} \,.
\label{eq:spectraltau}
\end{align}
Here $n=0,1,\ldots$ labels the harmonics for each branch $\beta$
(i.e.~$\omega_R\to\omega_{n,\beta}$), $\delta=1,2$ is the number of
supports,
\begin{equation}
 C_{n,\beta}= \begin{cases} 1 & \text{for }\beta=c,t \\
\left(\tanh^2\frac{k_{n,v/h} L}{2}\right)^{\left(-1\right)^n} &
\text{for }\beta=v,h
\end{cases}
\end{equation} 
and the bare resonant wavevectors $k_{n,\beta}\equiv
\omega^{-1}_{\beta}(\omega_{n,\beta})$ are determined by the TRE
solutions for clamped-clamped (clamped-free) boundary conditions in
the case of the bridge (cantilever) geometry \footnote{This yields
$k_{n,c/t}=\pi(n+\delta/2)/L$ and
$k_{n,v/h}\approx\pi(n+\delta-1/2)/L$ where the relative error is of
order $e^{-k_{n,v/h}L}$.}.  The first relation expresses the spectral
density in terms of $\tilde \rho_\beta(\omega)$. The second relation
involves the transmission coefficient $\tau_\beta(\omega)$ from a
semi-infinite beam into the support at a single junction which can be
interpreted as the probability for an incident phonon with frequency
$\omega$ to tunnel into the support.  As will be discussed further in
the next Section and derived rigorously in Subsec.~\ref{sub:ustar},
for phonon frequencies $\omega\to0$ the beam and the supports become
decoupled so that the junction plays a role analogous to a tunnel
barrier. Whence relation (\ref{eq:spectraltau}) allows to interpret
the losses to the supports in terms of \emph{phonon tunneling}. More
precisely, it provides a rigorous footing for the heuristic formula to
describe this dissipation mechanism set forth by Cross and Lifshitz in
Ref.~[\onlinecite{Cross01}] based upon Eq.~(\ref{eq:Q2}), namely,
\begin{equation}\label{heuristicQ}
\frac{1}{Q} \sim \frac{\delta}{2 L}
  \frac{\ud\omega_\beta}{\ud k}(k_n) \frac{\tau_\beta
  (k_n)}{\omega_\beta (k_n)} = \frac{\delta p_\beta}{2}
  \frac{\tau_\beta(k_n)}{k_n L}\,,
\end{equation}
where the implicit dimensionless prefactor is expected to be of order
unity and have the latter as its limit for $n\to\infty$ --- henceforth
we drop the branch index of the $k_n$, $\omega_n$.

The above approximation emerges from considering a phonon wavepacket
that bounces back and forth between both ends of the beam. Naturally,
it should be adequate for large $n$ and provides a simple intuitive
description of this dissipation mechanism in terms of phonon tunneling
at the junctions. Here we have added a factor of $\delta/2$, where
$\delta$ is the number of supports, to account for both the bridge and
cantilever geometries. Equation (\ref{heuristicQ}) can be obtained
from Eqs.~(\ref{eq:Q1}) and (\ref{eq:spectraltau}) which allow in
addition to determine that the corresponding dimensionless prefactor
is given by $C_{n,\beta}$ --- i.e. for the non-dispersive branches it
is \emph{exactly unity} for all $n$ and only deviates from unity for
the low-lying bending resonances.

The general relation between the transmission coefficient at a single
junction and the effective environmental DOS concomitant to the pair
of relations (\ref{eq:spectralrho}), (\ref{eq:spectraltau}) will be
proved in Sec.~\ref{s:singlejunc} where expressions for the functions
$\tau_\beta(\omega)$ and $\tilde\rho_\beta(\omega)$ are derived
explicitly for 3D supports
[cf.~Eqs.~(\ref{tautilde})-(\ref{eq:trans})]. The latter together with
Eqs.~(\ref{eq:spectralrho}) and (\ref{eq:Q1}) allow us to obtain
formulas for the $Q$-values of all the low-frequency resonances that
are given in the second column of Table \ref{table}. The first column
gives the resulting expression for the environmental force spectral
density which turns out to be ohmic in all cases except for the
torsional resonances.  The third and fifth columns give examples of
particular experimental relevance. Namely, a monolithic structure with
rectangular cross section for the beam of width $\mathsf{w}$ and
thickness $\mathsf{t}$ and a semiconducting suspended SWNT of radius
$R$ --- for the nanotube we use constants $\tilde c_\beta$ that
correspond to the ``continuum'' shell approximation for the rigidities
\cite{Robertson92,Yakobson96}. The validity of these results, which
are adequate for low frequencies, will be borne out in full detail in
the following Sections as we derive the Eqs.~(\ref{eq:spectralrho})
and (\ref{eq:spectraltau}). While the latter will hold in all
instances where any characteristic dimension of the supports is either
much larger than $L$ or at most of order $d_s\lesssim d$, Table
\ref{table} focuses on the case in which the limit $d_s\to 0$ yields a
3D support with no characteristic dimension (e.g.~an elastic
half-space). Within this context the second column is general with an
appropriate definition of the dimensionless displacements and angles
$\tilde u$ that only depend on the supports' material Poisson ratio
(cf.~Sec.~\ref{s:singlejunc} and App.~\ref{ap:exten}).  On the other
hand the specific $\tilde u$ used in the examples --- which will be
calculated in Sec.~\ref{s:singlejunc} --- correspond to the maximally
symmetric case of the half-space.

In the context of the applications already discussed, one focuses on a
specific resonance and it can be argued that a sound figure of merit
is afforded by the quantity $k_B T/\hbar\omega_R Q$. Therefore we
compare the different types of resonances (for different lengths $L$)
for the same resonant frequency $\omega_R$ (fourth and sixth
columns). It is clear that for specified materials these results for
the $Q$-values only depend on the ratios between the beam's
dimensions. Thus, for semiconductor heterostructure realizations the
corresponding formulas (second and third columns) hold all the way
from the nanoscale to the macroscopic regime and are also applicable
to micromechanical resonators. However, while for fabricated flexural
resonators with sub-micron transverse dimensions aspect ratios
$L/d\sim 10$ are not uncommon \cite{Cleland}, for typical
micromechanical resonators used in MEMS $L/d\gtrsim 100$ rendering the
above results an upper bound that would be hard to reach --- note for
comparison that the last two entries of the fourth column correspond,
respectively, to $L/\mathsf{t}=49$ and $L/\mathsf{w}=24$. Finally, we
note that for given 3D supports the above results are ``universal'' in
the specific sense that their only dependence on properties of the
beam is through $\mu_b$, $\langle r^2\rangle_S$, and the prefactor of
the TRE dispersion relations (\ref{disper}), quantities that specify
its low-frequency effective theory. In particular, the scaling laws
with the beam's length are completely general. Furthermore, for a
monolithic structure the dependence on the material's Poisson ratio in
the relevant range $1/5\lesssim\sigma\lesssim1/3$ is so weak
(cf.~Sec.~\ref{s:singlejunc}) that the $Q$-value of a given resonance
can be effectively regarded as a geometric property.

The other instance of interest to which Eqs.~(\ref{eq:spectralrho}),
(\ref{eq:spectraltau}) apply and that also yields simple scalings, is
the case of slab 2D supports of thickness $d_s=\mathsf{t}$ equal to the
thickness of the beam. The results we will obtain
[cf.~Sec.~\ref{s:singlejunc}, Eq.~(\ref{scale2Drho})] for the
frequency dependence of the $\tilde\rho_\beta(\omega)$ together with
Eqs.~(\ref{eq:spectralrho}), (\ref{eq:Q1}) imply the following
\begin{equation}
I_{n,\beta\neq v}=\frac{\omega_n}{Q_n} \qquad
I_{n,v}=\frac{\omega_n^2}{Q_n\omega} \,,
\end{equation}
which correspond to $1/f$ noise \cite{Dutta81,Paladino02}. We note
that the finite size of the slab provides a natural infrared cutoff.
The corresponding $Q$-values (which with the exception of horizontal
bending scale as the aspect ratio $L/\mathsf{w}$) have already been
derived in Ref.~[\onlinecite{Cross01}] up to the prefactor $C_n$
discussed above.

Standard fabrication procedures \cite{Cleland} normally result in an
undercut of the support of size $d_U$ at least comparable to the width
of the beam. Our analysis of the abrupt junction(s) with the
support(s) given in Sec.~\ref{s:singlejunc} and App.~\ref{ap:exten}
implies that it is quantitatively correct to use the 3D model for the
support(s) when $d_U\lesssim d$, and the 2D model discussed above when
$d_U\gg L$. In addition, based on heuristic considerations, the 3D
results for the $Q$-values given in Table \ref{table} are expected to
be qualitatively correct for \footnote{More precisely, they should
yield the correct order of magnitude provided the corresponding
resonance lies below the lowest infrared cutoff associated with the
phonon waveguide afforded by the undercut --- here we assume $d_U \sim
L \gg \mathsf{t}$.} $d_U<L$.

Finally, it should be noted that a substantial part of our derivation
of the spectral density for an isolated resonance \footnote{The
criterion underpinning the derivation is that the support-induced
frequency shift (see below) be smaller than the ``free spectral
range'' associated with the bare resonance.} is independent of the
geometry. More precisely Eqs.~(\ref{eq:spectralrho}) and
(\ref{eq:spectraltau}) can be viewed as a specific instance of the
following more general relation
\begin{align}
I(\omega)\approx \frac{\pi}{2 \rho_s^2 \omega_R \omega} \int_q  &
\left| \int_S \ud r^2\,\left( \bar{u}'_R \cdot \bv{\sigma}_q -
    \bar{u}_q \cdot \bv{\sigma}'_R \right) \cdot \hat{n} \right|^2
\nonumber \\  & \times\delta [\omega - \omega(q) ] \label{genI}
\end{align}
where $\bar{u}_q(\bar r)$ and $\bv{\sigma}_q(\bar r)$ are the
displacement and stress fields associated with scattering eigenmodes for
the whole structure, labeled by $q$ [eigenfrequencies $\omega(q)$],
and $\bar{u}'_R(\bar r)$, $\bv{\sigma}'_R(\bar r)$ are the analogous
fields for the resonator mode (cf.~Section \ref{II}). Here $S$ is the
contact surface between the resonator and its support(s) and $\rho_s$
is the mass density of the latter. As will become clear in
Subsec.~\ref{sub:Itau} the small parameter associated with the above
approximation is $|\Delta_I(\omega_R)|/2\omega_R$, where
$\Delta_I(\omega_R)/2$ is the support-induced shift of the resonant
frequency (cf.~App.~\ref{ap:I}).

\begin{figure}
\includegraphics[width=6.0cm]{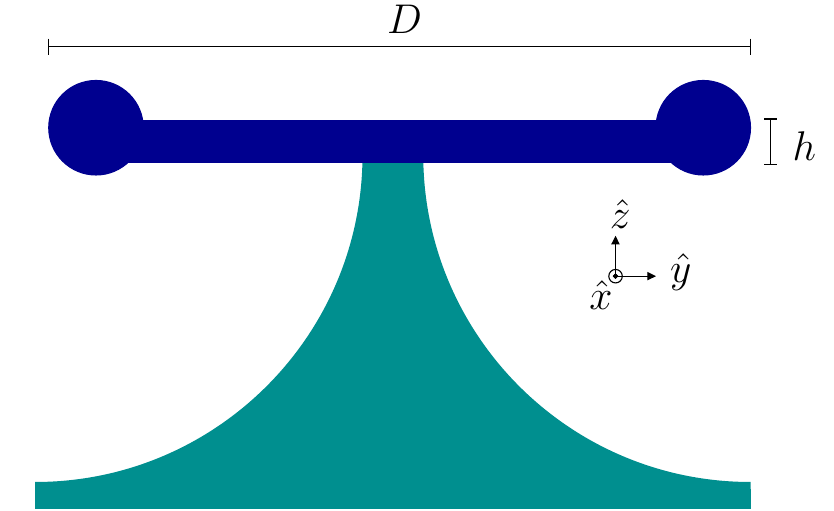}
\caption{Schematic diagram of a microtoroid structure with largest
  diameter $D$ and membrane thickness $h$ (the origin is set at the
  center of the contact area $S$ with the pedestal).
  \label{fig:torus}}
\end{figure}

In all natural scenarios $|\Delta_I(\omega_R)|/2\omega_R\ll 1$ arises
due to the abrupt nature of the junction(s) with the support(s) and
the condition $\sqrt{S}k_R\ll 1$ --- where $k_R$ is the typical
wavevector associated with the resonator mode. The behavior as $S\to0$
leads to two possibilities: (i) the limit is singular, or (ii) it
defines a well-behaved resonator geometry that can be described by 3D
elasticity. The beam geometry falls into case (i) for which
$\bar{u}'_R(\bar r)$ is specified by clamped boundary conditions at
$S$ (cf.~Subsec.~\ref{IIB}). In turn in case (ii) $\bar{u}'_R(\bar r)$
should satisfy free boundary conditions at the contact surface. An
experimentally relevant example of the latter is afforded by
\emph{microtoroids} \cite{Kippenberg07,Anetsberger08} (cf.~Figure
\ref{fig:torus}). For such structures an heuristic treatment of the
pedestal as a beam with \emph{adiabatically} varying cross section
allows us to obtain from Eq.~(\ref{genI}) the following
\begin{equation}\label{toroI}
I(\omega)\approx \frac{\sqrt{\rho_sE_s}}{m_R \omega_R} S
\,\tilde{u}^2_{R,z}(0)\,\omega 
\end{equation}  
for any axially symmetric isolated resonance. Here the resonator mode
is normalized so that the normal coordinate corresponds to the
elongation of the toroid's external radius, and $m_R$ and
$\bar{\tilde{u}}_R(\bar r)$ are, respectively, the corresponding
effective mass and mode profile (cf.~App.~\ref{ap:toro}). Equation
(\ref{eq:Q1}) then yields for the corresponding $Q$-value
\begin{equation}\label{toroQ}
  Q\approx \frac{m_R \omega_R}{\sqrt{\rho_sE_s} S
    \,\tilde{u}^2_{R,z}(0)}\,.
\end{equation}
Typical values for the radial breathing mode of state of the art
structures are \cite{Anetsberger08}: $\omega_R=2\pi\!\times\!50$MHz,
$m_R= 10^{-11}$Kg, $S=\pi/4\!\times\!(0.5\mu)^2$, and
$\tilde{u}_{R,z}(0)=1/2$, which (for an Si substrate) result in
$Q\approx 4\times10^3$. In addition to $\sqrt{S}k_R\ll 1$, the above
approximation assumes that there is perfect impedance-match between
the pedestal and the substrate, and that $\sqrt{S}\lesssim h$ --- here
$h$ is the smallest characteristic dimension of the resonator. It can
be argued that for pedestals with lengths at least comparable to
$2\pi/k_R$ deviations from this adiabatic scenario will only increase
the $Q$-value provided that $\omega_R$ does not coincide with a
resonance of the pedestal. The derivation of Eqs.~(\ref{genI}) and
(\ref{toroI}) is given in Appendix \ref{ap:toro}.

\section{Basic model and outline}\label{II}

\subsection{Normal modes of the supports, the beam, and the whole
structure}\label{sub:modes}

We will make two simplifying assumptions: ($i$) the supports have
typical dimensions that are much larger than the beam's length $L$ and
for a bridge both of them are identical, and ($ii$) both the supports
and the beam present reflection symmetries with respect to the $x z$
and $y z$ planes (cf.~Fig.~\ref{fig:beams}). The first assumption is
adequate to describe a wide range of realistic structures \footnote{If
there are support dimensions $d_s\lesssim d$ our derivations for the
lowest order in $d/L$ remain valid provided the limit $d\to0$ yields a
support satisfying assumption ($i$). Implicit in the latter is the
idealization that for a bridge the two supports are effectively
disconnected for phonons with wavelengths $\lesssim 2\pi/L$.}. The
second assumption is a theoretical simplification and has no impact on
our main results [Eqs.~(\ref{eq:spectralrho}), (\ref{eq:spectraltau})
and the first two columns of Table \ref{table}] as will be borne out
in App.~\ref{ap:exten}. These assumptions allow us to model the
supports as elastic half-spaces.

Thus, the whole structure presents reflection symmetries with respect
to the $x$ and $y$ axes that are associated, respectively, with the
operators $\hat R_x$ and $\hat R_y$ acting on the space of solutions
of the elastic wave equations. This allows us to find normal modes of
the whole structure that are eigenvectors of $\hat R_x$ and $\hat
R_y$. For a given normal mode $|u\rangle$ one can generate modes
$|u_{++} \rangle$, $|u_{+-}\rangle$, $|u_{-+} \rangle$, and $|u_{--}
\rangle$ with the desired reflection properties by the following
symmetrization procedure
\begin{equation}\label{eq:sym}
| u_{\mu\nu} \rangle\! =\! \hat S_{\mu\nu} | u \rangle
 \!\equiv\!\frac{1}{2}\!\left(\! | u \rangle + \mu \hat R_x | u \rangle
 + \nu \hat R_y | u \rangle+ \mu \nu\hat R_x \hat R_y | u \rangle
\! \right)
\end{equation}
with $\mu,\nu = \pm$, which guarantees:
\begin{align}\label{refsym}
&\hat R_x | u_{\mu\nu} \rangle = \mu | u_{\mu\nu} \rangle \,,
\nonumber\\ &\hat R_y | u_{\mu\nu} \rangle = \nu | u_{\mu\nu} \rangle
\, .
\end{align}
We note that
\begin{equation}
| u \rangle = \frac{1}{2} \sum_{\mu,\nu=\pm} |u_{\mu\nu}\rangle \, .
\end{equation}

Furthermore as in any standard scattering problem \cite{LaxPhyllips},
we can choose for these modes $\{|u\rangle \}$ incoming scattering
states that present simple asymptotic behavior for $t\to -\infty$
corresponding to the different modes that can propagate in \emph{the
  support(s)} --- i.e. the free modes of an elastic half-space
\cite{Landau,Graff}. Thus for each type of mode with well defined
reflection properties, there will be four types of asymptotic
behavior, namely: longitudinal bulk waves ($l$), transverse bulk waves
polarized along the plane of incidence (SV waves), transverse bulk
waves polarized perpendicular to the plane of incidence (SH waves),
and Rayleigh surface waves ($s$) --- also known as SAW. In all four
cases the corresponding \emph{unsymmetrized} solutions for a free
elastic half space can be written in the form:
\begin{equation}
\bar u^{(0)}_{\bar q,\gamma} (\bar r) =
\frac{1}{\left(2\pi\right)^{d_\gamma/2}}\left[\bar \varepsilon_0 e^{i
\bar q \cdot \bar r} + A_l \bar \varepsilon_l e^{i \bar q_l \cdot \bar
r} + A_t \bar \varepsilon_t e^{i \bar q_t \cdot \bar r}\right] \, .
\end{equation}
where $q_z>0$ [cf.~Fig.~\ref{fig:beams}] and $\gamma=l$, SV, SH, $s$
labels the type of mode. The first term corresponds to the incident
wave and the last two to the reflected longitudinal ($l$) and
transverse ($t$) waves. The dimensionality $d_\gamma$, polarizations
$\bar\varepsilon$, wavevectors $\bar q_{l/t}$ and amplitudes $A_{l/t}$
(which depend on $\bar q$ and $\gamma$) corresponding to each of these
types of modes are given in Appendix \ref{ap:halfspace}.  In addition
for the bridge geometry there will be right- and left-movers and
therefore eight types of modes. Thus, each mode $\bar u^{(0)}_{\bar
  q,\gamma} (\bar r)$ of the support on the left with
$q_x,\,q_y,\,q_z\,>0$ will generate four incoming ``right-moving''
symmetrized normal modes of the whole structure $\bar u_{\bar
  q,\gamma,\mu,\nu,\mathrm{R}} (\bar r)$ --- with analogous relations
and definitions for the left-movers and in the case of a cantilever.
In the following, to simplify the notation, we will absorb the
discrete indices in the $q$ [i.e.~$\bar u_{\bar
  q,\gamma,\mu,\nu,\mathrm{R/L}} (\bar r)\to \bar u_{q} (\bar r)$]
unless otherwise stated. Of course, for the bridge geometry we can
exploit the reflection symmetry with respect to the $x$-$y$ plane
through the midpoint of the beam to obtain the left-movers from the
right-movers so that only the latter need to be calculated.

On the other hand \emph{the beam} can be viewed essentially as a
phonon waveguide. Thus it presents a series of branches associated
with the size quantization of the transverse wavevector. An ubiquitous
feature of these systems is that there are four branches that lack an
infrared cut-off: two bending branches (vertical and horizontal) with
quadratic dispersion relations, and a torsional and a compression
branch with linear dispersion relations \cite{Landau,Graff}. Each of
these branches corresponds to a given type of symmetrized modes
$|u_{\mu\nu}\rangle$ --- $|u_{++} \rangle$ for compression ($c$),
$|u_{+-}\rangle$ for vertical bending ($v$), $|u_{-+} \rangle$ for
horizontal bending ($h$) and $|u_{--} \rangle$ for torsion ($t$). As
we are interested only in low frequency phonons these are the only
modes that can propagate inside the beam.  Henceforth we will identify
the discrete indices $\mu,\nu$ of the normal modes $\bar u_{q} (\bar
r)$ with the corresponding branch inside the beam by introducing a
suitable branch index $\beta=c,t,v,h$ (i.e.~$\{\mu,\nu\}\to\beta$). As
already discussed, to describe these branches at low frequencies one
can resort to TRE \cite{Landau}, that consists of an approximation in
which the small parameter is $kd$. More precisely it corresponds to
taking the lowest order terms in an expansion in $k d$ of the
transverse profile of the mode. The associated displacements are given
in Appendix \ref{ap:beam}. For a finite length beam, in addition to
these propagating solutions, it is necessary to also consider the role
of exponentially decaying evanescent solutions. These are of two
types: end corrections that decay over a length scale of order $d$ and
solutions associated with the bending branches with a decay length $1/k$
--- the latter can also be treated within TRE.

As we are interested in properties of the beam's motion our main task
will be to determine $\bar u_q(\bar r)$ inside the beam (and a
distance $\gg d$ away from the junctions). This problem is analogous
to the one of finding the electromagnetic field inside a Fabry-P\'erot
interferometer. Thus it can be approached by considering how the
elastic waves that can propagate inside the beam are reflected at the
abrupt junctions (and at the free end for the cantilever) and then
adding the infinite reflections. In the cases of the bending branches
for $k \lesssim 1/L$ it is clear that the real exponentials are not
simply end corrections and have to be taken into account in this
resummation process. Hence there are two ways of propagating for the
bending modes. This reduction of the problem to the reflection at a
single junction is presented in Section \ref{s:FP}.  In addition one
needs to consider how a wave incident from the support is transmitted
into the beam. These problems of reflection and transmission at a
single junction are tackled in Sec.~\ref{s:singlejunc}.

Heuristically, for very long wavelengths the abrupt junction is seen
from the inside of the beam as a clamped boundary condition, and from
the outside (i.e.~from the support) as a free boundary
condition. Let us consider for example the problem of reflection from
the inside. If one adopts as an approximation for the displacements
inside the beam the solution that corresponds to a clamped boundary
condition $\bar u_* (\bar r)$ it can be proved that the associated
relative error is at most of order $k d$
(cf.~Subsec.~\ref{sub:ustar}). An analogous treatment can be done for
the problem of transmission into the beam from the outside. It will be
proven in Subsec.~\ref{sub:rec} that the associated transmission
amplitudes can be exactly related by a reciprocity argument to the
total transmission coefficient into the support $\tau(\omega)$
associated with reflection from the inside of the beam. This and the
aforementioned approximation by free boundary conditions
(cf.~Subsec.~\ref{sub:ustar}) will allow us to establish simple
expressions for the functions $\tilde{\rho}(\omega)$ and
$\tau(\omega)$ valid to lowest order in $k d$.

Finally, in Sec.~\ref{s:spec} we will relate the environmental
spectral density $I(\omega)$ for each resonance with the normal modes
$\bar u_q (\bar r)$ and establish the fundamental relations
(\ref{eq:spectralrho}), (\ref{eq:spectraltau}) between the functions
$I(\omega)$ and $\tilde{\rho}(\omega)$, $\tau(\omega)$. As will be
discussed in the following Subsection, their derivation involves the
use of a canonical transformation that relates a closed quantum system
``scattering representation'' for the ``mechanical Fabry-P\'erot''
(furnished by the supported beam) with an open quantum system
``resonator-bath representation'' that singles out one of its
resonances.

\subsection{Closed system vs.~open system}\label{IIB}

To each normal mode $\bar u_q (\bar r)$ with frequency $\omega(q)$ we
can associate a bosonic annihilation operator $b(q)$. Thus, the normal
ordered Hamiltonian operator for the whole structure considered as a
closed quantum mechanical system adopts the simple form
\begin{equation}\label{eq:h-d}
H= \int_q \hbar \omega(q) b^\dag(q) b^{\vphantom{\dag}}(q) \, .
\end{equation}
Here $\int_q$ denotes $\int_0^\infty\ud q_x\ud q_y\ud q_z$ and
summation over the appropriate discrete indices detailed in
Subsec.~\ref{sub:modes}. In this diagonal representation the
resonances will emerge as poles of the propagator for the displacement
field
\begin{equation}\label{eq:u}
  \hat{\bar u} (\bar r) =\int_q \sqrt{\frac{\hbar}{2\rho_s\omega(q)}}\bar
  u_q (\bar r) b^\dag(q)+\text{H.c.}\,,
\end{equation}
and in this sense may be regarded as ``derived objects''. It is
important to note that when the beam is characterized by a different
density $\rho_b\neq\rho_s$ the above eigenmodes of the whole structure
are orthonormal in a non-trivial metric which defines the scalar
product that will be used in general (except in
Sec.~\ref{s:singlejunc} and Apps.~\ref{ap:beam} and \ref{ap:exten}):
\begin{equation}\label{eq:overlap}
  \langle v|u \rangle \equiv \int\!\!\ud {r}^3
  \,\frac{\rho(\bar r)}{\rho_s}\,\bar v^*(\bar r)\cdot \bar
  u^{\vphantom{*}}(\bar r) \,.
\end{equation}

On the other hand in the context of the applications already discussed
(cf.~Sec.~\ref{s:intro}), where the system is driven out of
equilibrium, it is normally convenient to isolate explicitly the
relevant degrees of freedom of the nanoresonator and treat them as an
open quantum system. In the simplest scenario there is one resonance
of interest so that the goal is to find a ``resonator-bath''
representation in which the Hamiltonian adopts the form
\begin{align}\label{eq:h-rb}
  H = \ &\hbar \omega^{\vphantom{'}}_R {b'_R}^{\!\!\!\dag}
  {b'_R}
  + \hbar \left({b'_R}^{\vphantom{\dag}} + {b'_R}^{\!\!\!\dag}\right)
  \int_{q} \left[
    \zeta(q)  b'(q) + \text{H.c.} \right] \nonumber \\
& + \int_{q} \hbar\omega(q)  {b'}^\dag\!(q) b'(q)   \,,  
\end{align}
with off-diagonal couplings $\zeta(q)$ that are sufficiently weak so
that the renormalization of the bare frequency $\omega_R$ is smaller
than the characteristic spacing between the resonances --- i.e.~the
free spectral range of the mechanical Fabry-P\'erot. In such a
representation the relevant resonance will correspond to the degree of
freedom described by the operators $b'_R$, ${b'_R}^{\!\!\!\dag}$ ---
which annihilate or create a quanta in the resonator mode $\bar u'_R
(\bar r)$ --- and in this sense can be viewed as a ``fundamental
object''. The discrete mode $\bar u'_R (\bar r)$ should be real,
localized in the beam, have finite averaged elastic energy density per
unit amplitude given by $\rho_s\,\omega^2_R/2$, and satisfy the
elastic wave equation inside the beam. In turn the $b'(q)$ will be
annihilation operators for a continuum of modes $\bar u'_q (\bar r)$
[the environment] that have support in the whole structure and are
labeled after the $\bar u_q (\bar r)$ with which they are in a one to
one correspondence \footnote{Note that the new set of modes $\{\bar
  u'_R (\bar r),\bar{u}'_{q}(\bar r)\}$ is chosen to be also orthonormal in
  the metric defined by Eq.~(\ref{eq:overlap}) --- which accounts for
  the discrete mode's elastic energy density being proportional to
  $\rho_s$ rather than $\rho_b$.}. More precisely, the classical state
$\bar u'_{q}(\bar r)$, evolved with the Hamiltonian density that
corresponds to setting $\zeta(q)=0$ in Eq.~(\ref{eq:h-rb}), and the
classical state $\bar u_q(\bar r)$, evolved with the elastic wave
equation for the whole structure, will share the same ``free''
asymptotic behavior for $t \to -\infty$. The form of the off-diagonal
term in which the environment only couples to the canonical coordinate
of the resonator $\hat{X}_R$ reflects the fact that the underlying
phonon Hamiltonian is diagonal in the momenta. This, after integrating
out the resulting equations of motion for the $b'(q)$ allows to obtain
the well known Equation \footnote{More precisely, its analog with
  $\omega_R^2\to\omega_R[\omega_R-\Delta_I(0)]$
  [cf.~Eqs.~(\ref{eq:I}), (\ref{eq:delta}) and Eq.~(\ref{omegaR})].
  This correction corresponds to the environment-induced
  renormalization of the resonator's frequency that does not cancel
  out given the lack of counterterm in Hamiltonian (\ref{eq:h-rb}).}
(\ref{motion}).

The Hamiltonian (\ref{eq:h-rb}) is just an instance of the thoroughly
studied problem of a single harmonic oscillator linearly coupled to a
field \cite{Hu92}. The dynamics of the resonator mode is completely
determined by the following environmental spectral density:
\begin{equation}\label{eq:I}
  I(\omega) \equiv 2 \pi \int_q |\zeta(q)|^2 \delta\left[\omega
    -\omega(q)\right] \, .
\end{equation}
This problem is integrable given that $H$ can be diagonalized by a
linear canonical transformation \cite{Ullersma1,Hakim85} that would
allow to determine the $\bar u_q (\bar r)$ given $\bar u'_q (\bar r)$,
$\bar u'_R (\bar r)$, $\omega_R$ and $\zeta(q)$. We are simply faced
with the inverse problem: the derivation of Hamiltonian
(\ref{eq:h-rb}), with $\bar{u}'_R (\bar r)$ and $\{\zeta(q)\}$
satisfying the aforementioned properties, from Hamiltonian
(\ref{eq:h-d}) \emph{given the normal modes} $\{\bar u_q (\bar r)\}$.
The solution is not unique \cite{Viviescas03} but quantities of
physical interest --- e.g.~the resonator's frequency dressed by the
environment, its $Q$-value and its relative oscillator strength in
physical spectra --- will remain invariant under different choices.

Our analysis of the long wavelength behavior of the propagation of
elastic waves inside the resonator (cf.~Subsec.~\ref{sub:ustar}) and
the fact that the resonances of interest have low frequencies that
correspond to wavelengths much larger than $d$, prompts us to choose
for $\bar u'_R (\bar r)$ an eigenmode of the elastic wave equation
satisfying clamped-clamped (clamped-free) boundary conditions for the
bridge (cantilever) geometries. It is straightforward to prove, using
the expression of the elastic energy density in terms of the
displacement field \cite{Landau,Graff}, that for this choice
$\omega_R$ will be given by the corresponding eigenvalue \footnote{We
note that while $\bar u'_R (\bar r) e^{-i \omega_R t}$ is only a
solution of the elastic wave equation inside the resonator, the
spatial average of the elastic energy density should be taken over the
whole structure when considering Hamiltonian $H$.  However, the
boundary term that arises due to the discontinuity at the junction
vanishes due to the choice of homogeneous boundary conditions. The
reader is referred to Ref.~[\onlinecite{Viviescas03}] where a related
treatment is given of the electromagnetic analog of our problem.}.  On
the other hand the diagonal representation afforded by the normal
modes $\bar u_q (\bar r)$ directly yields:
\begin{equation}\label{eq:fR}
\omega_R^2=\int_q \omega^2(q) |\langle u_q|u'_R \rangle|^2\,.
\end{equation}

The required canonical transformation from the diagonal representation
(\ref{eq:h-d}) to the ``resonator-bath'' representation of choice is
specified by:
\begin{align}\label{eq:tran}
  b(q) =& \frac{1}{2} \left[ \sqrt{\frac{\omega(q)}{\omega_R}} +
    \sqrt{\frac{\omega_R}{\omega (q)}}\right] \langle u_q | u'_R
  \rangle\, b'_R \nonumber
  \\
  & + \frac{1}{2} \left[ \sqrt{\frac{\omega(q)}{\omega_R}} -
    \sqrt{\frac{\omega_R}{\omega (q)}}\right] \langle u_q | u'_R
  \rangle \, {b'_R}^{\!\!\!\dag} \nonumber
  \\
  & + \int_{q'} \frac{1}{2} \left[ \sqrt{\frac{\omega(q)}{\omega
        (q')}} + \sqrt{\frac{\omega(q')}{\omega (q)}} \right] \langle
  u^{\vphantom{'}}_q | u'_{q'} \rangle \, b'(q') \nonumber
  \\
  & + \int_{q'} \frac{1}{2} \left[ \sqrt{\frac{\omega(q)}{\omega
        (q')}} - \sqrt{\frac{\omega(q')}{\omega (q)}} \right] \langle
  u^{\vphantom{'}}_q | {u'_{q'}}^{\!\!\!*} \rangle \, {b'}^\dag(q')\,,
\end{align}
and the analogous relation implied for the $b^\dag(q)$, with the
spatial overlap $\langle u^{\vphantom{'}}_q | u'_{q'}\rangle$ given by:
\begin{align}\label{eq:overqq}
  \langle u^{\vphantom{'}}_q | u'_{q'}\rangle =\,&\delta(q-q')
  +\frac{\langle u^{\vphantom{'}}_q | u'_R \rangle \langle u'_R |
    u_{q'}\rangle}{\Delta_S[\omega(q')]-i S[\omega(q')]} 
\nonumber \\  & \times 
\frac{1}{\omega^2(q)- \omega^2(q') - i \epsilon} \,.
\end{align}
Here we have defined, for any smooth function $f(\omega)$,
\begin{equation}\label{eq:delta}
  \Delta_f(\omega) \equiv \frac{1}{\pi}\; \mathscr{P}
  \!\!\int_0^\infty \!\!\ud \omega' \frac{2\omega'}{\omega^2-
    {\omega'}^2} \, f(\omega') \,, 
\end{equation}
and introduced the function
\begin{equation}\label{eq:S}
 S(\omega)\equiv \frac{\pi}{2\omega}\int_q |\langle u'_R | u_q \rangle|^2
 \delta[\omega-\omega(q)]\,.
\end{equation}
We note that this definition remains invariant if we replace the $\bar
u_q(\bar r)$ by any other basis of normal modes of the whole
structure, so that the function $S(\omega)$ is a property of the
resonator mode. In fact, it corresponds for $\omega>0$ to the spectral
density of the resonator mode's canonical coordinate at zero
temperature. The choice of $-i\epsilon$ in Eq.~(\ref{eq:overqq})
yields the desired asymptotic condition for $t \to -\infty$, already
discussed. The normalization of the resonator mode and
Eq.~(\ref{eq:fR}) imply two sum rules
\begin{equation}\label{eq:intS}
\int^\infty_0\!\!\!\omega S(\omega)\ud\omega=\frac{\pi}{2}\quad\
\text{and}\quad\ \int^\infty_0\!\!\!\omega^3
S(\omega)\ud\omega=\frac{\pi}{2}\omega^2_R \,,
\end{equation}
respectively, which will prove useful below. Here we have used that
for any smooth function $f(\omega)$
\begin{equation}\label{eq:ident}
\int_q |\langle u'_R | u_q \rangle|^2 f\left[ \omega(q) \right] =
\int^\infty_0 \! \ud \omega \frac{2\omega}{\pi}S(\omega)f(\omega) \,.
\end{equation}
On the other hand the overlap $\langle u^{\vphantom{'}}_q |
{u'_{q'}}^{\!\!\!*}\rangle = \langle u'_{q'} | u^*_q \rangle$ can be
expressed in terms of $\langle u^*_{q'} | u^{\vphantom{*}}_q \rangle$
and the overlap (\ref{eq:overqq}) inserting $\openone=\int_q |u_q
\rangle \langle u_q|$. In turn, as complex conjugation corresponds to
time inversion, $\langle u^*_{q'} | u^{\vphantom{*}}_q \rangle$ is
just the $S$-matrix for the corresponding elastic wave classical
scattering problem \cite{LaxPhyllips}. The facts that the $S$-matrix
only mixes states with the same frequency and that the resonator mode
$\bar u'_R (\bar r)$ is real can then be used to establish:
\begin{widetext}
\begin{equation}\label{overstar}
  \left[\sqrt{\frac{\omega(q)}{\omega(q')}} -
    \sqrt{\frac{\omega(q')}{\omega(q)}} \right]\langle
    u^{\vphantom{'}}_q | {u'_{q'}}^{\!\!\!*}\rangle =
    \left[\sqrt{\frac{\omega(q)}{\omega(q')}} -
    \sqrt{\frac{\omega(q')}{\omega(q)}} \right] 
    \frac{\langle u^{\vphantom{'}}_q | u'_R \rangle \langle u'_R |
    u_{q'}^*\rangle}{\Delta_S [\omega (q')]+i S[\omega(q')]} \;
    \frac{1}{\omega^2(q) -\omega^2 (q') +i \epsilon}\,.
\end{equation}

To proceed, one can first prove that the inverse transformation has
analogous form given by
\begin{align}\label{eq:inv}
  b'_R=& \int_{q'} \frac{1}{2} \left[
    \sqrt{\frac{\omega_R}{\omega (q')}} +
    \sqrt{\frac{\omega(q')}{\omega_R}} \right] \langle u'_R|
  u_{q'} \rangle\, b(q') 
+ \int_{q'} \frac{1}{2} \left[ \sqrt{\frac{\omega_R}{\omega
        (q')}} - \sqrt{\frac{\omega(q')}{\omega_R}} \right]
  \langle u'_R | u^*_{q'} \rangle \, {b^\dag(q')} \,,\nonumber
  \\
  b'(q) =& \int_{q'} \frac{1}{2} \left[ \sqrt{\frac{\omega(q)}{\omega
        (q')}} + \sqrt{\frac{\omega(q')}{\omega (q)}} \right] \langle
  u'_q | u^{\vphantom{'}}_{q'} \rangle \, b(q') 
+ \int_{q'} \frac{1}{2} \left[ \sqrt{\frac{\omega(q)}{\omega
        (q')}} - \sqrt{\frac{\omega(q')}{\omega (q)}} \right] \langle
  u'_q | u^*_{q'} \rangle \, b^\dag(q') \,.
\end{align}
\end{widetext}
We have done this by showing that substitution of Eq.~(\ref{eq:tran})
and its counterpart for ${b}^\dag(q)$ into the RHS of
Eq.~(\ref{eq:inv}) yields the identity, which reduces to a
straightforward calculation with the help of
Eqs.~(\ref{eq:overqq})-(\ref{overstar}).  Subsequently, in an analogous
fashion, one can use Eq.~(\ref{eq:inv}) to prove that the $b'(q)$,
${b'}^\dag(q)$ also satisfy canonical commutation relations. Then,
substitution of Eq.~(\ref{eq:tran}) and its counterpart for
${b}^\dag(q)$ into the diagonal Hamiltonian (\ref{eq:h-d}) leads to
the desired ``resonator-bath'' form embodied in Eq.~(\ref{eq:h-rb})
with off-diagonal couplings given by:
\begin{equation}\label{zeta}
\zeta(q) =  \frac{\langle u'_R | u_q \rangle}{2 \sqrt{\omega_R
    \,\omega(q)}}  \frac{1}{\Delta_S [\omega(q)] - i S[\omega(q)]}\,.
\end{equation}
Finally, substitution of this result into the definition (\ref{eq:I}) for
$I(\omega)$ allows us to obtain
\begin{equation}\label{Iex}
I(\omega)= \frac{1}{\omega_R} \frac{S(\omega)}{\Delta_S^2
  (\omega) +S^2 (\omega)}
\end{equation}
for $\omega>0$. This together with Eq.~(\ref{eq:S}) reduces the
derivation of the environmental spectral density to the calculation of
the overlaps $\langle u'_R | u_q \rangle$ and implies that it is
entirely determined by the choice for the resonator mode. We note that
both the $\bar u_q (\bar r)$ that diagonalize the Hamiltonian
(\ref{eq:h-rb}) and the $\bar u'_q (\bar r)$ that solve the inverse
problem for a given $\bar u'_R (\bar r)$ are fixed once their
asymptotic behavior is specified.

It is worth noting that within the above \emph{exact} treatment the
$\{\zeta_q\}$ describe the coupling to an environment that --- barring
selection rules arising from the symmetries --- includes the other
high-$Q$ resonances $\{\omega_n\}$ of the beam coupled to the
supports. Thus the naive expectation that $I(\omega)$ be smooth on the
frequency scale $\omega_R$ will fail in a neighborhood of
$\omega=\omega_n\neq\omega_R$ where the environment is structured and
$I(\omega)$ may exhibit interference effects. This issue and the
natural ultraviolet cutoff for $I(\omega)$, which is set by the
transverse dimension $d$, will be discussed further in
Sec.~\ref{s:spec} and App.~\ref{ap:I}.

\section{Reduction to a single junction}\label{s:FP}

It is clear from the analysis in the previous section that we will
eventually need an expression for the overlaps $\langle u'_R | u_q
\rangle$.  As $\bar u'_R (\bar r)$ only has support inside the beam,
knowledge of the normal mode $\bar u_q (\bar r)$ is only needed in
that region.  Furthermore, it is straightforward to realize that the
end corrections (cf.~App.~\ref{ap:beam}) will only yield contributions
to $\langle u'_R | u_q \rangle$ that are higher order in $d/L$ so that
we may focus on the four branches that can propagate at low
frequencies.  As already discussed in Subsec.~\ref{sub:modes}, our
system can then be regarded as a Fabry-P\'erot interferometer for
elastic waves and for the bridge we may concentrate on right-movers.
It proves useful to explicitly separate the transmission amplitude
$t_q$ for propagation inside the beam that would ensue for the
corresponding problem of a single junction
(cf.~Sec.~\ref{s:singlejunc}); since, given our model of the supports
(cf.~Subsec.~\ref{sub:modes}), in the limit $d/L\to 0$ the
combination $\bar u_q (\bar r)/t_q$ for $\bar r$ inside the beam only
depends on $q$ via $\omega_q$ and the branch index $\beta$. For each
branch the latter frequency will correspond to a wavevector $k(q)$ for
propagation inside the beam \footnote{Note that we consider the limit
$d/L\to 0$ with $k(q)L$ held constant.} and, as described in
App.~\ref{ap:beam}, the mode profile $\bar u_q (\bar r)/t_q$ can be
expressed in terms of an effective one-dimensional field
$\phi_{k(q),\beta}(z)$ [cf.~Eq.~\ref{def:phi}] --- henceforth in this
Section we will omit the $q$-dependence and the index $\beta$.

In the cases of compression and torsion only traveling-wave solutions
intervene (cf.~App.~\ref{ap:beam}).  Therefore
\begin{equation}\label{modetors}
  \phi_k (z) = A_+(k) e^{i k z} + A_-(k) e^{- i k z} \, ,
\end{equation}
where we take $0<z<L$, and $A_\pm(k)$ are the amplitudes of the right
and left moving components of $\phi_k$ (cf.~Fig.~\ref{fig:beams}). In
order to determine the above amplitudes we use the usual procedure for
Fabry-P\'erot interference, i.e.~we derive the amplitudes from the
infinite sum of contributions generated by consecutive reflections at
the ends. We find that they obey the following linear system of
self-consistent equations:
\begin{equation} \label{Ators}
\begin{pmatrix}
 -1\  & r^{(L)} \\  r^{(R)}\ & -1\  
\end{pmatrix}
\begin{pmatrix}
 A_+\vphantom{r^{(R)}} \\ A_-\vphantom{r^{(R)}} 
\end{pmatrix}
=
\begin{pmatrix}
 -1\vphantom{r^{(R)}} \\ 0\vphantom{r^{(R)}} 
\end{pmatrix}  \, ,
\end{equation}
where we have defined $r^{(R/L)}$ as the amplitudes for reflection at
the right/left ends of the beam, and the first entry on the RHS
corresponds to the displacements generated by the incoming wave from
the support. This system of equations is valid for both the bridge and
the cantilever.  For the bridge geometry we can use the reflection
symmetry with respect to the $x$-$y$ plane through the midpoint of the
beam to prove that the reflection amplitudes are related by
$r^{(R)}=r^{(L)} e^{i 2 k L}$. Furthermore, given our choice of
origin, dimensional analysis implies that for a specific support
material and type of beam the reflection amplitude $r^{(L)}$ is just a
function of $kd$ (analogous considerations will apply below to
$r^{(L)}_{\delta\eta}$, with $\delta,\eta = A,B$, and to the ratio
$b$).

On the other hand, as discussed in Sec.~\ref{sub:modes} and
App.~\ref{ap:beam}, in the cases of the two bending branches one needs
to also consider evanescent solutions that decay over a length scale of
order $1/k$.  In order to include these in our analysis, we generalize
the above procedure. The functions $\phi_k$ corresponding to the
bending modes will have four contributions with respective amplitudes:
$A_\pm(k)$ for propagation to the right/left, and $B_\pm(k)$ for decay
to the left/right. Hence we write
\begin{equation}\label{modebend}
  \phi_k (z) = A_+(k) e^{i k z} + A_-(k) e^{- i k z} +  B_+(k) e^{k z} 
  + B_-(k) e^{-k z} 
\end{equation} 
In formal terms the resummation of the successive ``reflections'' can
be viewed as an iterative procedure in which the $n$th contribution
when added to the $n-1$th matches the boundary conditions at the
right/left ends for odd/even $n$ (here $n=1,2\ldots$ with $n=0$
corresponding to the solution for a single junction). Thus,
``reflection'' at one end of the beam will generate both types of
contributions, i.e.~there are now eight reflection amplitudes.
We can define, for example, $r_{AB}^{(L)}$ as the amplitude for a
propagating mode to be reflected at the left junction into a decaying
mode and similarly for the other amplitudes. Hence the system of
self-consistent equations obeyed by the amplitudes which specify
$\phi_k (z)$ generalizes to
\begin{equation}\label{Abend}
\begin{pmatrix}
 -1\  & 0\  & r^{(L)}_{AA}\ & r^{(L)}_{BA} \\[5pt]
0\  & -1\  &r^{(L)}_{AB}\  & r^{(L)}_{BB} \\[5pt]
r^{(R)}_{AA}\  & r^{(R)}_{BA}\  & -1\  & 0 \\[5pt]
r^{(R)}_{AB}\  & r^{(R)}_{BB}\  & 0\  & -1
\end{pmatrix}
\begin{pmatrix}
A_+\vphantom{r^{(L)}_{AA}} \\[5pt]
B_-\vphantom{r^{(L)}_{AA}}  \\[5pt]
A_-\vphantom{r^{(L)}_{AA}}  \\[5pt]
B_+\vphantom{r^{(L)}_{AA}}  
\end{pmatrix}
=
\begin{pmatrix}
-1\vphantom{r^{(L)}_{AA}}  \\[5pt] 
-b\vphantom{r^{(L)}_{AA}}  \\[5pt] 
0\vphantom{r^{(L)}_{AA}} \\[5pt] 
0\vphantom{r^{(L)}_{AA}} 
\end{pmatrix} \, ,
\end{equation}
where $b$ corresponds to the decaying contribution generated by the
incoming wave from the supports. More precisely it is defined as the
ratio of the decaying contribution's amplitude to $t_q$ for the
corresponding problem of transmission from the support into the beam
at a single junction (see Sec.~\ref{s:singlejunc}). For the bridge
geometry we can reduce the eight reflection amplitudes to four using
again the corresponding reflection symmetry.  This yields
$r^{(R)}_{AA}=r^{(L)}_{AA} e^{i 2 k L}$, $r^{(R)}_{AB}=r^{(L)}_{AB}
e^{-(1-i) k L}$, $r^{(R)}_{BA}=r^{(L)}_{BA} e^{-(1-i) k L}$, and
$r^{(R)}_{BB}=r^{(L)}_{BB} e^{-2 k L}$.

Finally, we note that if one allows the ratio $b$ to depend on $q$,
the results of this Section for the modes $\bar u_q(\bar r)$ inside
the beam (and a distance $\gg d$ away from the junctions) are
applicable to the extent that contributions of order $\exp[-L/d]$
arising from the end corrections are negligible so that
Eqs.~(\ref{modetors})-(\ref{Abend}) are completely general for
$k(q)d\ll1$.

\section{Transmission through a 3D--1D junction}\label{s:singlejunc}

\subsection{Approximation by clamped and free boundary conditions}\label{sub:ustar}

We now turn to the analysis of the small $kd$ behavior of the
transmission amplitudes $t_q$, $b t_q$ and reflection amplitudes
$r^{(L)}_{c/t}$, $r^{(L)}_{v/h,\delta\eta}$ (with $\delta,\eta = A,B$)
that characterize a single 3D-1D abrupt elastic junction
(cf.~Subsec.~\ref{sub:modes}). To this effect we consider incoming
eigenstates that present simple asymptotic behavior for $t \to
-\infty$. In the case when the incoming wave is incident from the
support these eigenstates $|u_q\rangle$ will correspond to the
different free modes of an elastic half-space $\bar u^{(0)}_q (\bar
r)$ already discussed in Section \ref{sub:modes}.  Alternatively, for
incidence from the beam, they will correspond to the left-movers of an
infinite beam. Once again we consider modes with well defined
reflection symmetries. This implies for the eigenmodes incident from
the support the same labels $q,\gamma,\beta$ as for the modes of the
whole structure while the modes incident from the beam are specified
by $k,\beta$.

We focus on a specific eigenmode $\bar u (\bar r)$ and for simplicity
omit its labels. \emph{Inside the beam} [i.e.~for $z\geq 0$ and
$(x,y)$ in a beam's cross section] we decompose $\bar u (\bar r)$ into
\begin{equation}\label{eq:ustar}
\bar u (\bar r) \equiv \Delta\bar u_+ (\bar r) + \bar u_* (\bar r)
\end{equation}
where $\bar u_* (\bar r)$ is an approximation to the displacement
field specified by taking at $z=0$: ($i$) clamped boundary conditions
when the incoming wave \emph{incides from the beam} or ($ii$)
displacements specified by the corresponding solution for the free
elastic half-space [i.e.~$\bar u_* (x,y,0)=\hat S\,\bar u^{(0)}
(x,y,0)$] for \emph{incidence from the support}
[cf.~Eq.~(\ref{eq:sym}) and App.~\ref{ap:halfspace}]. The problem of
finding $\bar u (\bar r)$ inside the beam can then be formulated as an
integral equation for $\Delta\bar u_+ (\bar r)$:
\begin{equation}\label{eq:deltau}
\Delta\bar u_+ (\bar r) = \int_{S}\ud {r'}^2 \mathbf{G} (\bar r-\bar
r',\omega) \cdot \mathbf{F} \cdot \left[ \Delta \bar u_+ (\bar r') +
\bar u_* (\bar r') \right] \,,
\end{equation}
where $S$ is the beam's cross section at the origin, $\bar
r=(x,y,0)\in S$, and $z'\geq 0$. Here $\mathbf{F}$ and $\mathbf{G}$
are second rank tensors. The former is given by the linear
differential operator that maps $\bar u(\bar r')$ onto $\hat z\cdot
\bv{\sigma}(\bar r')$, where $\bv{\sigma}(\bar r')$ is the induced
stress tensor, and the latter by the retarded Green's function of the
free elastic half-space harmonically forced at its boundary
\cite{Graff}. More precisely, $G_{ij}(\bar r,\omega)$ is defined as
the $i$-th component of the outgoing displacement field generated at
point $\bar r$ by the harmonic stress source [frequency $\omega(k)$]
with amplitude specified at $z'=0$ by:
$\sigma_{jz}(x',y',0)=\delta(x')\delta(y')$, $\sigma_{lz}(x',y',0)=0$
for $l\neq j$. In order to establish that Eq.~(\ref{eq:deltau})
specifies the solution for $z>0$ we just need to show that the
expression (\ref{eq:ustar}) satisfies continuity of the displacement
and the stress with a solution in the half-space ($z<0$) having the
appropriate asymptotic boundary conditions. We define $\Delta\bar u_-
(\bar r)$ as the extension of the RHS of Eq.~(\ref{eq:deltau}) for
$z<0$ and arbitrary $x,y$. In case ($i$) this function directly gives
the required solution in the support, while in case ($ii$) the latter
is afforded by $\hat S \bar u^{(0)}(\bar r)+\Delta\bar u_- (\bar r)$.
In both cases the continuity of the stress follows trivially by
construction whereas Eq.~(\ref{eq:deltau}) enforces the continuity of
the displacement. It is understood that both $\Delta\bar u_+ (\bar r)$
and $\bar u_* (\bar r)$ are linear superpositions of the low frequency
harmonic solutions $|v^{(m)}_\beta(k)\rangle$ of a semi-infinite
elastic beam (cf.~App.~\ref{ap:beam}) with frequency $\omega(k)$
(specified by $k(q)$ for incidence from the support). Thus $\bar u
(\bar r)$ satisfies the elastic wave equation for $z>0$ and the only
traveling wave contributing to $\Delta\bar u_+ (\bar r)$ is the
right-mover corresponding to $k,\beta$. This yields
\begin{align}
\Delta\bar u_{+,\beta} (\bar r,k,d) =&\, \sum_m c_{\beta
  m}^{\vphantom{(*)}} (k,d) \nonumber \\ & \times \bar A_{\beta m}
  (x,y,d,\kappa_{\beta m}) e^{-\kappa_{\beta m}(k,d)z} \,, \nonumber
\end{align}
\begin{align}
  \bar u_{*,\beta} (\bar r,k,d) =& \,\bar u_{\mathrm{in},\beta} (\bar
  r, k,d) + \sum_m c_{\beta m}^{(*)} (k,d) \nonumber \\ & \times \bar
  A_{\beta m} (x,y,d,\kappa_{\beta m}) e^{-\kappa_{\beta m}(k,d)z} \,.
\label{eq:decomp}
\end{align}
Here we have reintroduced the ``symmetry index'' $\beta$
and eliminated the frequency in favor of $k$. The amplitudes in the
different harmonic solutions are given by $c_{\beta,m}(k,d)$ and
$c^{(*)}_{\beta,m}(k,d)$; while
$\bar{A}_{\beta,m}[x,y,d,\kappa_{\beta,m}(k,d)]$ and
$\kappa_{\beta,m}(k,d)$ are, respectively, the corresponding
transverse profiles and complex wavevectors of the latter. The
displacement field incident from the beam $\bar u_{\mathrm{in},\beta}
(\bar r,k,d)$ vanishes for case ($ii$) whereas for case ($i$) it is
given by $\bar A_{\beta,0}[x,y,d,ik] e^{-ikz}$ for \emph{propagating
  modes} and $\bar A_{v/h,1}[x,y,d,-\kappa_{v/h,1}]
e^{\kappa_{v/h,1}z}$ for the large decay length exponentials
associated with the bending branches (\emph{decaying modes}). This
``incident displacement'' and the terms with $m=0$ and with
$\beta=v/h, m=1$ yield the TRE part of the solution, while $\beta=c,t$
with $m>0$ and $\beta=v,h$ with $m>1$ correspond to the end
corrections. These are characterized by $\Re[\kappa_{\beta,m}(k,d)]
\gtrsim 1/d$ whereas
\begin{equation}
\kappa_{\beta,0}(k,d) \equiv -i k\, \qquad \kappa_{v/h,1}(k,d) = k
\left(1 + \mathcal{O} [k d] \right)\,.
\end{equation} 

To extract the small $kd$ behavior of the transmission and reflection
amplitudes we first prove that
\begin{equation}\label{eq:approxu}
  \frac{\left|\Delta\bar u_+ (\bar r)\right|}{\left|\bar u (\bar
  r)\right|} \lesssim \mathcal{O} [k d] \quad \text{for} \quad z\gg d\,.
\end{equation}
For this analysis we eliminate $k$ and $z$ in favor of $kd$ and $kz$,
respectively, which are then treated as independent variables
(henceforth we omit the resulting $d$ dependence of non-dimensionless
amplitudes).  If one substitutes Eq.~(\ref{eq:ustar}) into
$|\Delta\bar u_+ (\bar r)|/|\bar u (\bar r)|$ one can deduce that
\begin{equation}\label{eq:approxustar}
 \frac{\left|\Delta\bar u_+ (\bar r)\right|}{\left|\bar u_* (\bar
  r)\right|} \lesssim \mathcal{O} [k d] \ \Rightarrow \
  \frac{\left|\Delta\bar u_+ (\bar r)\right|}{\left|\bar u (\bar
  r)\right|} \lesssim \mathcal{O} [k d]\,,
\end{equation}
so that it suffices to prove the LHS of Eq.~(\ref{eq:approxustar}).
Furthermore, for $z\gg d$ the contributions of the end corrections are
exponentially suppressed so that it suffices to analyze the TRE
amplitudes [cf.~Eq.~(\ref{eq:decomp})]. In fact it can be proved that
the latter satisfy
\begin{align}
\left| c^{(*)}_{\beta,m}(kd)\right|\,\, & \lesssim\,\, \begin{cases}
    \mathcal{O} [kd] & \text{for case ($ii$) and}\ \beta=t \\
    \mathcal{O} [1] & \text{otherwise}
  \end{cases}  \label{eq:limcstar}\\
  \left| c_{\beta,m}(kd)\right|\,\, & \lesssim\,\, \begin{cases}
    \mathcal{O} \left[(kd)^2\right] & \text{for case ($ii$) and}\
    \beta=t \\ \mathcal{O} [k d] & \text{otherwise}
\end{cases} \label{eq:limc}
\end{align}
which then directly imply the LHS of Eq.~(\ref{eq:approxustar}).

First we establish the behavior (\ref{eq:limcstar}) of the starred
amplitudes which amounts to a rigorous derivation of the recipes used
in TRE to specify the boundary conditions for the effective one
dimensional field $\phi(z)$. To this effect we substitute
Eq.~(\ref{eq:decomp}) into
\begin{equation}\label{boundstar}
\bar u_{*,\beta} (x,y,0,k,d) =\begin{cases} 0 & \text{for case ($i$)}
  \\ \hat S_{\beta} \bar u^{(0)}_q (x,y,0) & \text{for case ($ii$)}
\end{cases} 
\end{equation}
and take on both sides the spatial averages $\langle\ldots\rangle_{S}$
and $\langle\ldots\rangle_{\mathrm{ang},S}$ that correspond,
respectively, to the displacement of the center of mass and the
spatially averaged angle $\bar \theta$ [as defined in
App.~\ref{ap:beam}; Eqs.~(\ref{ang}), (\ref{angnotation})] for the
cross section $S$ (henceforth we omit the latter label). If we consider
each component separately we have six equations that are linear in the
amplitudes $c^{(*)}_{\beta,m}(kd)$ with inhomogeneous terms arising
from $-\bar u_{\mathrm{in},\beta} (x,y,0,k,d)$ in case ($i$) and $\hat
S_{\beta}\bar{u}^{(0)}_q \!(x,y,0)$ in case ($ii$). The reflection
symmetries imply that in the cases of compression and torsion,
respectively, only the equation corresponding to $\langle u_z \rangle$
and to $\theta_z$ does not vanish trivially, while for vertical
(horizontal) bending the same applies to the two equations provided by
$\langle u_x \rangle$ ($\langle u_y \rangle$) and $\theta_y$
($\theta_x$). In each of these equations we solve for the TRE
amplitudes in terms of the end corrections and the inhomogeneous
terms. The small $kd$ behavior of the resulting expressions can be
extracted using the following properties of the modes' transverse
profiles:
\begin{align}
\langle A_{z,c m}\rangle \approx & 
\begin{cases}
\frac{1}{\sqrt{2\pi S}} & \text{for}\ m=0 \\ 
\mathcal{O}\left[(k d)^{2}\right] & \text{otherwise} 
\end{cases} \nonumber \\ 
\langle A_{x/y,v/h m}\rangle \approx & 
\begin{cases} 
\frac{1}{\sqrt{2\pi S}} & \text{for}\ m=0,1 \\ 
\mathcal{O}\left[(k d)^{4}\right] & \text{otherwise}
\end{cases} \nonumber \\  
\theta_{z,t m} \approx & 
\begin{cases} 
\frac{1}{\sqrt{2\pi I_z}} & \text{for}\ m=0 \\ 
\mathcal{O}\left[(k d)^{2}\right] & \text{otherwise}
\end{cases} \nonumber \\  
\frac{\theta_{y/x,v/h m}}{kd} \approx & 
\begin{cases}
\pm\frac{\left(i\right)^{m+1}}{d\sqrt{2\pi S}} & \text{for}\ m=0,1
\\ 
\mathcal{O}\left[(k d)^{3}\right] & \text{otherwise}
\end{cases} 
\nonumber\\ 
\left|\langle\int_{S}\ud {r'}^2 \mathbf{G} \left(\bar r - \bar r',
  \right.\right. & \left.\left.\!\!\!\omega[k]\right) \cdot
  \mathbf{F}\!\cdot
  \left\{\bar{A}_{\beta,m}\left[x',y',d,\kappa_{\beta,m}(k,d)\right]
  \right.\right.  \nonumber\\ \left.\left. \times
    e^{-\kappa_{\beta,m}(k,d)z'} \right\}\rangle\right.&
\left.\!\!\right| \,\approx\, \begin{cases} 
    \mathcal{O}\left[(kd)^{p_\beta}\right] &  \text{for}\
    \beta,m\!\in\!\mathrm{TRE} 
\\ \mathcal{O}\left[(kd)^{2p_\beta}\right] &  \text{otherwise} 
\end{cases} \nonumber
\end{align}
\begin{align}\label{eq:proptrans}
\left|\langle \int_{S}\ud {r'}^2 \mathbf{G} \left(\bar r - \bar
r',\right.\right. & \left.\left.\!\!\!\omega[k]\right) \cdot
\mathbf{F}\!\cdot
\left\{\bar{A}_{\beta,m}\left[x',y',d,\kappa_{\beta,m}(k,d)\right]
\right.\right.  \nonumber\\ \left.\left. \times
e^{-\kappa_{\beta,m}(k,d)z'} \right\}\rangle_\mathrm{ang}\right.&
\left.\!\!\right| \,\approx \, \begin{cases}
\mathcal{O}\left[(kd)^{p_\beta}\right] & \text{for}\
\beta,m\!\in\!\mathrm{TRE}  
\\ \mathcal{O}\left[(kd)^{2p_\beta}\right] & \text{otherwise}
\end{cases} 
\end{align}
which imply that the contributions of the end corrections scale at
most as the inertia (i.e.~as $\omega^2$). The above
Eqs.~(\ref{eq:proptrans}) follow directly from the universal
properties of the end corrections discussed in App.~\ref{ap:beam}
(Subsection \ref{sub:endcorr}) and the small $kd$ behavior of the TRE
solutions \footnote{For the last two properties, in the case of the
  TRE solutions, we have also used that $|\bar r| \mathbf{G}(\bar r -
  \bar r',\omega)$ is well behaved as $\omega\to 0$.} (cf.~Subsection
\ref{sub:tre}). Thus, with the help of the Taylor expansion of $\hat
S_{\beta}\bar{u}^{(0)}_q \!(x,y,0)$ at the origin, Eq.~(\ref{eq:sym}),
and Eq.~(\ref{angapprox}) we obtain for compression and torsion:
\begin{align}
c^{(*)}_{c,0}(0) & = \begin{cases} -1 & \text{for case ($i$)} \\
  2\sqrt{2\pi S} {u}^{(0)}_{z,q} (0) & \text{for case ($ii$)}
\end{cases} \nonumber 
\end{align}
\begin{align}
c^{(*)}_{t,0}(kd) & \approx \begin{cases} -1 & \text{($i$)} \\
    \sqrt{2\pi I_z}\left[\hat z \cdot \nabla \times \bar u^{(0)}_q(0)
      + 2\gamma_z u^{(0)}_{xy,q}(0)\right] & \text{($ii$)}
\end{cases}\label{eq:limcsct}
\end{align}
where $\gamma_z\equiv (I_y-I_x)/I_z$ and $u_{ij}$ is the strain
tensor. In the case of the bending branches for case ($i$) we get:
\begin{align}\label{eq:limcsvh1}
c^{(*)}_{v/h,0}(0) & = \begin{cases} i & \text{for propagating}\ \bar
  u_{\mathrm{in},\beta}  \\ -(1-i) & \text{for decaying}\ \bar
  u_{\mathrm{in},\beta} 
\end{cases} \nonumber\\
c^{(*)}_{v/h,1}(0) & = \begin{cases} -(1+i) & \text{for propagating}\
  \bar u_{\mathrm{in},\beta} \\ -i & \text{for decaying}\ \bar
  u_{\mathrm{in},\beta} 
\end{cases} 
\end{align}
while for case ($ii$) we obtain: 
\begin{equation}\label{eq:limcsvh2}
\begin{bmatrix} 1 & 1 \\[5pt] \pm i & \mp 1 \end{bmatrix} 
\cdot 
\begin{bmatrix} c^{(*)}_{v/h,0}(0) \\[5pt] c^{(*)}_{v/h,1}(0) \end{bmatrix} 
= \begin{bmatrix} 2\sqrt{2\pi S} {u}^{(0)}_{x/y,q}(0) \\[5pt] 0
\end{bmatrix} \,, 
\end{equation}
where the first equation corresponds to the average displacement
$\langle u_{x/y}\rangle$ and the second one to the angle
$\theta_{y/x}$. The fact that the latter does not contribute to lowest
order in this case is a consequence of the linear versus quadratic
dispersion relations that characterize the propagation of the relevant
modes ($\beta=v,h$) in the support and the beam,
respectively. Equation (\ref{eq:limcsvh2}) yields 
\begin{equation}\label{eq:limcsvh2s}
c^{(*)}_{v/h,0}(0) = \sqrt{2\pi S}(1-i) {u}^{(0)}_{x/y,q}(0) \quad
\frac{c^{(*)}_{v/h,1}(0)}{c^{(*)}_{v/h,0}(0)} = i \,.
\end{equation}

An analogous procedure can be followed to derive Eq.~(\ref{eq:limc}).
We substitute instead the decompositions (\ref{eq:decomp}) into
Eq.~(\ref{eq:deltau}) and now the averages $\langle\ldots\rangle_{S}$
and $\langle\ldots\rangle_{\mathrm{ang},S}$ yield linear equations for
the un-starred amplitudes with inhomogeneous terms arising from the
starred ones. Then Eq.~(\ref{eq:limc}) follows from
Eqs.~(\ref{eq:proptrans})-(\ref{eq:limcsvh2s}) completing our
derivation of Eq.~(\ref{eq:approxu}). Furnished with the latter it is
clear that the lowest order contributions in $kd$ to the transmission
amplitudes $t_q(kd)$, $b(kd)\, t_q(kd)$ and reflection amplitudes
$r^{(L)}_{c/t}(kd)$, $r^{(L)}_{v/h,\delta\eta}(kd)$ can just be
extracted from $\bar u_*(\bar r)$. Thus from Eq.~(\ref{eq:limcsct}) we
obtain for both compression and torsion $r^{(L)}(0)= -1$. In turn,
Eqs.~(\ref{eq:limcsvh1}) and (\ref{eq:limcsvh2s}) yield for the
two bending branches: $b(0)=i$, $r^{(L)}_{AA}(0)= i$,
$r^{(L)}_{AB}(0)= -(1+i)$, $r^{(L)}_{BA}(0) = - (1-i)$, and
$r^{(L)}_{BB}(0)= -i$.  The lowest order contribution $t_q^{(0)}$ to
each of the transmission amplitudes $t_q$ is provided by the
corresponding approximation for the starred amplitude
$c^{(*)}_{\beta,0}(kd)$ [cf.~Eqs.~(\ref{eq:limcsct}) and
(\ref{eq:limcsvh2s})]. The needed values at the origin of $\bar
u^{(0)}_q(\bar r)$ and its derivatives are straightforward to obtain
from the expressions given in App.~\ref{ap:halfspace}. We note that
the symmetries of the half-space imply
$u^{(0)}_{y;q,\theta,\varphi,\gamma}(0)=
u^{(0)}_{x;q,\theta,\pi/2-\varphi,\gamma}(0)$.

Finally, for the cantilever geometry a procedure analogous to the
above derivation of the starred amplitudes for case ($i$) yields the
standard TRE recipes for the reflection at the free end. The latter
given our choice of coordinate origin lead to: $r^{(R)}(0)=e^{i2kL}$
for both compression and torsion, and $r^{(R)}_{AA}(0)= ie^{i2kL}$,
$r^{(R)}_{AB}(0)= (1+i)e^{(i-1)kL}$, $r^{(R)}_{BA}(0) =
(1-i)e^{(i-1)kL}$, $r^{(R)}_{BB}(0)= -ie^{-2kL}$ for the two bending
branches.

\subsection{Reciprocity relations}\label{sub:rec}

Reciprocity is simply the relation $\langle v_+ | u_- \rangle =
\langle u^*_- |v^*_+ \rangle $ for a standard scattering problem in a
time reversal invariant theory\cite{Messiah,LaxPhyllips}. Here $v \in
V$ and $u \in U$ label freely propagating asymptotic states belonging
to the sets of interest $V$ and $U$, ``$-$'' and ``$+$'' denote
respectively incoming and outgoing scattering states and ``$*$''
denotes the time reversal operation. The inner product ``$\langle |
\rangle$'' is assumed to be preserved by the time evolution -
i.e.~the underlying theory is ``unitary''. However this
preserved inner product need not necessarily be the usual overlap as
in quantum mechanics where the ``unitarity'' corresponds to the
preservation of probability. In particular in our case of elasticity
theory it is defined so that $\langle w|w \rangle$ corresponds to the
energy carried by the solution $w$ and the unitarity corresponds
to energy conservation \cite{LaxPhyllips}.  This relation is quite
general but proves to be especially powerful when $V \sim U$, in the
sense that aside from possible discrete indices, the available free
eigenstates are essentially equivalent. In our specific context an
example satisfying this last requirement would be two different rods
(corresponding to $V$ and $U$) joined at an abrupt junction
\cite{Cross01}. Reciprocity directly implies that for unit incident
power in a traveling wave $\in U$ (with wave vector $k_U$) the power
transmitted into a traveling wave $\in V$ (with wave vector $k_V$) is
equal to the power transmitted into the wave corresponding to $-k_U$
for unit incident power in the one characterized by $-k_V$.

However our model for one junction does not satisfy the above
``asymptotic equivalence'' since it involves coupling a 1D system (the
beam) to 3D and 2D continua corresponding, respectively, to the bulk
and surface states (SAW) of the support. To overcome this difficulty
we first consider a model of the junction for which: ($i$) the support
is characterized by some finite dimension $D$, ($ii$) the support
states under scrutiny are equivalent to a phonon waveguide so that the
``asymptotic equivalence'' is satisfied, and ($iii$) for $D \to
\infty$ the support tends to a free elastic half-space. Then we apply
the reciprocity relations for finite $D$ and finally take the limit $D
\to \infty$. For bulk states (i.e.~$q$ with $\gamma \in
\{l,\mathrm{SV},\mathrm{SH}\}$) a suitable ``finite support'' is
afforded by another beam of square cross-section (side $D$) subject to
periodic boundary conditions on the external faces.  On the other hand
for surface states (i.e.~$\gamma = s$) a suitable construction is
given by a slab of thickness $D$ subject to periodic boundary
conditions at the external semi-infinite horizontal faces and a free
boundary condition at the finite vertical face. In both cases the
decoupled support is exactly solvable and the support states of
interest are equivalent to a phonon waveguide whose branches we index
with a single label $j$. Thus the free support eigenmodes
\footnote{These may be chosen to have well defined reflection
  symmetries as for the free elastic half space.} read
$|u^{(0)}_j(k',D)\rangle$ with dispersion relations $\omega_j(k',D)$
where $k'$ is the wavevector along the waveguide's axis. If we define
the amplitudes $t_{\beta,j}(k',D)$ such that
\begin{equation}\label{def:amp}
  \bar{u}^{(0)}_{j-}(k',D,\bar r)\longrightarrow
  t_{\beta,j}(k',D)\,\bar{v}^{(0)}_\beta[k_{\beta,j}(k',D),\bar r]
\end{equation}
asymptotically for $z\to\infty$, where $|v^{(0)}_\beta(k)\rangle$ are
the freely propagating TRE beam modes (cf.~App.~\ref{ap:beam}) with $k>0$
and $k_{\beta,j}(k',D)$ the wavevector in the beam's branch $\beta$
that corresponds to the frequency $\omega_j(k',D)$; we have
\begin{equation}\label{eq:lim1}
  \lim_{D\to\infty}|u^{(0)}_{j-}(k',D)\rangle = |u_q\rangle \quad
  \Rightarrow \quad \lim_{D\to\infty} t_{\beta,j}(k',D) = t_q  \,.
\end{equation}
Taking into account that the modes we consider are normalized (in the
standard Euclidean metric) in all 1D, 2D and 3D cases, it is then
straightforward to realize that
\begin{align}\label{eq:lim2}
  \lim_{D\to\infty} \int\dk' \sum_j |t_{\beta,j}(k',D)|^2 &
  \delta\left[\omega-\omega_j(k',D)\right] \nonumber\\ 
  & = \int_q |t_q|^2 \delta\left(\omega-\omega_q\right)
\end{align} 
where the $\{q\}$ run only over modes with the corresponding $\beta$
and the integration includes summation over the appropriate discrete
indices. On the other hand reciprocity directly implies 
\begin{equation}\label{eq:recipro1}
  \tau^{(j)}_\beta\left[k_{\beta,j}(k',D),D\right] =
  \frac{P_\beta\left[\omega_j(k',D)\right]}{P_j\left[\omega_j(k',D),D\right]}
  |t_{\beta,j}(k',D)|^2\,.    
\end{equation}
Here we have introduced $\tau^{(j)}_\beta(k,D)$ as the transmission
coefficient into the support branch $j$ for a traveling wave of type
$\beta$ incident from the beam with wavevector $-k$ and
$P_{\beta}(\omega)$ [$P_{j}(\omega,D)$] as the power carried by the
normalized free waveguide mode with frequency $\omega$ and branch
index $\beta$ [$j$]. The functions $P_{\beta}(\omega)$,
$P_{j}(\omega,D)$ have a universal expression in terms of the mass
density of the respective waveguide and the corresponding dispersion
relation. On the one hand, in complete analogy to the equivalent
scenario for the electromagnetic field, the power carried by the
waveguide mode is given by the product of the group velocity and the
corresponding energy per unit length (averaged over a period). On the
other hand the latter can be expressed in terms of the mass density
and the frequency by first considering the harmonic theory of the
underlying microscopic discrete lattice for a finite waveguide of
length $2\pi/k$ subject to periodic boundary conditions at its ends
and then taking the continuum limit. Thus we obtain
\begin{equation}\label{eq:power} 
  P_{\beta}(\omega)  =\frac{1}{4\pi}\,\rho_{b}\omega^2\, \frac{\ud
    \omega_{\beta}}{\ud k}\,, \quad
  P_{j}(\omega)  =\frac{1}{4\pi}\,\rho_{s}\omega^2 \,\frac{\ud
    \omega_{j}}{\ud k'}\,,
\end{equation}
where $\rho_{b}$ ($\rho_{s}$) is the density of the beam (support). We
can substitute Eqs.~(\ref{eq:recipro1}) and (\ref{eq:power}) into the
LHS of Eq.~(\ref{eq:lim2}) and perform the integration, the summation
and the limit to obtain for each branch $\beta$ the following
reciprocity relation:
\begin{equation}\label{recipro}
  \frac{\rho_b}{\rho_s} \int_q |t_q|^2\,\delta(\omega-\omega_q) =
 \tau_\beta[k_\beta(\omega)]\,
  \frac{\ud k_\beta}{\ud\omega} (\omega)
\end{equation}      
where $\tau_\beta(k)$ is the total transmission coefficient into the
support for a traveling wave of type $\beta$ incident from the beam
with wavevector $-k$, and the $\{q\}$ run only over modes with the
corresponding $\beta$. 

Finally we note that the above derivation can be extended to the cases
considered in App.~\ref{ap:exten} by suitable modifications of
requirement ($iii$) so that in all cases the limit $D\to\infty$ yields
the support under consideration (``3D asymmetric'' or slab).
Naturally, in the asymmetric cases the labels $q$ will no longer
relate to the beam branch index $\beta$ and in Eq.~(\ref{recipro}) we
will have $t_q\to t_{q,\beta}$ with the $\int_q$ running over all the
support modes. In the case of a SWNT for which we use the shell
``continuum'' model $\rho_{b}$ should be replaced by the surface
density of graphene $\sigma_G$, while for a nanowire or SWNT for which
the underlying model for the $|v^{(0)}_\beta(k)\rangle$ is discrete
the adequate ansatz reads: $\rho_{b}\to \mu_b/N_c$
(cf.~App.~\ref{ap:exten}).  In the case of a thin plate geometry
\cite{Cross01} both $\rho_{s}$ and $\rho_{b}$ should be replaced by
the surface density.  In fact it can be argued that
Eq.~(\ref{recipro}) does not depend on any specific properties of the
junction or the support and only relies on the phonon transport being
ballistic.

\subsection{Transmission coefficients for each branch}\label{sub:tcoeff}

\begin{figure}
\includegraphics[width=7cm]{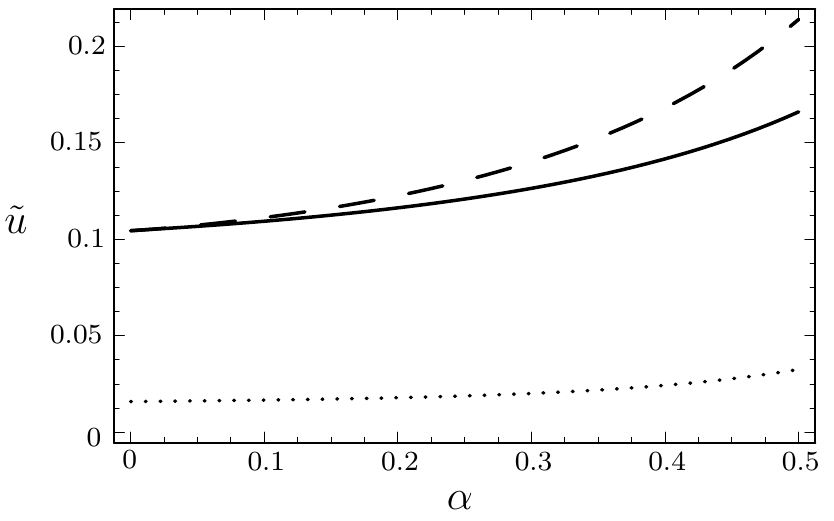}
\caption{Dimensionless displacements $\tilde{u}_c$ (dashed line),
  $\tilde{u}_{v/h}$ (solid line), and dimensionless angle
  $\tilde{u}_t^{(S)}$ (dotted line) as a function of the ratio
  $\alpha\equiv (c_t/c_l)^2$. For typical materials of interest
  $\sigma_s\approx 1/3$ which yields $\tilde{u}_c (1/4) = 0.13$,
  $\tilde{u}_t^{(S)} (1/4) = 0.019$, and $\tilde{u}_{v/h} (1/4) =
  0.12$. \label{fig:tildeu}}
\end{figure}

We turn now to the evaluation of the leading contribution in $kd$ to
the LHS of Eq.~(\ref{recipro}) for each branch $\beta$ which we define
as $\tilde{\rho}_\beta(\omega)$. On the one hand we will find in
Subsec.~\ref{sub:Itau} that the force spectral densities
$I_{n,\beta}(\omega)$ correct to lowest order in the reciprocal of the
aspect ratio $d/L$ only depend on the amplitudes $t_q^{(0)}$ through
these quantities $\tilde{\rho}_\beta(\omega)$. On the other hand it is
clear that if we substitute into Eq.~(\ref{recipro}) the
approximations $t_q^{(0)}$ given in Subsec.~\ref{sub:ustar} the
$\tilde{\rho}_\beta(\omega)$ determine the transmission coefficients
$\tau_\beta[k_\beta(\omega)]$ correct to lowest order in $kd$. It is
important to highlight that $\tilde{\rho}_\beta(\omega)$ is amenable
to reduction to a property of the free support at the origin that
directly relates to its DOS or, in the case of torsion, to its vacuum
spectrum for the angle [as defined in App.~\ref{ap:beam},
cf.~Eqs.~(\ref{ang})-(\ref{angnotation})] and geometrical properties
of the free beam ($S$, $I_z$, and $\gamma_z$). In fact we can consider
the displacement field operator for the free support $\hat{\bar
  u}^{(0)}(\bar r)$ and decompose it in terms of its normal modes
\footnote{This is just the analog for the decoupled support of
  Eq.~\ref{eq:u} and is obtained from it via the ansatz
  $\bar{u}\to\bar{u}^{(0)}$, $b\to b^{(0)}$.}  $\bar{u}^{(0)}_q(\bar
r)$. Then, the latter decomposition and Eqs.~(\ref{eq:limcsct}),
(\ref{eq:limcsvh2s}), and (\ref{recipro}) yield
\begin{align}
\tilde{\rho}_c (\omega) & = \frac{2\rho_b\omega}{\hbar}
S\int^{\infty}_{-\infty}\dt e^{i\omega t}
\langle\hat{u}^{(0)}_z(0,t)\hat{u}^{(0)}_z(0,0)\rangle\nonumber \\
\tilde{\rho}_t (\omega) & = \frac{2\rho_b\omega}{\hbar} I_z
\int^{\infty}_{-\infty}\dt e^{i\omega t} \lim_{S\to0}
\langle\hat{\theta}^{(0)}_z(t)\hat{\theta}^{(0)}_z(0)\rangle\nonumber\\
\tilde{\rho}_{v/h} (\omega) & = \frac{\rho_b\omega}{\hbar} S
\int^{\infty}_{-\infty}\dt e^{i\omega t}
\langle\hat{u}^{(0)}_{x/y}(0,t)\hat{u}^{(0)}_{x/y}(0,0)\rangle\,,
\end{align}
where we use Heisenberg operators and $\langle\ldots\rangle$ denotes
the vacuum expectation value \footnote{The limits $S\to0$ of the
  angles are taken keeping the shape of the cross section invariant.}.
This connection abets the interpretation of
$\tilde{\rho}_\beta(\omega)$ as an effective environmental DOS set
forth in the Introduction (cf.~Subsec.~\ref{sub:results}). We note
that the latter quantity has dimensions of linear density of states
and that for compression and bending it is proportional to the DOS of
the support times the area of the beam's cross section. We separate
the contributions of each type of support modes $\gamma$ so that
$\tilde{\rho}_\beta(\omega)\equiv
\sum_\gamma\tilde{\rho}_{\beta,\gamma}(\omega)$. In turn, given the
symmetrization of the modes, the wavevector integration in
Eq.~(\ref{recipro}) only involves $q_x,q_y,q_z>0$ while the symmetries
of the half-space imply for the two bending branches
$|t^{(0)}_{q,\theta,\varphi,\gamma,h}|=
|t^{(0)}_{q,\theta,\pi/2-\varphi,\gamma,v}|$ --- where we use spherical
coordinates for the wavevector $\bar q$ as in
App.~\ref{ap:halfspace}. One can then perform the substitutions
$\omega = c_\gamma q,v=\cos \theta$, integrate over $\omega$ and
$\varphi$, and eliminate $c_\gamma$ in favor of $c_t$ to obtain:
\begin{align}\label{tautilde}
  \tilde{\rho}_{c,\gamma}(\omega) & =\frac{\rho_b}{\rho_s}\!\frac{4S}{
    c_t^3}\,\tilde{u}_{c,\gamma}(\alpha)\,\omega^2 
\nonumber\\
    \tilde{\rho}_{t,\gamma}(\omega)
    & =\frac{\rho_b}{\rho_s}\!\frac{4I_z}{
    c_t^5}\left[\tilde{u}^{(A)}_{t,\gamma}+\gamma^2_z
    \tilde{u}^{(S)}_{t,\gamma}(\alpha)\right]\omega^4
\nonumber\\ \tilde{\rho}_{v/h,\gamma}(\omega) &
    =\frac{\rho_b}{\rho_s}\!\frac{2S}{
    c_t^3}\,\tilde{u}_{v/h,\gamma}(\alpha)\,\omega^2\,.
\end{align}
Here we have introduced the following dimensionless constants and
functions of the ratio $\alpha\equiv
(c_t/c_l)^2=(1-2\sigma_s)/2(1-\sigma_s)$ for the supports' material
($\sigma_s$ is the corresponding Poisson ratio):
\begin{widetext}
\begin{align}
  \tilde{u}_{c,l}(\alpha) =\, & \frac{\alpha^{3/2}}{2\pi}\int_0^1\dv
  \frac{(1 - 2 \alpha + 2 \alpha v^2)^2 v^2}{\left[ 4 \alpha^{3/2}
  \sqrt{1-\alpha+\alpha v^2} (1-v^2) v + (1 - 2 \alpha + 2 \alpha
  v^2)^2 \right]^2} \nonumber
\\
\tilde{u}_{c,\mathrm{SV}}(\alpha) =\,
  & \frac{2}{\pi} \int_0^{\sqrt{1-\alpha}}\dv \frac{(1-\alpha-v^2)
  (1-v^2) v^2 }{16 (1-\alpha-v^2) (1-v^2) v^2 + (2 v^2-1)^4} 
\nonumber\\ & \qquad 
+ \frac{2}{\pi}
  \int_{\sqrt{1-\alpha}}^{1}\dv\frac{(\alpha-1+v^2) (1-v^2)
  v^2}{\left[ 4 \sqrt{\alpha-1+v^2} (1-v^2) v + (2 v^2-1)^2 \right]^2}
  \nonumber
\end{align}
\begin{align}
 \tilde{u}_{c,\mathrm{SH}} =\, & 0\qquad
  \tilde{u}_{c,s}(\alpha) =
  \frac{1}{2\xi^3\!(\alpha)}\left[\sqrt{1-\alpha \xi^2\!(\alpha)} -
  \frac{1-\xi^2\!(\alpha)/2}{\sqrt{1-\xi^2\!(\alpha)}}
  \right]^{2}C^2(\alpha) \,, \label{utildec} 
\\
  \tilde{u}^{(A)}_{t,l} =\,&\tilde{u}^{(A)}_{t,\mathrm{SV}}
  =\,\tilde{u}^{(A)}_{t,s}=\,0 \qquad\qquad
  \tilde{u}^{(A)}_{t,\mathrm{SH}}=\,\frac{1}{12\pi}\,, \label{utildetA}\\
  \tilde{u}^{(S)}_{t,l}(\alpha) =\, &
  \frac{\alpha^{7/2}}{4\pi}\int_0^1\dv \frac{(1-\alpha+ \alpha v^2)
    (1-v^2)^2 v^2}{\left[4 \alpha^{3/2} \sqrt{1-\alpha +\alpha v^2}
      (1-v^2) v + (1-2\alpha +2\alpha v^2)^2 \right]^2}
  \nonumber\\
  \tilde{u}^{(S)}_{t,\mathrm{SV}}(\alpha) =\, & \frac{1}{16\pi}
  \int_0^{\sqrt{1-\alpha}}\dv \frac{(1-v^2)^2 v^2 (2 v^2 -1)^2}{16
    (1-\alpha-v^2) (1-v^2)^2 v^2 + (2 v^2-1)^4}  \nonumber\\ & \qquad +
  \frac{1}{16\pi}\int_{\sqrt{1-\alpha}}^{1}\dv \left[\frac{(1-v^2) v
      (2 v^2 -1)}{4 \sqrt{\alpha-1+v^2} (1-v^2) v + (2 v^2 -1)^2}
  \right]^2  \nonumber
\\
  \tilde{u}^{(S)}_{t,\mathrm{SH}} =\, & \frac{1}{24\pi} \qquad
  \tilde{u}^{(S)}_{t,s}(\alpha) =
  \frac{1}{64\xi(\alpha)}C^2(\alpha)\,,\label{utildet} \\
  \tilde{u}_{v/h,l}(\alpha) =\, & \frac{1}{4\pi}\int_0^1\dv \frac{4
    \alpha^{5/2} (1-\alpha+\alpha v^2) (1-v^2) v^2}{\left[ 4
      \alpha^{3/2} \sqrt{1-\alpha+\alpha v^2} (1-v^2) v + (1 -2\alpha
      +2\alpha v^2)^2 \right]^2} \nonumber
\\
 \tilde{u}_{v/h,\mathrm{SV}}(\alpha) =\, & \frac{1}{4\pi}
  \int_0^{\sqrt{1-\alpha}}\dv \frac{(2v^2-1)^2 v^2}{16 (1-\alpha-v^2)
  (1-v^2)^2 v^2 + (2 v^2-1)^4} \nonumber\\ & \qquad +
  \frac{1}{4\pi}\int_{\sqrt{1-\alpha}}^{1}\dv \frac{(2 v^2 -1)^2
  v^2}{\left[ 4 \sqrt{\alpha -1 + v^2} (1-v^2) v + (2 v^2 -1 )^2
  \right]^2} \nonumber\\ \tilde{u}_{v/h,\mathrm{SH}} =\, &
  \frac{1}{4\pi}\qquad \tilde{u}_{v/h,s}(\alpha) =
  \frac{\xi(\alpha)}{16}C^2(\alpha)\,, \label{utildeb}
\end{align}
\end{widetext}
where $\xi(\alpha)$ is the ratio of the velocity of propagation for
surface waves to $c_t$ and is a function of $\alpha$ that is always
less than unity \cite{Landau}. We note that for compression and
bending the $t_q$ have a non-vanishing limit for $k \to 0$ so that the
frequency dependence of the quantities $\tilde{\rho}$ follows directly
from the density of states of the 3D support \footnote{The 2D surface
  states (SAW) have a penetration depth that scales as $1/q_t$ so that
  their contribution to the DOS has the same scaling with $\omega$ as
  the bulk contribution.}.  Whilst in the case of torsion there is an
extra factor of $\omega^2$ given that the $t_q^{(0)}$ scale as the
derivatives of the displacement field $\bar u^{(0)}_q$.  In all cases
the contribution of SV support modes has two distinct terms: one
corresponding to polar angles $\theta$ below the critical angle
$\arccos\sqrt{1-\alpha}$ for which there is a reflected longitudinal
wave, and another corresponding to angles for which the longitudinal
component is evanescent. The corresponding results for the effective
environmental DOS for nanotubes and nanowires can be obtained from
Eq.~(\ref{tautilde}) via the ansatz $\rho_b S\to \mu_b$, $\rho_b
I_z\to \mu_b\langle r^2\rangle_S$ (cf.~App.~\ref{ap:exten}).  Finally,
we can use Eqs.~(\ref{recipro}), (\ref{tautilde})-(\ref{utildeb}) and
the TRE dispersion relations (\ref{disper}) to obtain after summing
over the index $\gamma$ the following expressions for the transmission
coefficients into the support
\begin{align}
  \tau_c(k) & = 4 \tilde{u}_c (\alpha) \left(\frac{3-4\alpha}{1-
      \alpha} \frac{E_b}{E_s} \right)^{3/2} \left(
      \frac{\rho_s}{\rho_b} \right)^{1/2} S k^2 \,,\nonumber
\\
\tau_{v/h}(k) & = 4 \tilde{u}_{v/h}(\alpha) \left(\frac{3-4\alpha}{1-
\alpha} \frac{E_b I_{y/x}}{E_s}\right)^{3/2}
\left(\frac{\rho_s}{\rho_b S}\right)^{1/2}\, k^5 \,,\nonumber
\end{align}
\begin{align}
\tau_t(k) & = 4\left[\frac{1}{12\pi} +
    \left(\frac{I_y-I_x}{I_z}\right)^2\tilde{u}^{(S)}_t(\alpha)
    \right]\left(\frac{3-4\alpha}{1-\alpha} \frac{C}{E_s}\right)^{5/2}
    \nonumber \\ & \quad \times \left(\frac{\rho_s}{\rho_b
    I_z}\right)^{3/2} k^4\,,\label{eq:trans}
\end{align}
where we have also substituted the definition of $\gamma_z$, used
$c_t=\sqrt{E_s/2\rho_s(1+\sigma_s)}$, and introduced
$\tilde{u}_\beta(\alpha) = \sum_\gamma \tilde{u}_{\beta,\gamma}
(\alpha)$. The latter functions are plotted in Fig.~\ref{fig:tildeu}
for all ratios $\alpha$ corresponding to physically allowed positive
values of the Poisson ratio $\sigma_s$.

Finally, we consider the frequency dependencies and the scalings with
the transverse dimensions for the analogous problem of an abrupt
junction between a rectangular beam of width $\mathsf{w}$ and a slab
(support) of the same thickness $\mathsf{t}\ll \mathsf{w}\equiv d$
[cf.~App.~\ref{ap:exten}]. The analogous expressions for the starred
amplitudes yield for the corresponding effective environmental DOS
\begin{align}\label{scale2Drho}
\tilde\rho_c(\omega) & \sim\mathsf{w}\omega \qquad
\tilde\rho_t(\omega)\sim \frac{\mathsf{w}^3}{\mathsf{t}^2}\omega
\nonumber\\ \tilde\rho_v(\omega) &\sim
\frac{\mathsf{w}}{\mathsf{t}}\qquad
\tilde\rho_h(\omega)\sim\mathsf{w}\omega
\end{align}
which via Eq.~(\ref{recipro}) allow us to recover the following
scalings with $kd$ for the transmission coefficients:
\begin{equation}\label{scale2Dtau}
\tau_{\beta\neq h}(k) \sim \mathsf{w}k \qquad \tau_{h}(k) \sim
(\mathsf{w}k)^3\,
\end{equation}
already derived in Ref.~[\onlinecite{Cross01}] by an alternative
method (cf.~the next Subsection). Here we have used that $\langle
r^2\rangle_S \sim \mathsf{w}^2$ and
$C/I_z\sim\mathsf{t}^2/\mathsf{w}^2$ for $\mathsf{t}\ll\mathsf{w}$.

\subsection{Multipole expansion method}\label{sub:mult}

The method followed above to derive the transmission coefficients
relies on using the \emph{transmission amplitudes from the support
into the beam} --- which given the results of Subsec.~\ref{sub:ustar}
can be extracted from the free modes of the supports --- and then
exploiting the reciprocity relation (\ref{recipro}). The naive
expectation is that an alternative method should be afforded by using
the far field of $\Delta\bar u_-(\bar r)$ [cf.~Eq.~(\ref{eq:deltau})]
corresponding to \emph{incidence from the beam} [case ($i$) in
Subsec.~\ref{sub:ustar}] to determine the power across the surface of
a hemisphere with $r \to\infty$ per unit incident power. 

The displacement field $\Delta\bar u_-(\bar r)$ can also be viewed as
the retarded solution for the problem of the free elastic half-space
harmonically forced at its boundary by the stress source
$\sigma_{iz}(\bar r)=\hat i\cdot\mathbf{F}\cdot\left[ \Delta \bar u_+
  (\bar r) + \bar u_* (\bar r) \right]$. As this radiation problem
involves a source with a typical dimension $d$ much smaller than the
wavelength associated with its frequency it is natural to use for
$\Delta\bar u_-(\bar r)$ its multipole expansion, whose successive
moments will yield terms of increasing order in $k d$. More
specifically, if we consider the asymptotic behavior of the
corresponding Green's function \cite{Graff} $G_{ij}(\bar r,\omega)$
for $q_t\,r \to\infty$ and the $n$th moment of the stress source
\footnote{We note that as the cross section $S$ lies at $z=0$ one has
  $M^{(n)}_{i_1,i_2\ldots i_{n+1}}=0$ for $i_m=3$ with $m\neq1$.}
\begin{equation}\label{eq:moment}
  M^{(n)}_{i_1,i_2\ldots i_{n+1}}\equiv
  S\langle\sigma_{i_1z}x_{i_2}x_{i_3} \ldots x_{i_{n+1}}\rangle_S
\end{equation}
with $i_m=1,2,3$, $n=0,1,\ldots$, one can show that the corresponding
contribution to the displacement field $\Delta\bar u^{(n)}_-(\bar r)$
scales as
\begin{equation}\label{eq:umn}
  \left|\Delta\bar u^{(n)}_-(\bar r)\right|\sim 
  \frac{q_t^n(k) M^{(n)}}{E_s r} \qquad q_t\,r\gg 1
\end{equation}
where we have introduced the norm $M^{(n)}\equiv
\{\sum_{\{i\}}[M^{(n)}_{i_1,i_2\ldots i_{n+1}}]^2/(n+1)\}^{1/2}$.
Naturally the source $\sigma_{iz}(\bar r)$ should also be expanded in
powers of $k d$. However our treatment of the solution inside the beam
(cf.~Subsec.~\ref{sub:ustar}) only provides us with the contributions
arising from the TRE part of $\bar u_*(\bar r)$ whose order is lower
than the inertia [$(kd)^{2p_\beta}$]. Thus the multipole expansion
will only be useful if it can be established that the former result in
contributions to the transmission coefficient that \emph{dominate over
those corresponding to the end corrections and to higher orders of the
TRE branches}. The relative orders follow directly from considering
the expansion in $kd$ of the RHS of Eq.~(\ref{eq:umn}).  A
straightforward analysis implies that in general
(i.e.~without invoking any symmetries) this is non-trivial
due \footnote{Though the universal properties discussed in
App.~\ref{ap:beam} imply that for the end corrections the total force
[zeroth order moment $M^{(0)}$] and total torque [antisymmetric part
of $M^{(1)}$] scale at most as the inertia, the symmetric part of
$M^{(1)}$ and the $M^{(2)}$ can be zeroth order in $k$.} to the role
played by the end corrections' symmetric part of the restriction of
$M^{(1)}_{i_1,i_2}$ to $i_1,i_2=1,2$ [$\mathbf{M}^{(S)}$] and by their
moments $\mathbf{M}^{(2)}$. We note that the leading contribution to
the far field of the compression, torsion, and bending TRE branches
scales, respectively, as $k$, $k^2$, and $k^3$. In the cases of
compression and bending these correspond to point sources given,
respectively, by a normal force and a tangential force applied at the
origin. On the other hand for torsion the leading contribution
corresponds to a normal torque applied at the origin only when the
corresponding $\mathbf{M}^{(S)}$ vanishes.  For the bending branches
the reflection symmetries are enough to cancel $\mathbf{M}^{(S)}$ and
$q_t^n(k)\sim k^2$ implies that higher order moments are irrelevant.
In the case of the torsional branch it is sufficient to have
$\mathbf{M}^{(S)}=\mathbf{M}^{(2)}=0$ which can be guaranteed by also
requiring symmetry under the swap $x\leftrightarrow y$ (i.e.~if the
cross section is symmetric under rotations by $\pi/2$ around $z$). In
contrast for compression augmenting the symmetry to that of a cylinder
does not guarantee the cancellation of the diagonal part of
$\mathbf{M}^{(S)}$ which could potentially give contributions to the
far field of order $k$. Thus, whilst for bending and torsion the
multipole expansion approach will provide a non-trivial corroboration
of the exact form of Eqs.~(\ref{eq:trans}) --- and together with
Eqs.~(\ref{tautilde}) of the reciprocity relation (\ref{recipro}) ---
for cross sections with the appropriate symmetries, for compression it
in principle only allows to check the $k$-dependence of these
relations.

Energy conservation implies that the evaluation of the contribution to
$\tau$ of the aforementioned leading multipole of the TRE branch can
be done at the origin. In general if we consider a harmonic point
source given by the superposition of a force $\bar F_\mathrm{out}$ and
a torque $\bar M_\mathrm{out}$ applied at the origin we have for the
power radiated into the half-space averaged over a period
\begin{equation}\label{pout}
P_\mathrm{out} = \tfrac{1}{2} \Re \left\{ \bar F_\mathrm{out}^* \cdot
  \dot{\bar u}_\mathrm{out}(0) +\bar M_\mathrm{out}^* \cdot \dot{\bar
    \theta}_\mathrm{out}(0) \right\} 
\end{equation}
where $\bar \theta_\mathrm{out}(0,t)= \lim_{S\to0}\langle \bar
u_\mathrm{out}(\bar r,t) \rangle_{\mathrm{ang},S}$ --- with $S$ a
circle on the free face at $z=0$ [cf.~Eq.~(\ref{angapprox})] --- and
$\bar u_\mathrm{out}(\bar r,t)$ is the generated displacement
field. If we now express the lowest order in $kd$ of the incident
power associated with $\bar u_\mathrm{in}(\bar r,t)$
[cf. Subsec.~\ref{sub:ustar}] in terms of the corresponding total
force $\bar F_\mathrm{in}$ and total torque $\bar M_\mathrm{in}$ using the TRE
transverse profiles given in App.~\ref{ap:beam} we get
\begin{equation}\label{taugral}
  \tau = \frac{P_\mathrm{out}}{P_\mathrm{in}}=\frac{\tfrac{1}{2} \Re
    \left\{ \bar F_\mathrm{out}^* \cdot \dot{\bar u}_\mathrm{out} (0) 
      +\bar M_\mathrm{out}^* \cdot \dot{\bar \theta}_\mathrm{out} (0)
    \right\}}{\tfrac{1}{2} \Re \left\{ \bar F_\mathrm{in}^* \cdot 
      \dot{\bar u}_\mathrm{in} (0) +\bar M_\mathrm{in}^* \cdot
      \dot{\bar \theta}_\mathrm{in} (0) \right\}}\,.
\end{equation}
Our results for the reflection coefficients $r^{(L)}_{c/t}(0)$,
$r^{(L)}_{v/h,\delta\eta}(0)$ imply
\begin{equation}\label{inout}
\left|\frac{F_\mathrm{out}}{F_\mathrm{in}}\right|^2 = 
\begin{cases}  
8 & \beta=v,h \\ 4 & \beta=c
\end{cases} 
\qquad
 \left|\frac{M_\mathrm{out}}{M_\mathrm{in}}\right|^2 = 
\begin{cases} 
8 & \beta=v,h \\ 4 & \beta=t
\end{cases}
\end{equation}
From the above Eqs.~(\ref{taugral}), (\ref{inout}), the properties of
the TRE solutions and the definition of the Green's function $\mathbf{G}$
it is straightforward to establish for the leading order of the
transmission coefficients
\begin{align}\label{taugreen}
&\tau_c(k) = 4 \Im \{G_{zz}[0,\omega(k)]\} F_c(k) \nonumber\\
&\tau_t(k) = 2 \Im \{\frac{\partial^2}{\partial x \partial
y}G_{yx}[0,\omega(k)]-\frac{\partial^2}{\partial
y^2}G_{xx}[0,\omega(k)]\} M_t(k) \nonumber\\ 
&\tau_{v/h}(k) = 4 \Im \{G_{xx/yy}[0,\omega(k)]\} F_{v/h}(k) \,.
\end{align}
The last factor in each of these equations $F_\beta(k)$ [$M_\beta(k)$]
is defined as the magnitude of the total force (torque) carried by the
TRE solution per unit amplitude. To derive Eq.~(\ref{taugreen}) for
$\beta=t$ we have also used that the stress point source corresponding
to a normal torque $M_\mathrm{out}$ at the origin is specified by
\begin{equation}
\sigma_{xz}=\tfrac{M_\mathrm{out}}{2} \delta(x)\delta'(y) \quad
\sigma_{yz}= - \tfrac{M_\mathrm{out}}{2} \delta'(x)\delta(y) \quad
\sigma_{zz}=0 \,,
\end{equation} 
which can be deduced from the source's symmetries, and the properties:
$G_{xy}[x,y,z,\omega(k)]=G_{yx}[y,x,z,\omega(k)]$,
$G_{yy}[x,y,z,\omega(k)]=G_{xx}[y,x,z,\omega(k)]$. Note that from the
latter we also have $\Im \{G_{xx}[0,\omega(k)]\} =\Im
\{G_{yy}[0,\omega(k)]\}$. If we now eliminate $k$ in favor of the
transverse wavevector in the support $q_t$ and define
\begin{equation}\label{tildegreen}
  \mathbf{G}(\bar r,\omega[q_t])\equiv
  \frac{2(1+\sigma_s)q_t}{E_s}\mathbf{\tilde{G}}(q_t\bar r)\,, 
\end{equation}
\emph{dimensional analysis} directly implies that the imaginary parts
of the function $\mathbf{\tilde{G}}$ and of its derivatives evaluated
at the origin are dimensionless functions of the Poisson ratio for the
support material \footnote{There is naturally a purely real singular
  part that corresponds to the quasistatic near field and over an
  oscillation period does not perform net work.}. This together with
\begin{equation}\label{FM}
  F_c(k)=E_b S k \quad M_t(k)= C k \quad F_{v/h}(k)=E_b I_{y/x} k^3
\end{equation}
and Eqs.~(\ref{taugreen}), (\ref{tildegreen}) yields scalings with $k
d$ for the transmission coefficients $\tau_\beta(k)$ that for
\emph{all branches} are consistent with Eqs.~(\ref{eq:trans})
providing a non-trivial check of the reciprocity relation
(\ref{recipro}).  We have also corroborated the prefactor for torsion
and bending in the cases where the aforementioned symmetries are met,
by explicitly calculating $\Im \{\tilde{G}_{xx}(0)\}$ and $\Im
\{\frac{\partial^2}{\partial x \partial
  y}\tilde{G}_{yx}(0)-\frac{\partial^2}{\partial
  y^2}\tilde{G}_{xx}(0)\}/2$ and comparing with 
\begin{align}\label{check}
  \frac{1}{2} \Im \{\frac{\partial^2}{\partial x \partial
    y}\tilde{G}_{yx}(0)-\frac{\partial^2}{\partial
    y^2}\tilde{G}_{xx}(0)\} = \frac{1}{12\pi}& =
  \tilde{u}^{(A)}_{t,\mathrm{SH}} \nonumber\\
  \Im\{\tilde{G}_{xx}(0,\alpha)\}& =\tilde{u}_{v/h}(\alpha)\,,
\end{align}
which can be obtained from Eqs.~(\ref{utildetA}), (\ref{utildeb}),
(\ref{eq:trans}), (\ref{taugreen}), (\ref{tildegreen}), (\ref{FM}),
and the expression for $q_t$ in terms of $k$ for each branch.  The
explicit derivation of $\Im \{\tilde{G}_{xx}(0)\}$ from its definition
(\ref{tildegreen}) is given in App.~\ref{ap:multipole}.  We have
checked numerically that the corresponding expressions for
$\tilde{u}_{v/h}(\alpha)$ given, respectively, by Eqs.~(\ref{imGxx})
and (\ref{utildeb}) coincide for all physical values of the sound
speed's ratio $\alpha$ corresponding to positive Poisson ratios
($0<\sigma_s<1/2$).  The analogous derivation for torsion (applicable
to a cross section symmetric under rotations by $\pi/2$) is greatly
simplified by the fact that only SH waves contribute --- leading to a
universal prefactor. Finally, we note that Eq.~(\ref{check}) and its
analog for compression allow us to interpret the $\tilde{u}$ as
dimensionless displacements and angles.

\section{Resonator -- bath representation and spectral densities
  \texorpdfstring{$I(\omega)$}{omega}}\label{s:spec}

\subsection{Resonator modes, scattering modes and their overlaps}
\label{sub:over}

We return now to the analysis of the normal modes of the whole
structure $\bar u_q (\bar r)$ (scattering modes) and the resonator
mode $\bar u'_R (\bar r)$ inside the beam for the purpose of
determining their overlaps to lowest order in $d/L$.  The prescription
for the resonator mode given in Subsec.~\ref{IIB} implies that the
procedures to be followed for the bridge and cantilever geometries
differ only in the boundary conditions: clamped boundary conditions at
the junctions, but free boundary conditions at the end of the
cantilever.  Of course the resulting Sturm-Liouville problem defines
an infinite set of resonator modes so that
$\omega_R\to\omega_{n,\beta}$ with $n=0,1,\ldots$ and $\beta=c,t,v,h$.
As we are interested in the regime $d/L\ll1$, a natural requirement to
identify these localized modes with the physical resonances of the
whole structure is $\tilde\omega_{n,\beta}\to\omega_{n,\beta}$ for
$d/L\to 0$, where the former are the real parts of the poles of the
propagator for the displacement field (\ref{eq:u}) --- this will be
borne out below. In this respect, it is worth noting once again that
our problem can be viewed as a mechanical lossy Fabry-P\'erot
(cf.~Subsec.~\ref{sub:modes} and Sec.~\ref{s:FP}). As already
discussed, we focus on low frequencies so that
$\omega_\beta[k(q)],\omega_{n,\beta}\ll\omega^{(\beta)}_*$ with
$k_\beta(\omega_*)\sim \pi/d$. In this regime the localized resonator
modes can be associated with effective one-dimensional fields
$\phi_{n,\beta}(z)$ defined in complete analogy to the fields
$\phi_{k(q),\beta}(z)$ describing the scattering modes
[cf.~Eqs.~(\ref{modetors}), (\ref{modebend}) and
Subsec.~\ref{sub:tre}]. As expected, the reflection symmetries imply
that resonator modes associated with a given branch $\beta$ are
orthogonal to scattering modes characterized by $\beta'\neq\beta$.

We now consider the results for the reflection amplitudes
$r^{(L)}_{c/t}(0)$, $r^{(L)}_{v/h,\delta\eta}(0)$, $r^{(R)}_{c/t}(0)$,
$r^{(R)}_{v/h,\delta\eta}(0)$ and for the ratio $b(0)$ obtained in the
previous Section (cf.~Subsec.~\ref{sub:ustar}) together with the
analysis for the propagation of low frequency modes inside the beam
performed in Sec.~\ref{s:FP} [cf.~Eqs.~(\ref{Ators}),
(\ref{Abend})]. It follows that to zeroth order in $d/L$ the
scattering effective one-dimensional field $\phi_{k(q),\beta}(z)$ and
its derivatives satisfy the same homogeneous boundary conditions as
those defining the $\phi_{n,\beta}(z)$ except for the value of the
field at the left junction \footnote{Note that the $\phi_{n,\beta}(z)$
are homogeneous solutions of the linear systems defined by the RHS of
Eqs.~(\ref{Ators}), (\ref{Abend}) in the limit
$d/L\to0$. \label{fullpole}} specified by
\begin{equation}\label{bound}
\phi_{k,c/t}(0)=1 \qquad \phi_{k,v/h}(0)=1+i \,.
\end{equation}
On the other hand all of these effective fields are solutions of the
TRE equation
\begin{equation}\label{TREeq}
\mathscr{D}_\beta\phi=\omega^2\phi
\end{equation}
with
\begin{equation}\label{TREeq:def}
\mathscr{D}_\beta \equiv
(-1)^{p_\beta}\tilde{c}^2_\beta\frac{\partial^{2p_\beta}}{\partial
z^{2p_\beta}}\,.
\end{equation}
It is then simple to use the aforementioned boundary conditions and
Eqs.~(\ref{TREeq}), (\ref{TREeq:def}) to obtain
\begin{align}\label{intpart}
& \langle \phi_{n,\beta}|\mathscr{D}_\beta|\phi_{k,\beta}\rangle 
=\omega^2_\beta(k)\langle \phi_{n,\beta}|\phi_{k,\beta}\rangle =
\omega^2_{n,\beta}\langle \phi_{n,\beta}|\phi_{k,\beta}\rangle
\nonumber\\ & + (-1)^{p_\beta}\tilde{c}^2_\beta
\frac{\partial^{2p_\beta-1}\phi_{n,\beta}}{\partial
z^{2p_\beta-1}}\phi_{k,\beta}|_{z=0}
\end{align}
where we have used integration by parts and that the
$\phi_{n,\beta}(z)$ are real (cf.~Subsec.~\ref{IIB}) to establish the
last equality. We have calculated the resonator modes
$\phi_{n,\beta}(z)$ (cf.~Ref.~[\onlinecite{Cleland,Graff}])
\emph{normalized them to the length} $L$ and computed their necessary
derivatives [note the $\pi$ phase freedom for the choice of
$\phi_{n,\beta}(z)$]. These together with Eqs.~(\ref{bound}),
(\ref{intpart}) and the dispersion relations for the different
branches, yield
\begin{equation}\label{Ov} 
\langle \phi_{n,\beta} | \phi_{k(\omega),\beta} \rangle\!=\!
 \sqrt{2 C_{n,\beta}}\, \frac{p_\beta}{k_n}\,
 \frac{\omega_n^2}{\omega^2-\omega_n^2} \times \begin{cases}
1 & \beta=c,t 
\\
e^{i\pi/4} & \beta=v,h
\end{cases}
\end{equation} 
Here $n=0,1,\ldots$, $C_{n,c/t}=1$, $C_{n,v/h}=\left(\tanh^2\frac{k_n
L}{2}\right)^{\left(-1\right)^n}$, and $p_\beta$ is the exponent of
the corresponding dispersion relation --- i.e.~$p_{c/t}=1$ and
$p_{v/h}=2$.

Equation (\ref{Ov}) will prove useful below when using the overlaps to
calculate the force spectral densities $I(\omega)_{n,\beta}$.
Naturally, its above derivation is invalid whenever $k$ is a zero of
the resolvent of the linear system --- Eq.~(\ref{Ators})
[Eq.~(\ref{Abend})] for $\beta=c,t$ [$\beta=v,h$] --- that determines
the scattering mode $\phi_{k,\beta}(z)$ which for $k=k_n$ (at a
generic value of $z$) diverges as $d/L\to0$. The behavior of the
overlaps and the spectral densities in the neighborhood of these
special points, which correspond to the resonances, is discussed
further in App.~\ref{ap:I}. Naturally the divergent behavior at
$\omega_n$ can be used to prove $\tilde\omega_n\to \omega_n$.

\subsection{Relations between \texorpdfstring{$I(\omega)$}{I(omega)}
  and \texorpdfstring{$\tilde \rho(\omega)$}{tilderho},
  \texorpdfstring{$\tau(\omega)$}{tay(omega)}}\label{sub:Itau}

In principle the overlaps calculated in the previous Subsection would
allow to obtain the leading contribution in $d/L$ to the environmental
spectral densities $I(\omega)$ from Eqs.~(\ref{eq:S}) and
(\ref{Iex}). However the latter exact expression has the drawback that
even for a generic value of $\omega$ the dispersive contribution
$\Delta_S (\omega)$ brings into play both the behavior at $\omega_R$
and at high frequency of the function $S(\omega)$, whilst the analysis
we have done of the scattering modes $\bar u_q (\bar r)$ and their
overlaps with the resonator modes fails at these frequencies. To
overcome this issue we first invert Eq.~(\ref{Iex}) to recover a well
known expression for the function $S(\omega)$ in terms of $I(\omega)$
\begin{equation}\label{S}
S(\omega) = \frac{\omega_R
  I(\omega)}{\left[\omega^2-\omega_R^2-\omega_R\Delta_I(\omega)
  \right]^2 + \omega_R^2I^2 (\omega)}\,,
\end{equation}
and subsequently derive from it the following approximate relation
\begin{equation}\label{Iapprox}
I(\omega) \approx \frac{S(\omega)}{\omega_R} \left[\left(\omega+
  \omega_R\right)^2 \left(\omega- \tilde\omega_R\right)^2+
  \omega_R^2 I^2(\tilde\omega_R)\right]\,.
\end{equation}
Here $\tilde\omega_R$ is the approximation to the renormalized frequency
afforded by the solution that the equation
\begin{equation}\label{omegaR}
\frac{\tilde\omega_R}{\omega_R} = 1 +
\frac{\Delta_I(\tilde\omega_R)}{\tilde\omega_R+\omega_R}\,
\end{equation}
has close to $\omega_R$ when $|\Delta_I(\omega_R)|/\omega_R\ll 1$. The
latter condition is necessary for the validity of the approximate
relation (\ref{Iapprox}) and the corresponding relative error is at
most of order $|\Delta_I(\omega_R)|/\omega_R$. The derivation of
Eqs.~(\ref{S}), (\ref{Iapprox}) is given in App.~\ref{ap:I} where we
also establish that $|\Delta_I(\omega_R)|/\omega_R\to0$,
$I(\omega_R)/\omega_R\to0$ for $d/L\to 0$ and give the analog of the
approximation (\ref{Iapprox}) for the couplings $\{\zeta_q\}$.  Hence,
to obtain $I(\omega)$ correct to lowest order in $d/L$ we can replace
the factor in square brackets in Eq.~(\ref{Iapprox}) by its limit for
$d/L\to 0$ which will be given by $\tilde\omega_R\to\omega_R$,
$I(\omega_R)\to0$. If we now consider the definitions of $t_q$ and
$\phi_k(z)$, and the corresponding expressions for the displacement
field given in Sec.~\ref{s:FP}, and substitute into the definition
(\ref{eq:S}) we obtain for the remaining factor
\begin{equation}\label{Stau}
  \frac{S(\omega)}{\omega_R} \approx \frac{\delta |\langle \phi_R |
  \phi_{k(\omega)} \rangle|^2}{4L\omega_R\,\omega}\,
  \frac{\rho_b}{\rho_s}\int_q |t_q|^2 \delta(\omega-\omega_q)\,,
\end{equation}
where we have used the definition (\ref{eq:overlap}) for the overlap
$\langle u'_R | u_q \rangle$, the expressions for $\bar u(\bar r)$ in
terms of $\phi(z)$ given in App.~\ref{ap:beam} and neglected the
contributions that involve derivatives of $\phi(z)$ which are higher
order in $d/L$ --- note that $\phi_n\to\phi_R$, defined in the
previous subsection, is normalized to the length while $|u'_R \rangle$
is normalized in the metric defined by Eq.~(\ref{eq:overlap}).  Here
$q$ runs only over modes with the branch index $\beta$ corresponding
to the appropriate resonator mode (frequency $\omega_R$) and we
have introduced the number of supports $\delta$ so that for the bridge
geometry $q$ is further restricted to right-movers. On the one hand
the last factor on the RHS of Eq.~(\ref{Stau}), correct to lowest
order in $kd$, is just given by the appropriate $\tilde
\rho_\beta(\omega)$ [Eq.~(\ref{tautilde}) summed over all $\gamma$]
which together with Eq.~(\ref{Ov}) for the overlaps and the relation
\begin{equation}\label{pbeta}
\left(\frac{p_\beta\,\omega_n}{k_n} \right)^2\frac{\ud
 k_\beta}{\ud\omega} (\omega)=\left(\frac{\omega_n}{\omega}
 \right)^{p_\beta-1}\frac{\ud\omega_\beta}{\ud k} (\omega) \,,
\end{equation}
valid for $p_\beta=1,2$, yields Eq.~(\ref{eq:spectralrho}), namely
\begin{equation*}
I_{n,\beta}(\omega) = \delta C_{n,\beta}\left[\frac{\ud\omega_\beta}{\ud
k} (\omega_n)\right]^2\,\frac{\tilde \rho_\beta(\omega)}{2L}\,
\frac{\omega_n}{\omega}\,.
\end{equation*}
On the other hand the last factor on the RHS of Eq.~(\ref{Stau}) can
also be expressed exactly in terms of the transmission coefficients
$\tau_\beta(\omega)$ using the reciprocity relation (\ref{recipro})
derived in Subsec.~\ref{sub:rec}. We can use this fact, Eq.~(\ref{Ov})
for the overlaps, and Eqs.~(\ref{pbeta}), (\ref{Stau}),
(\ref{Iapprox}), to finally obtain the relation (\ref{eq:spectraltau})
between the leading contribution to $I(\omega)$ for each
resonance and the transmission coefficients:
\begin{equation*} 
I_{n,\beta}(\omega) = \delta C_{n,\beta}\frac{\ud\omega_\beta}{\ud k}
(\omega)\,\frac{\tau_\beta(\omega)}{2L}\,
\left(\frac{\omega_n}{\omega}\right)^{p_\beta}\,.
\end{equation*}
The validity of Eqs.~(\ref{eq:spectralrho}) and (\ref{eq:spectraltau}),
to lowest order in $d/L$ is affected by the caveats discussed in the
previous Subsection, when deriving the overlaps, so that the
neighborhood of the \emph{other resonances} is excluded and $\omega$
should be much smaller than $\omega_\beta(\pi/d)$. Note that with the
judicious choice of dimensionless variables $d/L$,
$\omega/\omega_n$ the latter is not an additional approximation
as $\omega_\beta(\pi/d)/\omega_n\to\infty$ for $d/L\to 0$.
Clearly, the above procedure is questionable in a neighborhood of
$\omega_R$ where Eq.~(\ref{Ov}) for the overlaps diverges.
Nonetheless, $\omega_R$ is in fact included in the frequency range
where Eqs.~(\ref{eq:spectralrho}) and (\ref{eq:spectraltau}), are
valid. To resolve this issue it suffices to prove, without using the
Eqs.~(\ref{S}), (\ref{Iapprox}), that the function $I(\omega)$ remains
well behaved at this resonant frequency as $d/L\to0$
(cf.~App.~\ref{ap:I}).

In turn, it can be argued that $\omega_\beta(\pi/d)$ always sets the
order of the natural ultraviolet cutoff.  If we consider the exact
Eqs.~(\ref{eq:S}) and (\ref{Iex}) it is clear that the lattice
constant $a$ provides an obvious ultraviolet cutoff for the functions
$S(\omega)$ and $I(\omega)$ so that the case $d\sim a$ follows
trivially. If instead $d\gg a$, one finds that $\omega_\beta(\pi/d)$
provides a ``soft cutoff'' beyond which the functions $S(\omega)$ and
$I(\omega)$ decay as integer power-laws. In fact for
$\omega\gg\omega_\beta(\pi/d)$ one can argue that to analyze the
scaling of $S(\omega)$ it is permissible to replace in its definition
(\ref{eq:S}) the relevant modes $\bar u_q (\bar r)$ inside the beam
--- i.e.~those with $\omega_q\gg\omega_\beta(\pi/d)$ --- by the normal
modes the beam would present for $d\to\infty$. Thus the function
$\omega S(\omega)$ will scale as the DOS times the corresponding
Fourier transform of the resonator mode \cite{Graff}. The latter
yields a factor of $1/\omega^2$ for the longitudinal direction and a
factor of $1/\omega$ for each transverse direction. Hence we obtain
$S(\omega)\sim 1/\omega^m$, where $m=7$ for the case of a ``bulk cross
section'' and $m=6$ for the case of a SWNT modeled as a shell
\cite{Yakobson96}. A straightforward analysis yields in both cases
$1/\omega^2$ for the leading term of the asymptotic expansion of
$\Delta_S(\omega)$. The latter behavior together with Eq.~(\ref{Iex})
leads to $I(\omega)\sim 1/\omega^l$ with $l=3$ and $l=2$,
respectively. Finally, it is worth noting that the scaling for
$\omega\to 0$ can also be analyzed without resorting to the
approximation (\ref{Iapprox}). In fact one always has $\Delta_S(0)<0$
which together with Eqs.~(\ref{Stau}), (\ref{Ov}), (\ref{tautilde})
and (\ref{Iex}) imply that the functions $S(\omega)$ and $I(\omega)$
share the same scaling as a positive power-law detailed in the
introduction (Subsec.~\ref{sub:results}) for the different branches.

\section{Conclusions}\label{concl}

In summary, we have analyzed a generic beam geometry suspended from
structureless supports in the limit of large aspect ratio $d/L\to0$
and provided for each of its low frequency resonances a
Caldeira-Leggett model adequate to describe the associated quantum
Brownian motion induced by phonon tunneling losses. The corresponding
effective Hamiltonian for the low frequency vibrational degrees of
freedom is derived from the underlying microscopic physics performing
a controlled approximation in the natural small parameter $d/L$. This
yields the lowest order contribution in the aspect ratio to the
associated environmental force spectral densities. We find two general
formulas for these functions Eq.~(\ref{eq:spectraltau}) and
Eq.~(\ref{eq:spectralrho}) that involve, respectively, the
transmission coefficient at a single junction and an effective
environmental DOS. Whence providing two alternative pictures for this
dissipation mechanism in terms, respectively, of phonon tunneling
losses and support-induced modification of the DOS
[cf.~Eqs.~(\ref{S}), (\ref{eq:spectralrho}), and (\ref{tautilde})].
These yield for the spectral densities functions of frequency that
only depend on the length $L$ and on properties of the ``decoupled''
support and of the decoupled infinite beam. Furthermore they are
universal in the specific sense that they only depend on the
properties of the beam through the quantities that determine its low
frequency effective theory known as TRE.

These environmental spectral densities result in fundamental limits
for the intrinsic dissipation (i.e.~{\sl upper bounds} for the
$Q$-values) with structureless supports which are relevant for state
of the art mechanical resonators in the $0.1-1$GHz frequency range
(cf.~Table~\ref{table} and Ref.~\onlinecite{Ekinci05}). It is
important to note the \emph{fast degradation as the length is
  shortened} and that the fundamental limit of torsional resonators
greatly exceeds that of flexural ones (for comparable frequencies
\footnote{This disparity is just a trivial consequence of the
  dispersion relations (which imply $\omega_{0,t}/\omega_{0,v/h}\sim
  L/d$) and the strong dependence of the corresponding $Q$-values with
  the aspect ratio --- for similar dimensions the $Q$-values of the
  fundamental resonances of these branches are comparable.}). In fact,
for torsion this dissipation mechanism is likely to be negligible when
compared with other contributions to the mechanical damping
\cite{Ekinci05}. In addition we find that semiconducting SWNTs are far
more resilient to this dissipation mechanism than semiconductor
heterostructure realizations (typical values for the figure of merit
$k_B T/\hbar\omega_R Q$ are at least an order of magnitude larger).

In the case of 3D supports the environmental spectral densities are
ohmic for flexural and compressional resonances and superohmic for
torsional resonances, while in the case of 2D slab supports they yield
$1/f$ noise with an infrared cutoff provided by the size of the slab.
It is worth noting that this type of noise is normally associated with
interactions involving charge degrees of freedom
\cite{Dutta81,Paladino02} while here it arises in an insulator from
purely vibrational effects.

Naturally, supports can be engineered to suppress the phonon tunneling
losses. However in many feasible alternatives analysis of phonon
propagation in a beam geometry is essential to determine the improved
limits attainable \cite{Wang00}. In this context and also when
analyzing situations where the resonator mode couples to a pseudospin
the resonator-bath representation given in this paper may need to be
complemented with an approximation for the complete phonon propagator
inside the beam. In other words the effect of the ``environment''
cannot be ``lumped'' into the finite mechanical $Q$-value and there
may be interference effects between different resonances arising from
correlations between their effective noise sources.  Clearly, away
from the resonances an adequate approximation for the complete phonon
propagator is afforded by the lowest order contributions in $d/L$ to
the scattering modes [cf.~Eq~(\ref{eq:u})]. The latter are specified
by Eqs.~(\ref{modetors})-(\ref{Abend}) and our approximations for the
reflection and transmission amplitudes given in
Sec.~\ref{s:singlejunc} --- of particular interest is the behavior
that results for $\omega\to0$. On the other hand a satisfactory
solution for \emph{all frequencies} requires going beyond the lowest
order for the reflection amplitudes --- a matter pursued in detail
elsewhere \cite{WilsonRaetobepublished}.

Furthermore, we highlight the relevance of the precise connection,
given in App.~\ref{ap:beam}, between the effective one dimensional TRE
description and the underlying transverse mode profile to scenarios
where the resonator is manipulated by coupling to an embedded optical
emitter \cite{WilsonRae04}.

Finally, we have provided a general expression for the spectral
density of a given resonance in terms of the relevant elastic modes
[cf.~Eq.~(\ref{genI})] that allows to extend the treatment to other
geometries. We have illustrated this for an axially symmetric
resonator supported by a vertical pedestal of length at least
comparable to the resonator's size. A scenario which is relevant for
optomechanical systems based on microtoroids, microdisks or
microspheres.

\begin{acknowledgments}
  The author thanks W.~Zwerger for invaluable help and thorough
  revision of the manuscript and acknowledges helpful discussions with
  C.S.~Pe\c ca, A.~Imamo\u{g}lu, A.N.~Cleland, P.~Zoller, P.~Tamborenea,
  C.~Tejedor, T.J.~Kippenberg, and E.~Weig.
\end{acknowledgments}

\appendix

\section{Normal modes of the free elastic half-space}\label{ap:halfspace}

The free elastic half-space supports four types of normal modes that
can be classified based on the character of the incident wave:
longitudinal bulk waves, transverse bulk waves with two polarizations
(SV and SH), and Rayleigh surface waves\cite{Landau,Graff}. In all four
cases the corresponding eigenfunctions can be written in the general
form:
\begin{equation}
\bar u^{(0)}_{\bar q,\gamma} (\bar r) =
\frac{1}{\left(2\pi\right)^{d_\gamma/2}}\left[\bar \varepsilon_0 e^{i
\bar q \cdot \bar r} + A_l \bar \varepsilon_l e^{i \bar q_l \cdot \bar
r} + A_t \bar \varepsilon_t e^{i \bar q_t \cdot \bar r}\right] \, ,
\end{equation}
where $d_\gamma$ is the dimensionality ($d_\gamma=2$ for $\gamma=s$
and $d_\gamma=3$ for $\gamma\neq s$), $\bar \varepsilon_0$ and $\bar q$
correspond to the polarization and wavevector of the incident wave,
and $A_{t/l}$, $\bar \varepsilon_{t/l}$ and $\bar q_{t/l}$ to the
amplitude, polarization and wavevectors of the reflected
transverse/longitudinal waves. These depend on $\bar q$ and $\gamma$
and are given by

\begin{description}
\item[longitudinal waves:]

\begin{subequations}
\begin{align}
  \bar q &= \rho (\sin \theta \cos \varphi, \sin\theta
  \sin\varphi, \cos\theta) \, , \\
  \bar q_l &= \rho(\sin \theta
  \cos \varphi, \sin\theta \sin\varphi, - \cos\theta) \, ,  \\
  \bar q_t &= \rho \left(\sin \theta \cos \varphi, \sin\theta
    \sin\varphi,
    - \sqrt{1/\alpha-\sin^2\theta}\right) \, , \\
  \bar\varepsilon_0 &= \frac{\bar q}{\rho} \, , \quad 
  \bar \varepsilon_l = \frac{\bar q_l}{\rho} \, ,  \\
  \bar\varepsilon_t &= \left(-\cos \varphi
    \sqrt{1-\alpha\sin^2\theta}, -\sin\varphi
    \sqrt{1-\alpha\sin^2\theta}, \right.
  \nonumber \\  & \quad \left. -\sqrt{\alpha}
      \sin\theta \vphantom{\sqrt{\sin^2}} \right)   \, , \\ 
  A_l &= \frac{2 \alpha^{3/2} \sin\theta \sqrt{1-\alpha \sin^2\theta}
    \sin 2\theta - (1-2 \alpha \sin^2\theta)^2}{2 \alpha^{3/2}
    \sin\theta \sqrt{1-\alpha
      \sin^2\theta} \sin 2\theta + (1-2 \alpha \sin^2\theta)^2}  \, , \\
  A_t &= \frac{2 \sqrt{\alpha} \sin 2\theta (2 \alpha
    \sin^2\theta-1)}{2 \alpha^{3/2} \sin\theta \sqrt{1-\alpha
      \sin^2\theta} \sin 2\theta + (1-2 \alpha \sin^2\theta)^2} \,; 
\end{align}
\end{subequations}

\item[SV waves:]

\begin{subequations}
\begin{align}
  \bar q & = \rho (\sin \theta \cos \varphi, \sin\theta \sin\varphi,
  \cos\theta) \,,
\\ 
\bar q_l &= \rho \left(\sin \theta \cos \varphi, \sin \theta
  \sin\varphi, - \sqrt{\alpha- \sin^2\theta + i\epsilon}\right) \, ,
\\
\bar q_t &= \rho (\sin \theta \cos \varphi, \sin \theta \sin\varphi,
-\cos \theta) \, , 
\end{align}
\begin{align}
\bar \varepsilon_0 &= (\cos \theta \cos \varphi, \cos \theta
\sin\varphi, -\sin\theta) \,,
\\
\bar \varepsilon_l &= \frac{\bar q_l}{\rho\sqrt{\alpha}} \, ,
\\
\bar \varepsilon_t &= - (\cos \theta \cos \varphi, \cos \theta
\sin\varphi, \sin\theta) \,,
\\
A_t &= \frac{2 \sqrt{\alpha-\sin^2\theta + i\epsilon} \sin\theta \sin
  2\theta - \cos^2 2\theta}{2 \sqrt{\alpha-\sin^2\theta + i\epsilon}
  \sin\theta \sin 2\theta + \cos^2 2\theta} \, ,
\\
A_l &= \frac{2 \sqrt{\alpha} \sin 2\theta \cos 2\theta}{2
  \sqrt{\alpha-\sin^2\theta + i\epsilon} \sin\theta \sin 2\theta +
  \cos^2 2\theta} \,;
\end{align}
\end{subequations}

\item[SH waves:]

\begin{subequations}
\begin{align}
  \bar q &= \rho (\sin \theta \cos \varphi, \sin\theta \sin\varphi,
  \cos\theta) \, , \\
  \bar q_t &= \rho (\sin \theta \cos \varphi,
  \sin\theta \sin\varphi, - \cos\theta) \, , \\
  \bar \varepsilon_0 &= \bar \varepsilon_t = (\sin
  \varphi, - \cos\varphi, 0) \, , \\
  \bar \varepsilon_l &=0 \, , \quad A_l = 0 \, ,\quad A_t =1 \,;
\end{align}
\end{subequations}

\item[surface waves:]

\begin{subequations}
\begin{align}
  \bar q &= \rho (\cos \varphi, \sin \varphi) \\ \bar q_l &= \rho
  \left(\cos \varphi, \sin \varphi, -i \sqrt{1-\alpha \xi^2}\right) \,
  , \\ \bar q_t &= \rho \left(\cos \varphi, \sin \varphi, -i \sqrt{1-
  \xi^2}\right) \, , \\ \bar \varepsilon_0 &= 0 \, , \quad \bar
  \varepsilon_l = \frac{\bar q_l}{\rho \, \xi \sqrt{\alpha}} \, , \\
  \bar \varepsilon_t &= \frac{1}{\xi} \left(i \sqrt{1-\xi^2}
  \cos\varphi, i \sqrt{1-\xi^2} \sin \varphi,1\right) \, , \\ A_l &= -
  \xi \sqrt{2 \alpha \rho}C(\alpha) \,, \quad A_t = i \xi
  \sqrt{2\rho}\frac{\xi^2/2-1}{\sqrt{1 - \xi^2}} C(\alpha)\,, \\
  C(\alpha) & = \left[
  \frac{2-\xi^2\!(\alpha)}{\left[1-\xi^2\!(\alpha)\right]^{3/2}}
  \left[\xi^4\!(\alpha)/4+\xi^2\!(\alpha)-1\right] \right. \nonumber
  \\ &\quad \left. + \frac{2-\alpha\xi^2\!(\alpha)}{\sqrt{1-\alpha
  \xi^2\!(\alpha)}} \right]^{-1/2}\,.
\end{align}
\end{subequations}
\end{description}
Here we have defined $\sqrt{\alpha}\equiv c_t/c_l$, where $c_t$ and
$c_l$ are the velocities of propagation of transverse and longitudinal
waves in the elastic medium, and adopted spherical coordinates for the
wavevector $\bar q$ (cf.~Fig.~\ref{fig:beams}). The parameter $\xi$ is
the ratio of the velocity of propagation for surface waves to $c_t$
and is a function of $\alpha$ that is always less than unity
\cite{Landau}. We note that for SV waves with polar angles $\theta >
\cos^{-1}\sqrt{1-\alpha}$ the longitudinal component is evanescent.

\section{Normal modes of the beam}\label{ap:beam}

\subsection{TRE solutions}\label{sub:tre}

At low frequencies there are four branches of propagating modes in an
infinite beam: two bending branches (vertical and horizontal), a
torsional and a compressional branch. An heuristic way of
understanding how these four types of motion arise is to decompose the
associated displacements of the points of each cross section into an
overall translation, an overall rotation, and a ``residual''
deformation. The vicinity of free boundary conditions and the fact
that the wavelength $2\pi/k$ of these low frequency modes is much
longer than $d$ determine that the deformation is higher order in the
small parameter $k d$, when compared with the rotation and
translation. Thus each cross section can be seen as a slightly
deformed rigid body. Clearly if we considered a chain of coupled rigid
bodies, each unit would have six degrees of freedom that would lead to
six branches. However only \emph{four} of them will have a vanishing
dispersion relation and respect the requirement that the cross
sections remain only slightly deformed as they move together, namely,
those associated with the three possible translations and to rotation
around the beam's axes. As these considerations are quite general,
though throughout this Section we focus on the case when the low
frequency effective theory can be derived from 3D elasticity
[cf.~Eq.~(\ref{disper})], in the case of nanowires and SWNTs the
expressions for the TRE modes that follow are warranted up to the
highest order for which the strain vanishes --- i.e.~first order for
bending and zeroth order for compression and torsion --- with discrete
coordinates $x,y$ and up to a prefactor.

We turn now to the analysis of the small $kd$ behavior of the
corresponding normalized eigenmodes for these four propagating
branches ($\beta=c,t,v,h$ with $m=0$). We may focus on a right mover
given by $|v^{(0)}_\beta(k)\rangle\equiv\bar A_{\beta,0}[x,y,d,-ik]
e^{ikz}$ with eigenvalue $\omega^2_{\beta,0}(k,d)$ and obtain the left
movers by reflection symmetry and (in the case of bending) the large
decay length exponentials $\bar A_{v/h,1}$ by analytic continuation.
The problem of determining the Taylor expansions of $\bar A_{\beta,0}$
and $\omega^2_{\beta,0}(k,d)$ in powers of $k$ can be formulated as the
search for an harmonic solution $\bar u_\beta(\bar r,t)$ of the 3D
elastic wave equations for the beam \cite{Graff,Landau} via the
ansatz:
\begin{equation}\label{eq:expphi}
  \bar u_\beta(\bar r,t) = \sum_{n=0}^\infty \bar
  A_{\beta,0}^{(n)}(x,y) \frac{\partial^n\phi}{\partial z^n}(z,t) , 
\end{equation}
with $\phi (z,t) = e^{i (k z-\omega t)}$. If one neglects in
Eq.~(\ref{eq:expphi}) all orders that scale at least as the inertia
(i.e.~as $\omega^2$) one obtains the TRE low frequency effective
theory \cite{Graff,Landau}.  Then the function $\phi(z,t)$ will
correspond to the effective one dimensional field.  On the other hand,
the lowest order for $\omega^2$ yields the corresponding approximate
dispersion relation $\omega \propto k^{p_\beta}$. It is
straightforward to realize by substitution of Eq.~(\ref{eq:expphi})
into the elastic wave equations for the beam \cite{Graff,Landau} that
the result of this truncation at order $2p_\beta-1$ will yield an
exact solution of the corresponding static equations ($\omega\to 0$)
provided that $\frac{\partial^{2p_\beta}\phi}{\partial z^{2p_\beta}}$
vanishes. Thus the leading terms of the Taylor expansions of the
transverse profiles $\bar A_{\beta,0}$ can be extracted from the
corresponding static solutions. These are given by

\begin{description}
\item[compression:]

\begin{align}\label{eq:profilec}
u_x (x,y,z) &= - \sigma x \frac{\partial \phi}{\partial z}(z) \,
,\nonumber \\ u_y (x,y,z) &= - \sigma y \frac{\partial \phi}{\partial
z}(z) \, ,\nonumber \\ u_z (x,y,z) &= \phi(z) \, ;
\end{align}

\item[torsion:]

\begin{align}\label{eq:profilet}
u_x (x,y,z) &= -y \phi(z)  \, , \nonumber\\ 
u_y (x,y,z) &= x \phi(z)  \, , \nonumber\\
u_z (x,y,z) &= \psi (x,y) \frac{\partial \phi}{\partial z}(z) \, ; 
\end{align}

\item[vertical bending:]

\begin{align}\label{eq:profileb}
  u_x (x,y,z) &= \phi(z) + \frac{\sigma}{2} (x^2 - y^2)
    \frac{\partial^2 \phi}{\partial z^2}(z) \, , \nonumber\\ u_y
    (x,y,z) &= \sigma x y \frac{\partial^2 \phi}{\partial z^2}(z) \, ,
    \nonumber\\ u_z (x,y,z) & = -x \frac{\partial \phi}{\partial z}(z)
    + \sigma \chi_v (x,y) \frac{\partial^3 \phi}{\partial z^3}(z)\,.
\end{align}
\end{description}
In the above, the coordinate system is oriented along the principal
axes of inertia, $(x,y)=0$ corresponds to the center of mass of the
cross section, and $\sigma$ is the Poisson ratio of the beam's material.
The functions $\psi(x,y)$ and $\chi_v (x,y)$ are determined by the
static equation $\nabla\cdot\bv{\sigma}=0$, the free boundary
conditions, and the condition $\langle u_z\rangle_S=0$. In the case of
bending these yield
\begin{align}\label{eq:chi}
  \frac{\partial \chi_v}{\partial x}(x,y) &= -\frac{\partial
    \tilde\chi_v}{\partial y}(x,y) +
    \frac{x^2+y^2}{2\sigma}\,,\nonumber\\ \frac{\partial
    \chi_v}{\partial y}(x,y) &=\frac{\partial \tilde\chi_v}{\partial
    x}(x,y) + \frac{xy}{\sigma}\,,\nonumber\\ \langle \chi_v
    (x,y)\rangle_S &=0\,,
\end{align}
where $\tilde\chi_v$ is an harmonic function that at the cross section's
boundary $(x[l],y[l])$ satisfies
\begin{align}\label{eq:chitilde}
  \tilde\chi_v(l)= &
  \frac{1}{6\sigma}\left[3(1+\sigma)x^2y+(1-\sigma)y^3\right]^l_0 \nonumber\\ 
 & - \frac{2(1+\sigma)}{\sigma}\int^l_0 xy\frac{\ud x}{\ud l'}\ud l'.
\end{align}
The analogous relations for the case of torsion are given in Reference
[\onlinecite{Landau}] while the transverse profile of the displacement
field for \emph{horizontal bending} can be obtained from the RHS of
Eq.~(\ref{eq:profileb}) and Eqs.~(\ref{eq:chi}), (\ref{eq:chitilde})
via the ansatz $x \leftrightarrow y$,
$\chi_v(y,x)\rightarrow\chi_h(x,y)$,
$\tilde\chi_v(y,x)\rightarrow\tilde\chi_h(x,y)$.

Finally, the desired approximation for $\bar A_{\beta,0}[x,y,d,-ik]$
is obtained from Eqs.~(\ref{eq:profilec})-(\ref{eq:profileb}) via the
replacement
\begin{equation}
  \frac{\partial^n\phi}{\partial z^n}\to \begin{cases}
      \frac{1}{\sqrt{2\pi S}} (ik)^n & \text{for}\  \beta=c,v,h \\[5pt]
 \frac{1}{\sqrt{2\pi I_z}} (ik)^n & \text{for}\  \beta=t
\end{cases} 
\end{equation}
where we have taken into account that the 1D continuum modes
$|v^{(0)}_\beta(k)\rangle$ are normalized in the standard Euclidean
metric. Whence, we define
\begin{equation}\label{def:phi}
 \phi_{k(q),\beta}(z)= \begin{cases} \frac{\sqrt{2\pi
      S}}{t_q}\,u_{z,q}(0,0,z) & \text{for}\ \beta=c \\[5pt]
      \frac{\sqrt{\pi I_z}}{\sqrt{2} t_q}\,\hat{z}\cdot\nabla\times \bar
      u_{q}(0,0,z) & \text{for}\ \beta=t \\[5pt] \frac{\sqrt{2\pi
      S}}{t_q}\,u_{x/y,q}(0,0,z) & \text{for}\ \beta=v/h\,.
\end{cases} 
\end{equation}
The corresponding approximate dispersion relations (\ref{disper}) are
specified by
\begin{align}\label{tildec}
 \tilde{c}_{c} = \sqrt{\frac{E_b}{\rho_b}}\,, && \tilde{c}_{t} =
 \sqrt{\frac{C}{\rho_b I_z}}\,, && \tilde{c}_{v/h} = \sqrt{\frac{E_b
     I_{y/x}}{\rho_b S}}\,.
\end{align}
Here $I_i$ are the moments of inertia with respect to the principal
axes of a cross section per unit surface density, $C$ is the torsional
rigidity of the beam and $E_b$ its Young's modulus.

\subsection{End corrections}\label{sub:endcorr}

If one considers now a semi-infinite beam ($z>0$) it is clear that
complex values for $k$ (the wavevector along $z$) are physically
meaningful provided that their imaginary part is positive. The
corresponding solutions can be understood as the analytic continuation
of the ``traveling wave'' eigenmodes of an indefinite beam. More
precisely, $\omega^2(k,d)$ and the transverse profile $\bar
A(x,y,k,d)$ will be multivalued analytic functions of $k$ that yield
harmonic solutions on the paths in the complex plane specified by
$\Re[\omega^2(k,d)]\geq 0,\Im[\omega^2(k,d)]=0$. This can be viewed
explicitly for a cylindrical cross section given that the resulting
problem is separable (see Ref.~[\onlinecite{Graff}] and references
therein) but will hold in general provided that the boundary of the
cross section is well behaved. We assume for simplicity that the
latter is characterized by a single length scale $d$ but all our
considerations follow for an arbitrary cross section (within the
caveats that follow) provided that we reinterpret $d$ as the largest
chord.  Heuristic considerations imply that for a generic geometry the
finite complex zeroes of $\omega^2(k,d)$ will have a real part that is
at least of order $1/d$. We will further assume that at these zeroes
$\omega^2(k,d)$ is non-singular (i.e.~they are not branch points) and
has non-zero derivative; and that the associated path yielding
physical solutions admits a parametrization in terms of $\Re[k]$ or
$\Im[k]$ that can be Taylor expanded. Thus each of these non-trivial
zeros yields a branch of end corrections (with labels \footnote{For
  asymmetric cross sections the $\beta$ index is conventional.}
$\beta,m$ as defined in Subsec.~\ref{sub:ustar}) specified by the
transverse profiles $\bar
A_{\beta,m}[x,y,d,\kappa_{\beta,m}(\omega^2,d)]$ and the ``wavevectors
along $z$'' $\kappa_{\beta,m}(\omega^2,d)\equiv -ik$, where the
functions $\bar A_{\beta,m}$, $\kappa_{\beta,m}$ can be expanded in
natural powers of $\omega^2$.

To analyze the problem of transmission at a 3D-1D junction it proves
useful to consider the overall displacement and rotation of a given
cross section. The former is simply defined as the displacement of the
cross section's center of mass. In turn, a natural definition
of a spatially averaged angle for a neighborhood $V$ undergoing
harmonic motion is afforded by considering the maximum total angular
momentum $\bar L_V$ with respect to the center of mass
$\bar r_V$ of $V$ over a period:
\begin{equation}
  \langle \bar u(\bar r) \rangle_{\mathrm{ang},V} \equiv \frac{1}{\omega}\,
  \mathbb{I}^{-1} \bar L_V
\end{equation}
where $\mathbb{I}$ is the equilibrium inertia tensor for $V$. If we
orient the axes along the principal axes of inertia and set the origin
at $\bar r_V$, we obtain:
\begin{equation}\label{ang}
\langle \bar u(\bar r) \rangle_{\mathrm{ang},V}\cdot\hat i =
    \frac{\int_V \ud r^3 \,\bar r \times \bar u (\bar r)\cdot\hat
    i}{\int_V \ud r^3 (r^2-x_i^2)}.
\end{equation}
If $V$ is small enough this reduces to
\begin{equation}\label{angapprox}
\langle \bar u(\bar r) \rangle_{\mathrm{ang},V}\cdot\hat i \approx
\frac{1}{2} \nabla \times \bar u (0) \cdot\hat i + \gamma_i
u_{jk} (0) \ \text{with}\ \epsilon_{ikj}=1
\end{equation}
where $\gamma_i\equiv \sum_{j,k} \epsilon_{ijk}(I_k-I_j)/2I_i$ and the
relative order of the corrections is at most
$u(0)d/\min\{|u_{ij}(0)|\}$ with $d$ the typical dimension of the
neighborhood $V$. We note that throughout Sec.~\ref{s:singlejunc} we
use the notation
\begin{equation}\label{angnotation}
\bar \theta\equiv\langle \bar u(\bar r) \rangle_{\mathrm{ang},S}
\end{equation}
where $S$ is the beam's cross section at the origin.

In the limit $\omega\to 0$ each branch of end corrections yields an
exponentially decaying solution ($\propto
e^{-\kappa_{\beta,m}(0,d)z}$) of the corresponding static problem
(i.e.~a \emph{static} end correction). To establish
Eqs.~(\ref{eq:proptrans}) underpinning our analysis of a 3D-1D
junction, in addition to the aforementioned analyticity properties, we
just need to use the following exact \emph{universal} properties of
static end corrections:
\begin{align}\label{eq:propend1}
\!\!\!\! \langle \mathbf{F}\cdot
\left\{\bar{A}_{\beta,m}\left[x,y,d,\kappa_{\beta,m}(0,d)\right]
e^{-\kappa_{\beta,m}(0,d)z} \right\}\rangle_{S} & =\, 0
\\\label{eq:propend2} \langle \mathbf{F}\cdot
\left\{\bar{A}_{\beta,m}\left[x,y,d,\kappa_{\beta,m}(0,d)\right]
e^{-\kappa_{\beta,m}(0,d)z} \right\}\rangle_{\mathrm{ang},S} & =\, 0
\\ \label{eq:propend3} \langle \bar
A_{\beta,m}\left[x,y,d,\kappa_{\beta,m}(0,d)\right]\rangle_{S} & =\, 0
\\ \label{eq:propend4} \langle \bar
A_{\beta,m}\left[x,y,d,\kappa_{\beta,m}(0,d)\right]\rangle_{\mathrm{ang},S}
& =\, 0 \\ \label{eq:propend5} \langle\int_{S}\ud {r'}^2 \,\mathbf{G}
\left(\bar r - \bar r',0\right) \cdot \,\,\mathbf{F}\cdot
\left\{\bar{A}_{\beta,m}
\left[x',y',d,\kappa_{\beta,m}\right.\right. &
\left.\left.\!\!\!\!(0,d)\right]\right.  \nonumber \\ \left.\times\,
e^{-\kappa_{\beta,m}(0,d)z'} \right\}\rangle_{S} & =\, 0 \\
\label{eq:propend6} \langle \int_{S}\ud {r'}^2 \,\mathbf{G} \left(\bar
r - \bar r',0\right) \cdot\,\, \mathbf{F}\cdot
\left\{\bar{A}_{\beta,m}\left[x',y',d,\kappa_{\beta,m}\right.\right. &
\left.\left.\!\!\!\!(0,d)\right]\right.  \nonumber\\ \left.\times\,
e^{-\kappa_{\beta,m}(0,d)z'} \right\}\rangle_{\mathrm{ang},S} & =\,
0\,,
\end{align}
where now $\mathbf{G}(\bar r,0)$ is the static Green's function of a
free elastic half-space loaded at the free surface. Equations
(\ref{eq:propend1}) and (\ref{eq:propend2}) correspond, respectively,
to the total force and torque applied at $S$ and can be established by
simply stating the equilibrium conditions for a given finite segment
of the beam. In fact the exponential dependence on $z$ implies that
the total external force and total external torque applied to the
finite segment result, respectively, proportional to the L.H.S of
Eqs.~(\ref{eq:propend1}) and (\ref{eq:propend2}).

On the other hand Equations (\ref{eq:propend3}) and
(\ref{eq:propend4}) can be reduced to the following Lemma:

\textbf{Lemma:} \textit{For a finite beam of length $L$ in equilibrium
  subject to specified displacements at $z=0$ and free boundary
  conditions at $z=L$ the average displacements and angles at both
  ends coincide,} \\[5pt]
by taking the displacements at $z=0$ specified by $\bar
A_{\beta,m}[x,y,d,\kappa_{\beta,m}(0,d)]$ and then sending
$L\to\infty$. In turn, the above Lemma can be understood using a
variational argument given that the elastic energy density is positive
definite in the relative distances between the material points of the
beam (see Ref.~[\onlinecite{Graff}] and references therein). Equations
(\ref{eq:propend1}) through (\ref{eq:propend4}) underly the standard
recipes for the boundary conditions in the TRE treatment of static
small deflections (i.e.~``linearized-strain'' theory)
\cite{Graff,Landau}. In this respect, it is important to note that for
a beam in equilibrium the corresponding TRE solutions, already
discussed, are exact within the linearized-strain 3D theory.  Thus the
validity of the latter recipes hinges on whether the strain can be
linearized and not on the smallness of $d/L$.

We now consider the cantilever geometry (cf.~Fig.~\ref{fig:beams}) for
an \emph{arbitrary} length $L$ and apply the stress source associated
to a given static end correction ($\beta,m$) at $z=L$, the usual free
boundary conditions at the other surfaces and fixed displacement
boundary conditions for $z\to -\infty$ in the support (the latter
ensues for $\omega\to 0$ given the boundary conditions satisfied by
the support's Green's function).  Equations
(\ref{eq:propend1})-(\ref{eq:propend2}) imply that the TRE part of the
corresponding static solution inside the beam
[cf.~Eq.~(\ref{eq:profilec})-(\ref{eq:profileb})] is at most an
overall displacement and rotation. However a variational argument
analogous to the one underlying the above Lemma implies that this
displacement and rotation of the beam should vanish so that the
solution is given solely by ``end corrections''. Then, continuity at
the junction yields Eqs.~(\ref{eq:propend5})-(\ref{eq:propend6}) if we
take the limit $L\to 0$. We note that this derivation will be valid
for any well behaved support \emph{irrespective of any symmetries}.

\section{Generalizations to asymmetric structures, 2D slab
  supports, and SWNTs}\label{ap:exten}

The derivations in Subsec.~\ref{sub:ustar} rely on the assumption that
both the support and the beam are symmetric under reflections with
respect to the $x$ and $y$ axes. It is straightforward to extend these
results to the case where only the beam's cross section presents these
symmetries so long as we assume that the (now asymmetric) support is
still characterized by a \emph{structureless continuum} of free
eigenmodes $\bar u^{(0)}_q(\bar r)$ that has an effective 3D density
of states \footnote{Any surface or edge states resulting from the free
  boundaries have a penetration depth of order $1/q_t$.}.  More
specifically each mode $\bar u^{(0)}_q(\bar r)$ should be
characterized by a single length scale of order $1/q_t$ (where
$q_t(\omega_q)$ is the wavevector for propagation of transverse
elastic waves in the support's material). Thus the analogous property
will hold for the corresponding Green's function $\mathbf{G} (\bar
r,\bar r',\omega)$ ($\bar r'$ lies at the boundary where the normal
stress vanishes when the support is unloaded). A simple example would
be an elastic quarter-space or any fraction of an elastic space with
origin at the junction subtending a solid angle of order $\pi/2$. In
this more general scenario there will be mode mixing between the
branches due to the scattering at the support. Equation
(\ref{eq:deltau}) will still be valid with the analogous definition
for $\bar u_*$ but now in Eq.~(\ref{eq:decomp}) the decompositions of
the fields $\Delta\bar u_+$ and $\bar u_*$ in terms of the eigenmodes
of the beam will involve an additional sum over all the branches
$\beta$ (i.e.~$\sum_m \to \sum_{m,\beta}$). Concomitantly these fields
will no longer have a branch index except for $\bar u_*$ in case ($i$)
where $\beta'$ will denote the character of the associated incident
field $\bar u_{\mathrm{in},\beta'}$, and the analog of
Eq.~(\ref{boundstar}) will no longer involve a symmetrization operator
$\hat S_{\beta}$. The support mode labels $q$ no longer bear any
relation with the beam branch index $\beta$ and now a convenient
parametrization of the complete solution is afforded by $q_t$ instead
of $k$.

In complete analogy to the symmetric case we substitute the
\emph{modified} decompositions (\ref{eq:decomp}) into the
\emph{modified} Eq.~(\ref{boundstar}) and Eq.~(\ref{eq:deltau}) and
take on both sides the spatial averages $\langle\ldots\rangle_{S}$ and
$\langle\ldots\rangle_{\mathrm{ang},S}$. Note that the beam's
symmetries imply that on the LHS of each of the resulting equations
for the amplitudes $c_{\beta m}^{(*)}$ and $c_{\beta m}$ only the term
with the relevant $\beta$ --- i.e. the one associated with the TRE
branch for which the corresponding average does not vanish --- will
survive. Once again we solve for the TRE starred amplitudes in the
linear system arising from the analog of Eq.~(\ref{boundstar}) and
substitute them into the linear system arising from the analog of
Eq.~(\ref{eq:deltau}), whence we solve for the un-starred TRE
amplitudes. As before the end corrections yield terms that scale at
most as the inertia $(q_t d)^2$. Thus, we find that --- with the
noticeable exception of the un-starred torsional TRE amplitude for
case ($ii$) --- the results for the amplitudes $c_{\beta m}^{(*)}$ and
$c_{\beta m}$ are given by Eqs.~(\ref{eq:limc})
and (\ref{eq:limcsct})-(\ref{eq:limcsvh2s}) with the following
straightforward modifications: $kd$ is replaced by $\sqrt{q_t d}$,
$\bar u^{(0)}_q(\bar r)$ denotes now the free modes of the ``generic''
support, and in case ($i$) the RHS of the equations for the starred TRE
amplitudes are now multiplied by $\delta_{\beta,\beta'}$.

The caveat for incidence from the support [case ($ii$)] is that now
$c_{t,0}(q_t d)$ has a contribution comparable to $c^{(*)}_{t,0}(q_t
d)$ of order $q_t d$ arising from the other TRE branches. Thus in this
case Eq.~(\ref{eq:approxu}) is no longer valid. However the correction
to the approximate transmission amplitude $t_{q,t}^{(0)}$ (whose
scaling as $q_t d$ is preserved) arising from $c_{t,0}(q_t d)$ can be
readily added to yield
\begin{align}\label{corrt}
t_{q,t}^{(0)}\,=\, & \frac{\partial c^{(*)}_{t,0}}{\partial q_t}(0)\,
q_t\, + \sum_{\{\beta\neq t,m\}\in \mathrm{TRE}}\sqrt{2\pi I_z} \hat z
\nonumber\\ & \cdot \langle\int_{S}\ud {r'}^2 \mathbf{G} \left(\bar
r,\bar r',0\right)\cdot\hat z\,f^{(0)}_{\beta,m}(\bar
r')\rangle_{\mathrm{ang},S}\, c_{\beta m}^{(*)}(0)
\end{align}
with
\begin{align}
f^{(0)}_{v/h,m}(x_1,x_2) & = (-1)^m \frac{E_b}{\sqrt{2\pi
S}}x_{1/2}k^2_{v/h}(q_t) \quad m=0,1 \nonumber\\
f^{(0)}_{c,0}(x_1,x_2) & = i\frac{E_b}{\sqrt{2\pi S}} k_c(q_t)
\,,
\end{align}
where we have used the transverse profiles of the TRE modes given in
App.~\ref{ap:beam} and $\mathbf{G} \left(\bar r,\bar r',0\right)$ is
now the static Green's function for the asymmetric support under
consideration. 

The aforementioned results for the starred amplitudes imply that for
incidence from the beam to lowest order in $q_t d$ there is no mixing
between the branches so that the reflection amplitudes for
$\beta\neq\beta'$ vanish and those for $\beta=\beta'$ coincide with
the $r^{(L)}_{c/t}[k(q_t)d]$, $r^{(L)}_{v/h,\delta\eta}[k(q_t)d]$
already found for the symmetric case. On the other hand for incidence
from the support the solution is a superposition of all four branches
with approximate amplitudes $t_{q,\beta}^{(0)}$ given by
Eq.~(\ref{corrt}) and for $\beta\neq t$ the same results as in
Subsec.~\ref{sub:ustar} in terms of the modified $\bar u^{(0)}_q(\bar
r)$. If we now revise the treatment of the modes of the whole
structure given in Sec.~\ref{s:FP}, the fact that the branches do not
mix implies that to lowest order the displacement field inside the
beam (and away from the junctions) is just given by adding coherently
the four contributions of the TRE branches which are given by similar
expressions as in the symmetric case but with modified transmission
amplitudes $t_{q,\beta}^{(0)}$ (i.e.~the reflection amplitudes and the
ratio $b$ remain the same). Furthermore, the symmetries of the beam
imply that a resonator mode of branch $\beta$ only has non-vanishing
overlap with the contribution to a scattering mode of the same branch
so that the quantities $\langle u'_R | u_q \rangle$ (to lowest order
in $q_t d$) coincide with those for the symmetric supports except for
the modification in the prefactor $t_{q,\beta}^{(0)}$. Thus, in
general the results for the force spectral densities
(cf.~Sec.~\ref{sub:Itau}) that we will extract from these overlaps
will be completely analogous to those for the fully symmetric case.

The only caveat is when there are degeneracies between resonances of
different branches, i.e.~$\omega_{m,\beta}=\omega_{n,\beta'}=\omega_R$
with $\beta\neq\beta'$. Unlike the fully symmetric case there is now
mode mixing which, albeit higher order in $d/L$, may nonetheless
invalidate our approximations for the functions $I_{m,\beta}(\omega)$,
$I_{n,\beta'}(\omega)$ in a neighborhood of $\omega_R$
(cf.~Sec.~\ref{sub:Itau} and App.~\ref{ap:I}). The situation is
completely equivalent to the failure of our treatment for
$I_{n,\beta}(\omega_{m,\beta})$ with $n\neq m$ with the difference
that now the syndrome occurs precisely where the function is more
relevant. On the other hand it is simple to realize that this
breakdown is completely consistent with the nature of our
approximation which is perturbative in $d/L$. If the asymmetry removes
the selection rules that prevented the supports from coupling the two
degenerate modes, the Caldeira-Leggett Hamiltonian provided by
Eq.~(\ref{eq:h-rb}) with a single discrete resonator mode no longer
affords a convenient representation as $d/L\to0$. In particular, this
will affect the two bending branches for cross sections symmetric
under rotations by $\pi/2$ whenever the supports break the $\hat{R}_y$
symmetry [cf.~Eq.~(\ref{refsym}) and Fig.~\ref{fig:beams}]. Thus, in a
realistic scenario (i.e.~at finite $d/L$ and finite mode-splitting)
our results for the bending resonances $\omega_{n,v/h}$ will only be
applicable to the case of SWNTs and nanowires (discussed below)
provided the mode-splitting is negligible compared with the natural
linewidth $\omega_{n,v/h}/Q_{n,v/h}$ induced by the phonon tunneling.

We have also analyzed the case when the cross section is also
asymmetric following the analogous procedure. The key point in this
further extension is that the lowest order contributions in $kd$ to
the TRE mode profiles given in App.~\ref{ap:beam} satisfy the
reflection symmetries for arbitrary cross section (with the axes
oriented along the principal axes of inertia). Thus when the starred
amplitudes are zeroth order in $q_t d$ the overlaps $\langle u'_R |
u_q \rangle$ will once again coincide [to lowest order in $q_t d$ and
up to a modification in the prefactor $t_{q,\beta}^{(0)}$] with those
for the symmetric supports. It follows that the extension to
asymmetric cross sections carries over for the results that will ensue
for the force spectral densities (cf.  Sec.~\ref{sub:Itau}) except for
torsional resonances --- in which case the relevant transmission
amplitudes are higher order --- and the aforementioned case of
degeneracies.

Furthermore, we have used the same framework to analyze the analogous
problem of an abrupt junction between a rectangular beam of width
$\mathsf{w}$ and a slab (support) of the same thickness \cite{Cross01}
$\mathsf{t}\ll \mathsf{w}\equiv d$. Thus, we can use instead of 3D
elasticity the 2D effective theory provided by ``thin plate
elasticity'' \cite{Cross01,Landau} adequate for phonon wavelengths
much larger than $\mathsf{t}$. A completely analogous treatment is
feasible based on the 2D analog of Eqs.~(\ref{eq:deltau}),
(\ref{eq:decomp}), with: ($i$) $\bar u^{(0)}_q(y,z)$ and
$\mathbf{G}(y,z,\omega)$ denoting now, respectively, the free modes
and retarded Green's function of the two dimensional elastic
half-plane (i.e. a semi-infinite thin plate lying on the $yz$-plane in
our notation), and suitable redefinitions of ($ii$) the operator
$\mathbf{F}$ and ($iii$) the beam's mode profiles
$\bar{A}_{\beta,m}[y,d,\kappa_{\beta,m}(k,d)]$ which are now
one-dimensional. The reflection symmetries also apply in this case so
that the vertical bending and torsional branches of the beam only
couple to the flexural modes of the plate while its horizontal bending
and compressional branches only couple to the in-plane modes of the
latter. Thus one can prove the same results for the reflection
amplitudes and obtain results analogous to Eqs.~(\ref{eq:limcsct}),
(\ref{eq:limcsvh2s}) for the transmission amplitudes $t_q$, $b t_q$ in
terms of the averaged displacements and angles of the slab support ---
the redefinitions ($iii$) imply that in the corresponding
cross-section-dependent prefactors one has to replace $S\to
\mathsf{w}$ and $I_z\to \mathsf{w}\langle r^2\rangle_S$
(cf.~Subsec.~\ref{sub:results}). The only caveat is that for vertical
bending the second relation for the starred amplitudes in
Eq.~(\ref{eq:limcsvh2}) is no longer homogeneous since now the
averaged angles $\theta_y$ scale as $\sqrt{\omega}$ for the relevant
``free modes'' of \emph{both} the beam and the plate (flexural modes).
This leads to a modification of the ratio $b$ for vertical bending
that now may acquire a $q$-dependence. As a consequence of the above
analogy the relations (\ref{eq:spectralrho}) [(\ref{eq:spectraltau})]
between the spectral densities $I_{n,\beta}(\omega)$ and the effective
environmental DOS of the support $\tilde\rho_\beta (\omega)$
[transmission coefficients $\tau_\beta(\omega)$] will be the same (in
terms of the analogous functions for the thin plate ``support'') as in
the 3D case with the exception of the expression for $C_{n,v}$ that
will be modified for small $n$ --- for large $n$ the large decay
length exponentials can be neglected and the dependence on the ratio
$b$ of $I_{n,\beta} (\omega)$ becomes negligible.

Finally, we turn to the case of SWNTs and nanowires. One should note
that as there is no valid underlying ``microscopic'' theory applicable
to both the supports and the beam the precise analog of
Eq.~(\ref{eq:deltau}) will in general be unknown, depending on details
of the clamping procedure. However, one can argue that insofar as
there is no dissipation \emph{inside} the ``junction'' the above
results for the reflection and transmission amplitudes will still hold
--- with the straightforward replacements in Eqs.~(\ref{eq:limcsct}),
(\ref{eq:limcsvh2}), (\ref{eq:limcsvh2s}) that result from the
redefinition of the transverse mode profiles: $S\to N_c$, $I_z\to
N_c\langle r^2\rangle_S$, with $N_c$ the number of atoms in the unit
cell \footnote{For a SWNT if we use the shell model the appropriate
  ansatz reads instead $S\to 2\pi R$, $I_z\to 2\pi R\langle
  r^2\rangle_S$ --- naturally, the results for the functions $\tilde
  \rho_\beta(\omega)$ are independent of the ``microscopic model''
  used [cf.~Eqs.~(\ref{eq:limcsct}),
  (\ref{eq:limcsvh2s}),(\ref{recipro})].} (of the appropriate
equivalent 1D chain in the case of a nanowire \cite{Segall02}). In
particular, one should note that the analog of the universal
properties of the end corrections (cf.~App.~\ref{ap:beam}) are
expected to hold if the SWNT or nanowire is to be regarded as
``clamped''.

\section{Extension to microtoroids and other geometries}\label{ap:toro}

The facts that the resonator mode $|u'_R\rangle$ and the scattering
modes $|u_q\rangle$ are solutions of the time-independent elastic wave
equations (cf.~Sec.~\ref{II}), and that $\bar{u}'_R(\bar r)$ is real,
directly imply \cite{Landau}:
\begin{align}
\omega_q^2 \langle u'_R|u_q\rangle & = - \frac{1}{\rho_s} \int_{V_R}
\!\!\! \ud r^3 \, u'_{R,i} \frac{\partial \sigma_{q,ij}}{\partial  x_j}  
 \label{ovIa:tor}\\
\omega_R^2 \langle u'_R|u_q\rangle & = - \frac{1}{\rho_s} \int_{V_R}
\!\!\! \ud r^3 \, u_{q,i} \frac{\partial \sigma'_{R,ij}}{\partial x_j} 
 \label{ovIb:tor}
\end{align}
where we use Einstein's sum convention and that $\bar{u}'_R(\bar r)$
only has support in $V_R$. The above can be re-expressed as
\begin{align}
\omega_q^2 \langle u'_R|u_q\rangle & =  - \frac{1}{\rho_s} \int_{V_R}
\!\!\! \ud  r^3 \left[ \frac{\partial}{\partial x_j} \left( u'_{R,i}
    \sigma_{q,ij} \right) -  u'_{R,ij} 
  \sigma_{q,ij} \right]
 \label{ovIIa:tor}
\end{align}
\begin{align}
\omega_R^2 \langle u'_R|u_q\rangle & =  - \frac{1}{\rho_s} \int_{V_R}
\!\!\! \ud r^3 \left[ \frac{\partial}{\partial x_j} \left( u_{q,i}
    \sigma'_{R,ij} \right) -  u_{q,ij} 
  \sigma'_{R,ij} \right] \label{ovIIb:tor}
\end{align}
by using that the stress tensor is symmetric and that the contraction
of a symmetric tensor with an antisymmetric one vanishes. As the
mapping of the strain $u_{ij}$ onto the stress $\sigma_{ij}$
corresponds to a symmetric matrix (given that its entries are the
second derivatives of the elastic energy with respect to the strain
components) we have
\begin{equation}
u'_{R,ij}\sigma_{q,ij}=u_{q,ij}\sigma'_{R,ij}\,. \label{symov}
\end{equation}
Substracting Eq.~(\ref{ovIIb:tor}) from Eq.~(\ref{ovIIa:tor}) and
using Eq.~(\ref{symov}), the divergence theorem, and the free boundary
conditions satisfied by the displacement fields at the free surfaces
of the resonator we obtain
\begin{equation}\label{ovgen}
 \langle u'_R|u_q\rangle = \frac{\int_S \ud r^2\,\left( \bar{u}'_R
    \cdot \bv{\sigma}_q - \bar{u}_q \cdot \bv{\sigma}'_R \right) \cdot
    \hat{n}}{\rho_s \left[\omega^2_R - \omega^2(q)\right]}
\end{equation}
which together with Eqs.~(\ref{eq:S}) and (\ref{Iapprox}) finally
yields Eq.~(\ref{genI}), namely
\begin{align}
I(\omega)\approx \frac{\pi}{2 \rho_s^2 \omega_R \omega} \int_q  &
\left| \int_S \ud r^2\,\left( \bar{u}'_R \cdot \bv{\sigma}_q -
    \bar{u}_q \cdot \bv{\sigma}'_R \right) \cdot \hat{n} \right|^2
\nonumber \\  & \times\delta [\omega - \omega(q) ]\,. \nonumber
\end{align}
The above is very general as it only assumes
$|\Delta_I(\omega_R)|/\omega_R\ll 1$ to ensure the validity of
Eq.~(\ref{Iapprox}). 

We now focus on an axially symmetric structure consisting of a
``resonator volume'' supported by a vertical pedestal. Concrete
relevant realizations of this geometry are microtoroids
(cf.~Fig.~\ref{fig:torus}), microdisks and microspheres. We consider
the regime in which the contact area $S$ between the resonator
volume and the pedestal satisfies $S\ll h D$ and $\sqrt{S}\lesssim
h$, where $D$ and $h$ are, respectively, the largest diameter and
smallest characteristic dimension of the former. Thus there will be
\emph{axially symmetric} resonances with typical wavevector
\footnote{In instances where there are several disparate
  characteristic wavelengths --- as is the case for the radial
  breathing mode of a microtoroid that hybridizes appreciably with the
  flexural modes of the corresponding membrane \cite{Anetsberger08}
  --- $k_R$ should be taken as the largest of the corresponding
  wavevectors.}  $k_R$ that satisfy $\sqrt{S}k_R\ll1$. We focus on one
  of them \footnote{An example of particular relevance in the context
  of optomechanics is the radial breathing mode of a microtoroid
  \cite{Kippenberg07,Anetsberger08}.}  and assume that it is
  ``isolated'' (cf.~Subsec.~\ref{sub:results}). The limit $S\to 0$
  yields a finite resonator volume subject to free boundary conditions
  on its whole surface, which specifies the natural choice of boundary
  conditions for the corresponding resonator mode \footnote{In the
  case of microtoroids with membrane thickness
  \cite{Kippenberg07,Anetsberger08} $h$ much smaller than the smallest
  pedestal radius one should instead define the resonator volume as
  the undercut region, so that $S$ corresponds to the vertical rim at
  the smallest radius of the undercut. Thus, this scenario falls into
  the singular case ($i$) as defined below Eq.~(\ref{genI}) and
  $\bar{u}'_R$ should instead be specified by clamped boundary
  conditions at $S$.}  $\bar{u}'_R$. This directly implies that the
  terms involving $\bv{\sigma}'_R$ vanish in Eqs.~(\ref{ovgen}),
  (\ref{genI}) while property $\sqrt{S}k_R\ll1$ allows us to factor
  $\bar{u}'_R$ out of the integral which reduces then to the total
  force across $S$ associated with the scattering mode $q$.

We assume that for studying the propagation of modes with wavevector
$q\sim k_R$ the pedestal can be modeled as an infinite beam with
slowly varying cross section $q S(z)/S'(z)\gg1$. This should capture
the adiabatic limit of perfect impedance-match with the substrate for
the purpose of studying the dissipation. In turn, our prior treatment
of a 3D-1D junction (cf.~Sec.~\ref{s:singlejunc}) can be extended to
our present context (with the roles of support and resonator
interchanged). Now the 3D object (resonator volume) presents finite
dimensions comparable to the relevant wavevectors. Naturally,
providing rigorous derivations (as the ones given in
Sec.~\ref{s:singlejunc}) is now complicated by the lack of an explicit
expression for the Green's function of the 3D object. Nonetheless,
heuristic considerations imply that: (i) in the limit $\sqrt{S}q\ll1$
with $\sqrt{S}\lesssim h$ the scattering mode $\bar{u}_{q}$ can be
asymptotically approximated in the pedestal by $\bar{u}_{*,q}$ (which
corresponds to an incoming wave satisfying clamped boundary conditions
at $S$) provided $\omega(q)$ is not close to a resonance $\omega_n$ of
the resonator volume, and (ii) once again $I(\omega)$ is smooth in a
neighborhood of $\omega_R$. The latter implies (as for the beam
geometry) that the restriction to modes $\bar{u}_{q}$ with
$\omega(q)\neq\omega_n$ is immaterial as it only invalidates our
result for $I(\omega)$ at $\omega=\omega_n\neq\omega_R$. In turn, it
is straightforward to extend the formalism in App.~\ref{ap:beam} to a
beam with adiabatically varying cross section. Thus the universal
properties of the end corrections imply that to lowest order in
$\sqrt{S}q$ only the TRE part of $\bar{u}_{q}$ contributes to the
aforementioned total force. The axial symmetry of the resonance
further implies that only \emph{compressional modes} yield a finite
contribution \footnote{It is straightforward to extend the derivation
to resonances with other reflection symmetries $\hat R_x$, $\hat R_y$
--- as for the symmetric beam geometry each resonance only couples to
the TRE branch with the same symmetries.}. Thus we obtain
(cf.~Fig.~\ref{fig:torus})
\begin{equation}\label{totF}
\int_S \ud r^2\,\bar{u}'_R \cdot \bv{\sigma}_q \cdot \hat{n} \approx -
E_s\sqrt{\frac{S}{2\pi}} \bar{u}'_R(0)
\cdot\hat{z}\frac{\partial\phi_q}{\partial z}(0)
\end{equation}
where the corresponding effective one
dimensional field satisfies \cite{Graff}
\begin{equation}\label{adiabcross}
\frac{\partial^2\phi_q}{\partial z^2}+
\frac{S'}{S}\frac{\partial\phi_q}{\partial z} +
\frac{\rho_s}{E_s}\omega^2(q)\phi_q=0\,
\end{equation}
and the approximation by $\bar{u}_{*,q}$ implies $\phi_q(0)=0$,
$\phi^{(-)}_q(0)=1$ --- here we define $\phi^{(-)}_q$ [$\phi^{(+)}_q$]
as the incoming (outgoing) components of the scattering mode
\footnote{To understand the normalization and the support's
  one-dimensional DOS (needed below to integrate over $q$) it is
  instructive to consider the infinite length limit of the elastic
  object generated by reflecting the pedestal with respect to $S$ and
  identifying the resulting endpoints --- with an adequate smoothening
  of the possible discontinuities in $S'(z)$.} $q$. We adopt for
  simplicity an exponential dependence $S(z)=S e^{-\Gamma z}$ which
  leads to the following
\begin{align}
  \phi_q(z) & = \phi^{(-)}_q(z) - \phi^{(+)}_q(z)\,,
\label{adiabphi}\\ \phi^{(\mp)}_q(z) & = e^{(\frac{\Gamma}{2}\pm i q)
z} \qquad \textrm{with} \qquad q= \sqrt{\frac{\rho_s}{E_s}\omega^2(q)
- \frac{\Gamma^2}{4}}\,, \nonumber\\ \frac{\ud q}{\ud\omega} & =
\frac{\rho_s\omega(q)}{E_s\sqrt{\frac{\rho_s}{E_s}\omega^2(q) -
\frac{\Gamma^2}{4}}}\,.\label{adiabDOS}
\end{align}

Then, substitution of Eqs.~(\ref{totF}), (\ref{adiabphi}),
(\ref{adiabDOS}) into Eq.~(\ref{genI}) allows us to obtain
\begin{equation}
I(\omega)\approx
\sqrt{\frac{E_s}{\rho_s}}\frac{S}{\omega_R}{u'}^2_{\!R,z}
\sqrt{\omega^2-\omega^2_I}\, \Theta\!\left(\omega-\omega_I\right)\,.
\end{equation}
Clearly the infrared cutoff $\omega_I = \sqrt{E_s/\rho_s}\Gamma/2$ is
an artifact of the adiabatic assumption $q S(z)/S'(z)\gg1$ which
implies $\omega\gg\omega_I$. In the latter appropriate limit we finally
obtain Eq.~(\ref{toroI}) for the spectral density where
$\tilde{u}_{R,z}(0)=\sqrt{m_R/\rho_s}u'_{R,z}(0)$ corresponds to the
resonator mode normalized to the ``relevant coordinate''
\footnote{Note that $\bar{u}'_R$ is normalized using the non-trivial
  metric defined by Eq.~(\ref{eq:overlap}).} (in our specific context
  $\Delta D/2$). Note that, as expected, the result is independent of
  $\Gamma$. As for the beam geometry, for typical materials the $Q$
  will be mostly a size-independent geometric property.

\section{Derivation of \texorpdfstring{$\Im\{\tilde
    G_{xx}(0)\}$}{the green function} for the elastic
  half-space}\label{ap:multipole}

In the following we define $\bar u \equiv \mathbf{\tilde{G}}\cdot \hat{x}$
and use dimensionless variables setting $q_t\equiv 1$. 
We need to solve for the half space ($z<0$):
\begin{equation}
\nabla^2 \bar u + \frac{1}{1-2 \sigma_s} \nabla (\nabla\cdot \bar u) +
 \bar u =0 \,,
\end{equation}
subject to the following boundary conditions at $z=0$:
\begin{align}
  2 u_{x z} &= \delta(x)\delta(y)\,, \\
  2 u_{y z} &= \frac{2}{1-2 \sigma_s} \left[ (1-\sigma_s) u_{z z} + \sigma_s
    (u_{x x}+u_{y y}) \right] = 0\,.
\end{align}
One can can construct the solution as a superposition of longitudinal
($j=1$), SH ($j=2$) and SV ($j=3$) waves:
\begin{align}\label{uhalf}
  \bar u (x,y,z) =& \int_{-\infty}^{+\infty}\!\!
  \int_{-\infty}^{+\infty} \frac{\ud q_x \ud q_y}{4 \pi^2} \bar
  \varepsilon_j
  (q_x,q_y) \tilde{u}_j (q_x,q_y) \nonumber \\
  & \times e^{i \left(q_x x + q_y y -\sqrt{\alpha_j - q_x^2 - q_y^2} z
    \right)}\,,
\end{align}
where we have defined
\begin{align}
\alpha_1 &= \frac{c_t^2}{c_l^2} = \frac{1-2 \sigma_s}{2 (1-\sigma_s)} =
\alpha (\sigma_s), & \quad \alpha_2 = \alpha_3 = 1\,, 
\end{align}
\begin{align} 
\bar \varepsilon_1 &= (q_x, q_y, q_{z,1}), & \bar \varepsilon_2 = (q_y, -
q_x, 0)\,, 
\\
\bar\varepsilon_3 &= (-q_{z,3} q_x, -q_{z,3} q_y, q_x^2+q_y^2)\,.
\end{align}
We note that $q_{z,j}\equiv -\sqrt{\alpha_j-q_x^2-q_y^2}$ can be
imaginary, we do not need to normalize the $\bar \varepsilon_j$ for
our purposes and $\alpha<1$.

In order to obtain the outgoing solution one can introduce damping,
then calculate (\ref{uhalf}) that will correspond to the steady
state solution, and, finally, take the limit of damping coefficient
going to zero. If we decompose the total displacement field into
transverse and longitudinal components $\bar u = \bar u_\perp + \bar
u_\parallel$, the modified equations of 3D elasticity can be written
as
\begin{align}
  & \frac{\partial^2 \bar u_\eta}{\partial t^2} + \epsilon
  \frac{\partial \bar u_\eta}{\partial t} - c_\eta^2 \nabla^2 \bar
  u_\eta = 0\,, \qquad \eta = \perp,\parallel \\ & \nabla \times \bar
  u_\parallel = 0\,,  \qquad\qquad   
\nabla \cdot \bar u_\perp = 0 \,,
\end{align}
with $\epsilon>0$, and $c_\perp = q_t c_t$, $c_\parallel = q_t
c_l$. We look for solutions $\bar u (\bar{r},t)= \bar 
\varepsilon_j e^{i \left( \bar q_j \cdot \bar r - \omega t
\right)}$, which leads to:
\begin{equation}
q_j^2 = \frac{\omega^2}{c_j^2} \left( 1 + i \frac{\epsilon}{\omega}
\right)\,,
\end{equation}
with $c_1=c_\parallel$ and $c_{2,3}=c_\perp$. These solutions can be
obtained from the undamped ones by the replacement $\alpha_j \to
\alpha_j (1 +i \epsilon)$. The analysis of the limiting procedure
$\epsilon \to 0$ allows the determination of the adequate integration
contour $\cal{C}$ in the complex plane in the standard way.

If we substitute expression (\ref{uhalf}) and Fourier transform we
arrive at: 
\begin{widetext}
\begin{equation}
\begin{bmatrix} 
2 q_x q_{z,1} & q_y q_{z,2} & q_x (q_x^2+q_y^2- q_{z,2}^2) \\
2 q_y q_{z,1} & - q_x q_{z,2} & q_y (q_x^2+q_y^2- q_{z,2}^2) \\
(1-\sigma_s) q_{z,1}^2 + \sigma_s (q_x^2+q_y^2) & 0 & (1-2\sigma_s) q_{z,2}
(q_x^2+q_y^2) 
\end{bmatrix} \begin{pmatrix}  
\tilde{u}_1 \\ \tilde{u}_2 \\ \tilde{u}_3 
\end{pmatrix}
= 
\begin{pmatrix}  
-i \\ 0 \\ 0
\end{pmatrix}
\,.
\end{equation}
We adopt polar variables $q_x = \rho \cos \varphi$, $q_y = \rho \sin
\varphi$ and solve for the $\tilde u_j$
\begin{align}
  \tilde{u}_1 (\rho, \varphi) &= i (1-2 \sigma_s) \frac{\rho
    \sqrt{1-\rho^2+i\epsilon} \cos \varphi}{D(\rho)} & \tilde{u}_2
  (\rho, \varphi) = i  \frac{ \sin
    \varphi}{\rho \sqrt{1-\rho^2+i\epsilon}} \nonumber\\
  \tilde{u}_3 (\rho, \varphi) &= i 
  \frac{ \left[ (1-\sigma_s) \alpha(\sigma_s) +
      (2\sigma_s -1) \rho^2\right] \cos \varphi}{\rho D(\rho)}
\end{align}
where
\begin{align}
   D(\rho)  = & 2 (1-2 \sigma_s ) \left[ \rho^2 \sqrt{1-\rho^2+i\epsilon}
    \sqrt{\alpha(\sigma_s) (1+i\epsilon)-\rho^2} 
+ \left(\frac{1}{2}-\rho^2\right)
    \left(\frac{1+i\epsilon}{2}-\rho^2\right)\right]\,.
\end{align}
If we consider the integral of expression (\ref{uhalf}) evaluated at
$z=0$ over the contour $\cal{C}$ we can identify a regular and a
singular contribution to the displacement at the free boundary. We are
only interested in the imaginary part of the $x$ component to which
only the regular part contributes. Finally, evaluation at the origin
$x=y=0$ leads to:
\begin{equation}\label{imGxx}
  \Im\{\tilde G_{xx} (0,\alpha)\} = \frac{1}{\pi} \mathscr{P} \left\{
    \int_0^1 \ud v \frac{v^2 (v^2-1/2)^2}{p(1-v^2)} -
    \int_0^{\sqrt{\alpha}} \ud v \, v^2 \frac{(\alpha-v^2)
      (v^2+1-\alpha)}{p(\alpha-v^2)} \right\} + R[\xi(\alpha),\alpha] +
  \frac{1}{4 \pi}\,, 
\end{equation}
\end{widetext}
where we have eliminated $\sigma_s$ in favor of $\alpha$, the
contribution of the Rayleigh pole associated with the surface waves
(SAW) is given by
\begin{equation}
 R[\xi,\alpha] = - \frac{1}{16 \xi} \frac{(1-\xi^2/2)^2
   \sqrt{1-\xi^2} + (1-\xi^2) \sqrt{1-\alpha \xi^2}}{6 (\alpha-1) + 2 (3-2
   \alpha) \xi^2 - \xi^4} 
\end{equation}
and we have defined
\begin{equation}\label{poly}
 p(v) \equiv 16 (\alpha -1) v^3 + 8 (3-2 \alpha) v^2 -8 v +1 \, .
\end{equation}
We note that the parameter $\xi$, already introduced in
App.~\ref{ap:halfspace}, satisfies $\xi=1/\sqrt{v_*}$, where $v_*$ is
the only real root of (\ref{poly}) greater than unity.

\section{Derivation of Equation \texorpdfstring{(\ref{Iapprox})}{
  }}\label{ap:I}

We will first invert Eq.~(\ref{Iex}) to obtain Eq.~(\ref{S}). To this
effect we define a complex function $G(\omega)$ on the real axis such
that for $\omega>0$
\begin{equation}\label{GIR}
  \Im [G(\omega)]= -S(\omega) \quad \text{and} \quad \Re [G(\omega)]= 
  \Delta_S(\omega)
\end{equation}
and the extension to $\omega<0$ is specified by
\begin{equation}
  G(-\,\omega)=G^*(\omega),
\end{equation}
which, after rearranging the integral for $\Delta_S(\omega)$ using the
partial fraction expansion of $2\omega'/(\omega^2- {\omega'}^2)$,
yields for the whole real axis
\begin{equation}\label{Gomega}
  G(\omega)=\frac{1}{\pi}\; \mathscr{P}
  \!\!\int_{-\infty}^\infty \!\!\ud \omega' \frac{\sgn(\omega')
    S(|\omega'|)}{\omega - \omega'} - i\,\sgn(\omega)S(|\omega|) .
\end{equation}
Given that $S(\omega)$ is well behaved ($C_\infty$ for $\omega>0$) and
given its behavior for $\omega\to0$ and $\omega\to\infty$ discussed at
the end of Subsec.~\ref{sub:Itau}, Eq.~(\ref{Gomega}) implies that the
real and imaginary parts of $G(\omega)$ are Hilbert transforms of each
other.  Then, it follows from Titchmarsh's theorem \cite{Titchmarsh}
that $G(z)$ is analytic in the complex upper half-plane and that the
integral over a semicircular contour in the latter tends to zero as
the radius is increased. Thus we arrive at
\begin{equation}\label{Gz}
  G(z)= \frac{1}{2\pi i} \int^{\infty}_{-\infty}\!\!\ud\omega 
  \frac{G(\omega)}{\omega-z}\,,
\end{equation}
valid for $\Im(z)>0$. We note that $G(\omega)$ gives the propagator of the
resonator mode's canonical coordinate at zero temperature so that in
physical terms the above properties of $G(z)$ can be understood as a
consequence of causality --- the inverse Fourier transform of
$G(\omega)$ only has support for $t>0$. If we now substitute
Eq.~(\ref{Gomega}) into Eq.~(\ref{Gz}), interchange the order of the
integrations over $\omega$ and $\omega'$, perform the former, and
rearrange the latter we obtain
\begin{equation}\label{GS}
  G(z)= \frac{1}{\pi}\int_0^\infty \!\!\ud \omega' \frac{2\omega'}{z^2-
    {\omega'}^2} \, S(\omega') \,, 
\end{equation}
which with the help of Eq.~(\ref{eq:intS}) can be re-expressed as
\begin{equation}\label{Gasympt}
  G(z)= \frac{1}{z^2} +  \frac{\omega_R^2}{z^4} + \frac{1}{\pi
    z^4}\int_0^\infty \!\!\ud \omega
  \frac{2\omega^5}{z^2- \omega^2} \, S(\omega) \,.
\end{equation}
To proceed we define the function 
\begin{equation}\label{F}
  \Sigma(z)\equiv z^2 - \omega_R^2 - \frac{1}{G(z)}\,,
\end{equation}
which using Eqs.~(\ref{Gomega}), (\ref{GIR}) and
(\ref{Iex}) can be shown to have the property
\begin{equation}\label{ImF}
  \Im[\Sigma(\omega)]=\frac{-\sgn(\omega)S(|\omega|)}{\Delta_S^2(\omega)
    + S^2 (|\omega|)}=-\omega_R\,\sgn(\omega)I(|\omega|)\,.
\end{equation}
As already discussed in Subsec.~\ref{sub:Itau}, $S(\omega)$ has a
natural ultraviolet cutoff, given by
$\omega_*\sim\omega_\beta(\pi/d)$, beyond which it is bounded by a
power law $1/\omega^m$ with $m\geq 6$.  From this behavior one can
derive that the last term in the RHS of Eq.~(\ref{Gasympt}) is bounded
by $\ln|z/\omega_*|/|z|^6$ if $m=6$ and by $1/|z|^6$ if $m>6$. This
property together with Eqs.~(\ref{F}) and (\ref{Gasympt}) directly
implies that
\begin{equation}\label{Fasympt}
  |\Sigma(z)| \lesssim
  \mathcal{O}\!\!\left[\frac{\ln|z/\omega_*|}{|z|^2}\right] \quad
  \text{or} \quad |\Sigma(z)| \lesssim
  \mathcal{O}\!\!\left[\frac{1}{|z|^2}\right],
\end{equation}
respectively, for $z\to\infty$.  On the other hand Eq.~(\ref{GS})
implies that the propagator $G(z)$ has no zeros with $\Im(z)>0$ so
that $\Sigma(z)$ is analytic in the upper half-plane. This property
and Eq.~(\ref{Fasympt}) allow us to obtain
\begin{equation}\label{Fint}
  \Sigma(\omega)= \frac{1}{\pi i} \; \mathscr{P}
  \!\!\int^{\infty}_{-\infty}\!\!\ud\omega'
  \frac{\Sigma(\omega')}{\omega'-\omega}\,,
\end{equation}
using the Cauchy integral formula. If we now take the real part of
this equation and use Eqs.~(\ref{F}), (\ref{Gomega}), (\ref{GIR}) and
(\ref{ImF}) we obtain after some simple rearrangements
\begin{equation}\label{DeltaI}
  \Re[\Sigma(\omega)]=\omega_R \Delta_I(\omega)= \omega^2 - 
  \omega_R^2 - \frac{\Delta_S(\omega)}{\Delta_S^2(\omega) +S^2 (\omega)}
\end{equation}
for $\omega>0$. If we assume that the functions $\Delta_I(\omega)$ and
$I(\omega)$ are known, Eq.~(\ref{DeltaI}) together with
Eq.~(\ref{Iex}) provide a system of algebraic equations for the
unknowns $\Delta_S(\omega)$ and $S(\omega)$. Thus we can solve for
$S(\omega)$ to finally obtain the desired result given by
Eq.~(\ref{S}).

If we now consider Eqs.~(\ref{DeltaI}) and (\ref{ImF}), we have
\begin{equation}\label{Fomega}
  \Sigma(\omega) = \omega_R\left[\Delta_I(\omega) - 
    i\,\sgn(\omega)I(|\omega|)\right]
\end{equation}
that together with Eqs.~(\ref{eq:h-rb}) and (\ref{eq:I}) allow us to
interpret the function $\Sigma(\omega)/2\omega_R$ as the resonator's
self-energy induced by its interaction with the environment.  Given
the properties of $I(\omega)$ discussed below we may assume that
$\Sigma(z)$ is analytic in a neighborhood of $z=\omega_R$. Thus, it
follows from Eq.~(\ref{F}) that any pole of $G(z)$ in that
neighborhood is given by a root $z_R$ of the equation
\begin{equation}\label{pole}
 z^2 - \omega_R^2 - \Sigma(\omega_R) - 
  \sum_{n=1}^\infty \frac{\partial^n \Sigma}{\partial\omega^n} 
  \left(\omega_R\right) \left(z - \omega_R\right)^n = 0\,.
\end{equation}
In turn, the behavior of the function $S(\omega)$ discussed 
below [cf.~Eq.~(\ref{Lorentzian})] and Eq.~(\ref{Gomega}) imply that
for $d/L\to 0$ the propagator $G(z)$ has indeed a pole in the lower
half-plane close to $\omega_R$ that tends to $\omega_R$. If we now
consider the asymptotic expansion of $(z_R - \omega_R)/\omega_R$ and
$\Sigma(\omega)/\omega_R^2$ as a function of $d/L$ for $d/L\to 0$ and
substitute it into Eq.~(\ref{pole}), it is straightforward to realize
[cf.~Eqs.~(\ref{ImF}), (\ref{DeltaI})] that $(z_R -
\omega_R)/\omega_R\to 0$ directly implies the properties
$|\Delta_I(\omega_R)|/\omega_R\to0$, $I(\omega_R)/\omega_R\to 0$ used
in Subsec.~\ref{sub:Itau} and in the derivation of Eq.~(\ref{Iapprox})
pursued below. Heuristically, this vanishing of the self-energy can be
viewed as an unavoidable consequence of the behavior of the
transmission coefficients $\tau(\omega)$ for $d/L\to 0$ unveiled in
Sec.~\ref{s:singlejunc}. In fact the interaction with the bath giving
rise to the self-energy can be understood in terms of phonon tunneling
between the beam and its supports, i.e.~a mechanism that is suppressed
as $\tau(\omega)\to 0$.

To proceed we reformulate Eq.~(\ref{Iapprox}) by defining
$I_*(\omega)$ as its RHS. Thus, it is enough to prove
\begin{equation}\label{aprox1}
  I(\omega)=I_*(\omega)\left|1 \, + \,
    \mathcal{O}\left[\frac{|\Delta_I(\omega_R)|}{\omega_R}
    \right]\right|^2 \,.
\end{equation}
This can also be written as the condition
$|E(\omega)|<|\Delta_I(\omega_R)|/\omega_R$ where the function
$E(\omega)$ is defined by
\begin{widetext}
\begin{equation}\label{E}
  E(\omega) \equiv \frac{\Delta_I (\omega) - \Delta_I (\tilde\omega_R)
  - i \left[I(\omega) - I(\tilde\omega_R)\right]-
  \frac{\omega\,-\,\tilde\omega_R}{\omega_R\, +\,\tilde\omega_R}
  \Delta_I (\tilde\omega_R)}{\omega-\tilde\omega_R + i \,
  \frac{\omega_R}{\omega_R\, +\,\omega} \, I (\tilde\omega_R)}\,,
\end{equation}
\end{widetext}
[cf.~Eq.~(\ref{omegaR})] so that given Eq.~(\ref{S}) we have
\begin{equation}\label{aprox2}
  \frac{I(\omega)}{I_*(\omega)} = \left|1 \, -
    \,\frac{\omega_R}{\omega_R\, +\,\omega}\,E(\omega)\right|^2 \,.
\end{equation}
We note that Eq.~(\ref{omegaR}) is just the real part of
Eq.~(\ref{pole}) restricted to the real axis. If we use again its
asymptotic expansion in terms of the aspect ratio $d/L$ it is
straightforward to derive from $\tilde\omega_R \to \omega_R$ when $d/L
\to 0$ that $2(\tilde\omega_R - \omega_R)/\Delta_I (\omega_R)\to
1$, which together with Eq.~(\ref{omegaR}) implies that $\Delta_I
(\tilde\omega_R)$ and $\Delta_I (\omega_R)$ coincide to lowest order
in the aspect ratio. It is worth noting that though the real part of
the exact pole of $G(z)$ differs from $\tilde\omega_R$, as the imaginary
part of Eq.~(\ref{pole}) comes into play, analogous considerations
yield $2(z_R - \omega_R)/\Delta_I (\omega_R)\to 1$ so that
$\Delta_I (\omega_R)/2$ provides the lowest order contribution
to the renormalization of the bare frequency.

In order to ensure the validity of Eq.~(\ref{Iapprox}) we find that in
addition to $|\Delta_I (\omega_R)|/\omega_R\ll 1$ the
following assumptions (whose validity is discussed below) are needed:
\begin{equation}\label{assum1}
  \left|\frac{\partial^n
      \Delta_I}{\partial\omega^n}\left(\omega\right)\right|
  \sim\frac{\left|\Delta_I(\omega)\right|}{\omega^n} \, , \quad 
   \left|\frac{\partial^n I}{\partial\omega^n}\left(\omega\right)\right|\sim 
  \frac{I(\omega)}{\omega^n}\, , 
\end{equation}
for $n=1,2$ and frequencies smaller than the ultraviolet cutoff
$\omega_*\gg\omega_R$. Heuristic considerations imply that from
Eq.~(\ref{assum1}) and the behaviors of $I(\omega)$ for $\omega\to0$
and $\omega\to\infty$ (discussed at the end of
Subsec.~\ref{sub:Itau}), it follows that the only relevant frequency
scale when considering the profile of $I(\omega)$ is $\omega_*$ near
which this function attains its maximum. In turn this implies that
\begin{equation}\label{assum2}
|\Delta_I(\tilde\omega_R)|\approx|\Delta_I(0)|\sim I(\omega_*)\sim
\mathrm{max} \{I(\omega)\}  \,,
\end{equation}
and that in the interval $(0,\omega_*)$, $\frac{\partial
  I}{\partial\omega}\left(\omega\right)>0$.  On the other hand we can
use the mean value theorem to write
\begin{align}\label{mvalue}
  \pi\Delta_I(\omega)=&-\int_{\omega/2}^{3\omega/2}\ud\omega'
  \left[\frac{\partial I}{\partial\omega}(\Omega[\omega',\omega])+
    \frac{1}{\omega + \omega'}I(\omega')\right] \nonumber \\ &+
  \left[\int_0^{\omega/2}+\int_{3\omega/2}^{\infty}\right]
  \frac{\omega'}{\omega-\omega'}\,\frac{2}{\omega+\omega'}I(\omega')
  \,\ud\omega' \, ,
\end{align}
where $\Omega[\omega',\omega]\in(\omega,\omega')$. Equations
(\ref{assum1})-(\ref{mvalue}) imply
\begin{equation}\label{assum3}
  \frac{|\Delta_I\left(\omega\right)|}{\omega} \, , \,
  \frac{I\left(\omega\right)}{\omega} 
  \lesssim \frac{|\Delta_I\left(\tilde\omega_R\right)|}{\tilde\omega_R}\, , 
\end{equation}
for $\tilde\omega_R \leq \omega < \omega_*$, and 
\begin{equation}\label{assum4}
  \Delta_I\left(\omega\right) \approx
  \Delta_I\left(\tilde\omega_R\right), \qquad 
  I\left(\omega\right) < I\left(\tilde\omega_R\right)\lesssim
  |\Delta_I\left(\tilde\omega_R\right)| \, ,  
\end{equation}
for $\omega < \tilde\omega_R$.

Furnished with relations (\ref{assum1}), (\ref{assum3}) and
(\ref{assum4}), we turn now to the analysis of $|E(\omega)|$ defined
in Equation (\ref{E}). First we consider the case
$|\omega-\tilde\omega_R|\gtrsim\tilde\omega_R$. For
$\omega>\tilde\omega_R$ we use $|\omega-\tilde\omega_R|\sim\omega$ and
(\ref{assum3}), while for $\omega<\tilde\omega_R$ we use
(\ref{assum4}). In both instances it is simple to establish
$|E(\omega)|\lesssim
|\Delta_I\left(\omega_R\right)|/\omega_R$. To analyze the
remaining case $|\omega-\tilde\omega_R|\ll\tilde\omega_R$ we consider
the Taylor expansions of the functions $\Delta_I\left(\omega\right)$
and $I\left(\omega\right)$ around the frequency $\tilde\omega_R$.
Equation (\ref{assum1}) implies that substituting into Eq.~(\ref{E})
the linear parts of these expansions results in a relative error for
$E(\omega)$ of order $|\omega-\tilde\omega_R|/\tilde\omega_R\ll 1$.
Thus we may write
\begin{align}\label{Eapprox}
  E(\omega) \approx &\left[\frac{\partial
      \Delta_I}{\partial\omega}\!\left(\tilde\omega_R\right) - i
    \frac{\partial I}{\partial\omega}\!\left(\tilde\omega_R\right) -
    \frac{\Delta_I (\tilde\omega_R)}{\omega_R\,
      +\,\tilde\omega_R}
  \right]\nonumber\\
  & \times\frac{\omega\,-\,\tilde\omega_R}{\omega-\tilde\omega_R + i \,
    \frac{\omega_R}{\omega_R\, +\, \omega} \, I(\tilde\omega_R)}\,,
 \end{align}
 which using again Eqs.~(\ref{assum1}) and (\ref{assum3}) leads to the
 desired result for $E(\omega)$. This completes the derivation of
 Eq.~(\ref{Iapprox}) for the frequency range in which it is used in
 Subsection \ref{sub:Itau}. It is worth noting that a straightforward
 derivation using Eqs.~(\ref{zeta}), (\ref{Iex}), (\ref{DeltaI}), and
 (\ref{E}) leads to
\begin{equation}
  \left|\frac{\zeta(q)-\zeta_* (q)}{\zeta_* (q)}\right|  = \left|
    \frac{\omega_R}{\omega_R\, +\,\omega}\,E(\omega)\right| 
\end{equation}
with
\begin{equation}\label{zetaapprox}
\zeta_* (q)\equiv \frac{\langle
  u'_R|u_q\rangle}{2\sqrt{\omega_R\omega_q}}
  \left[\left(\omega_q+\omega_R\right)
  \left(\omega_q-\tilde\omega_R\right) + i\omega_R
  I(\tilde\omega_R)\right]\,
\end{equation}
which can prove useful when analyzing the manipulation of a specific
resonance by coupling to a pseudospin \cite{WilsonRae04}.

Finally, to warrant our derivation of Eqs.~(\ref{eq:spectralrho}) and
(\ref{eq:spectraltau}) in Subsection \ref{sub:Itau} we also need to
establish that the function $I(\omega)$ remains well behaved at
$\omega=\omega_R$ as $d/L\to0$. This can be accomplished by using the
exact expression (\ref{Iex}), Eq.~(\ref{eq:S}), and the behavior of
the overlaps for small but finite $d/L$. To understand the latter it
is essential to consider the qualitative behavior of corrections to
the scattering modes $\phi_{k(q),\beta}(z)$. This can be done
\cite{WilsonRaetobepublished} using: ($i$) that each of the reflection
amplitudes admits an expansion in powers of $k d$ and those
corresponding to a junction relate to the finite transmission
coefficient via energy conservation and ($ii$) the exact reduction to
a single junction performed in Sec.~\ref{s:FP}. We note that for our
present purposes the precise form of the coefficients of the expansion
($i$), which will be studied elsewhere \cite{WilsonRaetobepublished},
is not important.  Propositions ($i$), ($ii$), and Eq.~(\ref{eq:S})
allow us to write the following Lorentzian approximation $L\equiv1$:
\begin{align}\label{Lorentzian}
S(\omega) & \approx A_R (d) \frac{\Gamma_R (d)/2}{\left[ \omega -
    \tilde\omega_R (d) \right]^2 + \Gamma^2_R (d)/4} \,,
\nonumber\\
\Delta_S(\omega)  &\approx  A_R (d) \frac{\omega -
    \tilde\omega_R (d)}{\left[ \omega -
    \tilde\omega_R (d) \right]^2 + \Gamma^2_R (d)/4} + B_R(d)\,,
\end{align} 
where the corrections are higher order in $d$ for all frequencies in a
small neighborhood of $\omega_R$, and to establish
\begin{equation}\label{limAR}
  \lim_{d\to0} A_R(d)=\frac{1}{2\omega_R},\qquad\lim_{d\to0} B_R(d)=0.
\end{equation}
Inserting Eq.~(\ref{Lorentzian}) into Eq.~(\ref{Iex}) we obtain
\begin{align}\label{LorentzianI}
I(\omega) \approx & \frac{1}{\omega_R A_R(d)} \frac{\Gamma_R (d)}{2} 
\left( 1+ 2 \frac{B_R(d)}{A_R(d)} \left[ \omega - \tilde\omega_R (d) \right] 
\right. \nonumber \\ & \ \left. 
  + \frac{B_R^2(d)}{A_R^2(d)} \left\{ \left[ \omega - \tilde\omega_R (d)
    \right]^2 + \frac{\Gamma_R^2 (d)}{4}
\right\} \right)^{-1} \,,
\end{align} 
which together with Eq.~(\ref{limAR}) leads to $I(\omega)
\approx\Gamma_R (d)$ for frequencies close to $\omega_R$ implying that
(as expected on physical grounds) the resonator mode's environment is
\emph{structureless} at the characteristic resonant frequency
$\omega_R$ (cf.~Subsec.~\ref{IIB}).  

An analogous procedure allows to analyze the qualitative behavior at
the \emph{other resonances} $\omega=\omega_n\neq\omega_R$ where we
find that $S(\omega)$ presents Fano profiles that result in
corresponding features in the function $I(\omega)$ that have
negligible relative spectral weight as $d/L\to0$. Thus, though the
latter do not affect the behavior at other frequencies, they
invalidate Eqs.~(\ref{assum3}), (\ref{assum4}) at these special
frequencies precluding in their neighborhoods the use of
Eq.~(\ref{Iapprox}). To conclude, we point out that from
Eqs.~(\ref{Lorentzian}), (\ref{Stau}), and (\ref{Ov}) one can obtain
an approximation for the function $S(\omega)$ [in terms of
$\Gamma_R(d)$] adequate for all low frequencies other than
$\omega_n\neq\omega_R$. The latter approximation together with the
asymptotic behavior discussed at the end of Subsec.~\ref{sub:Itau} and
the exact relation (\ref{Iex}) allow for an independent heuristic
justification of Eq.~(\ref{assum1}) --- i.e.~without resort to the
approximation (\ref{Iapprox}) or the properties of the spectral
densities expected on physical grounds.


\begin{thebibliography}{83}
\expandafter\ifx\csname natexlab\endcsname\relax\def\natexlab#1{#1}\fi
\expandafter\ifx\csname bibnamefont\endcsname\relax
  \def\bibnamefont#1{#1}\fi
\expandafter\ifx\csname bibfnamefont\endcsname\relax
  \def\bibfnamefont#1{#1}\fi
\expandafter\ifx\csname citenamefont\endcsname\relax
  \def\citenamefont#1{#1}\fi
\expandafter\ifx\csname url\endcsname\relax
  \def\url#1{\texttt{#1}}\fi
\expandafter\ifx\csname urlprefix\endcsname\relax\def\urlprefix{URL }\fi
\providecommand{\bibinfo}[2]{#2}
\providecommand{\eprint}[2][]{\url{#2}}

\bibitem[{\citenamefont{Craighead}(2000)}]{Craighead00}
\bibinfo{author}{\bibfnamefont{H.~G.} \bibnamefont{Craighead}},
  \bibinfo{journal}{Science} \textbf{\bibinfo{volume}{290}},
  \bibinfo{pages}{1532} (\bibinfo{year}{2000}).

\bibitem[{\citenamefont{Roukes}(2001)}]{Roukes01}
\bibinfo{author}{\bibfnamefont{M.}~\bibnamefont{Roukes}},
  \bibinfo{journal}{Physics World} \textbf{\bibinfo{volume}{14}},
  \bibinfo{pages}{25} (\bibinfo{year}{2001}).

\bibitem[{\citenamefont{Ekinci and Roukes}(2005)}]{Ekinci05}
\bibinfo{author}{\bibfnamefont{K.~L.} \bibnamefont{Ekinci}} \bibnamefont{and}
  \bibinfo{author}{\bibfnamefont{M.~L.} \bibnamefont{Roukes}},
  \bibinfo{journal}{Review of Scientific Instruments}
  \textbf{\bibinfo{volume}{76}}, \bibinfo{pages}{061101}
  (\bibinfo{year}{2005}).

\bibitem[{\citenamefont{Milburn et~al.}(1994)\citenamefont{Milburn, Jacobs, and
  Walls}}]{Milburn94}
\bibinfo{author}{\bibfnamefont{G.~J.} \bibnamefont{Milburn}},
  \bibinfo{author}{\bibfnamefont{K.}~\bibnamefont{Jacobs}}, \bibnamefont{and}
  \bibinfo{author}{\bibfnamefont{D.~F.} \bibnamefont{Walls}},
  \bibinfo{journal}{Phys. Rev. A} \textbf{\bibinfo{volume}{50}},
  \bibinfo{pages}{5256} (\bibinfo{year}{1994}).

\bibitem[{\citenamefont{Bocko and Onofrio}(1996)}]{Bocko96}
\bibinfo{author}{\bibfnamefont{M.~F.} \bibnamefont{Bocko}} \bibnamefont{and}
  \bibinfo{author}{\bibfnamefont{R.}~\bibnamefont{Onofrio}},
  \bibinfo{journal}{Rev. Mod. Phys.} \textbf{\bibinfo{volume}{68}},
  \bibinfo{pages}{755} (\bibinfo{year}{1996}).

\bibitem[{\citenamefont{Schwab and Roukes}(2005)}]{Schwab05}
\bibinfo{author}{\bibfnamefont{K.~C.} \bibnamefont{Schwab}} \bibnamefont{and}
  \bibinfo{author}{\bibfnamefont{M.~L.} \bibnamefont{Roukes}},
  \bibinfo{journal}{Physics Today} \textbf{\bibinfo{volume}{58}},
  \bibinfo{pages}{36} (\bibinfo{year}{2005}).

\bibitem[{\citenamefont{Sidles et~al.}(1995)\citenamefont{Sidles, Garbini,
  Bruland, Rugar, Z\"{u}ger, Hoen, and Yannoni}}]{Sidles95}
\bibinfo{author}{\bibfnamefont{J.~A.} \bibnamefont{Sidles}},
  \bibinfo{author}{\bibfnamefont{J.~L.} \bibnamefont{Garbini}},
  \bibinfo{author}{\bibfnamefont{K.~J.} \bibnamefont{Bruland}},
  \bibinfo{author}{\bibfnamefont{D.}~\bibnamefont{Rugar}},
  \bibinfo{author}{\bibfnamefont{O.}~\bibnamefont{Z\"{u}ger}},
  \bibinfo{author}{\bibfnamefont{S.}~\bibnamefont{Hoen}}, \bibnamefont{and}
  \bibinfo{author}{\bibfnamefont{C.~S.} \bibnamefont{Yannoni}},
  \bibinfo{journal}{Rev. Mod. Phys.} \textbf{\bibinfo{volume}{67}},
  \bibinfo{pages}{249} (\bibinfo{year}{1995}).

\bibitem[{\citenamefont{Thomas et~al.}(1996)\citenamefont{Thomas, Lionti,
  Ballou, Gatteschi, Sessoli, and Barbara}}]{thomas96}
\bibinfo{author}{\bibfnamefont{L.}~\bibnamefont{Thomas}},
  \bibinfo{author}{\bibfnamefont{F.}~\bibnamefont{Lionti}},
  \bibinfo{author}{\bibfnamefont{R.}~\bibnamefont{Ballou}},
  \bibinfo{author}{\bibfnamefont{D.}~\bibnamefont{Gatteschi}},
  \bibinfo{author}{\bibfnamefont{R.}~\bibnamefont{Sessoli}}, \bibnamefont{and}
  \bibinfo{author}{\bibfnamefont{B.}~\bibnamefont{Barbara}},
  \bibinfo{journal}{Nature} \textbf{\bibinfo{volume}{383}},
  \bibinfo{pages}{145} (\bibinfo{year}{1996}).

\bibitem[{\citenamefont{Cleland}(2003)}]{Cleland}
\bibinfo{author}{\bibfnamefont{A.~N.} \bibnamefont{Cleland}},
  \emph{\bibinfo{title}{Foundations of Nanomechanics}}
  (\bibinfo{publisher}{Springer, Berlin}, \bibinfo{year}{2003}).

\bibitem[{\citenamefont{Gassmann et~al.}(2004)\citenamefont{Gassmann, Choi, Yi,
  and Bruder}}]{Gassmann04}
\bibinfo{author}{\bibfnamefont{H.}~\bibnamefont{Gassmann}},
  \bibinfo{author}{\bibfnamefont{M.-S.} \bibnamefont{Choi}},
  \bibinfo{author}{\bibfnamefont{H.}~\bibnamefont{Yi}}, \bibnamefont{and}
  \bibinfo{author}{\bibfnamefont{C.}~\bibnamefont{Bruder}},
  \bibinfo{journal}{Phys. Rev. B} \textbf{\bibinfo{volume}{69}},
  \bibinfo{pages}{115419} (\bibinfo{year}{2004}).

\bibitem[{\citenamefont{Walls and Milburn}(1994)}]{Walls}
\bibinfo{author}{\bibfnamefont{D.~F.} \bibnamefont{Walls}} \bibnamefont{and}
  \bibinfo{author}{\bibfnamefont{G.~J.} \bibnamefont{Milburn}},
  \emph{\bibinfo{title}{Quantum Optics}} (\bibinfo{publisher}{Springer,
  Berlin}, \bibinfo{year}{1994}).

\bibitem[{\citenamefont{Raimond et~al.}(2001)\citenamefont{Raimond, Brune, and
  Haroche}}]{Haroche01}
\bibinfo{author}{\bibfnamefont{J.~M.} \bibnamefont{Raimond}},
  \bibinfo{author}{\bibfnamefont{M.}~\bibnamefont{Brune}}, \bibnamefont{and}
  \bibinfo{author}{\bibfnamefont{S.}~\bibnamefont{Haroche}},
  \bibinfo{journal}{Rev. Mod. Phys.} \textbf{\bibinfo{volume}{73}},
  \bibinfo{pages}{565} (\bibinfo{year}{2001}).

\bibitem[{\citenamefont{Leibfried et~al.}(2003)\citenamefont{Leibfried, Blatt,
  Monroe, and Wineland}}]{Leibfried03}
\bibinfo{author}{\bibfnamefont{D.}~\bibnamefont{Leibfried}},
  \bibinfo{author}{\bibfnamefont{R.}~\bibnamefont{Blatt}},
  \bibinfo{author}{\bibfnamefont{C.}~\bibnamefont{Monroe}}, \bibnamefont{and}
  \bibinfo{author}{\bibfnamefont{D.}~\bibnamefont{Wineland}},
  \bibinfo{journal}{Rev. Mod. Phys.} \textbf{\bibinfo{volume}{75}},
  \bibinfo{pages}{281} (\bibinfo{year}{2003}).

\bibitem[{\citenamefont{Stenholm}(1986)}]{Stenholm86}
\bibinfo{author}{\bibfnamefont{S.}~\bibnamefont{Stenholm}},
  \bibinfo{journal}{Rev. Mod. Phys.} \textbf{\bibinfo{volume}{58}},
  \bibinfo{pages}{699} (\bibinfo{year}{1986}).

\bibitem[{\citenamefont{LaHaye et~al.}(2004)\citenamefont{LaHaye, Buu,
  Camarota, and Schwab}}]{Schwab04}
\bibinfo{author}{\bibfnamefont{M.~D.} \bibnamefont{LaHaye}},
  \bibinfo{author}{\bibfnamefont{O.}~\bibnamefont{Buu}},
  \bibinfo{author}{\bibfnamefont{B.}~\bibnamefont{Camarota}}, \bibnamefont{and}
  \bibinfo{author}{\bibfnamefont{K.~C.} \bibnamefont{Schwab}},
  \bibinfo{journal}{Science} \textbf{\bibinfo{volume}{304}},
  \bibinfo{pages}{74} (\bibinfo{year}{2004}).

\bibitem[{\citenamefont{Knobel and Cleland}(2003)}]{Knobel03}
\bibinfo{author}{\bibfnamefont{R.~G.} \bibnamefont{Knobel}} \bibnamefont{and}
  \bibinfo{author}{\bibfnamefont{A.~N.} \bibnamefont{Cleland}},
  \bibinfo{journal}{Nature (London)} \textbf{\bibinfo{volume}{424}},
  \bibinfo{pages}{291} (\bibinfo{year}{2003}).

\bibitem[{\citenamefont{Weig et~al.}(2004)\citenamefont{Weig, Blick, Brandes,
  Kirschbaum, Wegscheider, Bichler, and Kotthaus}}]{Weig04}
\bibinfo{author}{\bibfnamefont{E.~M.} \bibnamefont{Weig}},
  \bibinfo{author}{\bibfnamefont{R.~H.} \bibnamefont{Blick}},
  \bibinfo{author}{\bibfnamefont{T.}~\bibnamefont{Brandes}},
  \bibinfo{author}{\bibfnamefont{J.}~\bibnamefont{Kirschbaum}},
  \bibinfo{author}{\bibfnamefont{W.}~\bibnamefont{Wegscheider}},
  \bibinfo{author}{\bibfnamefont{M.}~\bibnamefont{Bichler}}, \bibnamefont{and}
  \bibinfo{author}{\bibfnamefont{J.~P.} \bibnamefont{Kotthaus}},
  \bibinfo{journal}{Phys. Rev. Lett.} \textbf{\bibinfo{volume}{92}},
  \bibinfo{pages}{046804} (\bibinfo{year}{2004}).

\bibitem[{\citenamefont{Armour et~al.}(2002)\citenamefont{Armour, Blencowe, and
  Schwab}}]{Armour02}
\bibinfo{author}{\bibfnamefont{A.~D.} \bibnamefont{Armour}},
  \bibinfo{author}{\bibfnamefont{M.~P.} \bibnamefont{Blencowe}},
  \bibnamefont{and} \bibinfo{author}{\bibfnamefont{K.~C.}
  \bibnamefont{Schwab}}, \bibinfo{journal}{Phys. Rev. Lett.}
  \textbf{\bibinfo{volume}{88}}, \bibinfo{pages}{148301}
  (\bibinfo{year}{2002}).

\bibitem[{\citenamefont{Martin et~al.}(2004)\citenamefont{Martin, Shnirman,
  Tian, and Zoller}}]{Martin04}
\bibinfo{author}{\bibfnamefont{I.}~\bibnamefont{Martin}},
  \bibinfo{author}{\bibfnamefont{A.}~\bibnamefont{Shnirman}},
  \bibinfo{author}{\bibfnamefont{L.}~\bibnamefont{Tian}}, \bibnamefont{and}
  \bibinfo{author}{\bibfnamefont{P.}~\bibnamefont{Zoller}},
  \bibinfo{journal}{Phys. Rev. B} \textbf{\bibinfo{volume}{69}},
  \bibinfo{pages}{125339} (\bibinfo{year}{2004}).

\bibitem[{\citenamefont{Wilson-Rae et~al.}(2004)\citenamefont{Wilson-Rae,
  Zoller, and Imamoglu}}]{WilsonRae04}
\bibinfo{author}{\bibfnamefont{I.}~\bibnamefont{Wilson-Rae}},
  \bibinfo{author}{\bibfnamefont{P.}~\bibnamefont{Zoller}}, \bibnamefont{and}
  \bibinfo{author}{\bibfnamefont{A.}~\bibnamefont{Imamoglu}},
  \bibinfo{journal}{Phys. Rev. Lett.} \textbf{\bibinfo{volume}{92}},
  \bibinfo{pages}{075507} (\bibinfo{year}{2004}).

\bibitem{Hopkins03-Clerk08}
\bibinfo{author}{\bibfnamefont{A.}~\bibnamefont{Hopkins}},
  \bibinfo{author}{\bibfnamefont{K.}~\bibnamefont{Jacobs}},
  \bibinfo{author}{\bibfnamefont{S.}~\bibnamefont{Habib}}, \bibnamefont{and}
  \bibinfo{author}{\bibfnamefont{K.}~\bibnamefont{Schwab}},
  \bibinfo{journal}{Phys. Rev. B} \textbf{\bibinfo{volume}{68}},
  \bibinfo{pages}{235328} (\bibinfo{year}{2003});
%
\bibinfo{author}{\bibfnamefont{J.}~\bibnamefont{Eisert}},
  \bibinfo{author}{\bibfnamefont{M.~B.} \bibnamefont{Plenio}},
  \bibinfo{author}{\bibfnamefont{S.}~\bibnamefont{Bose}}, \bibnamefont{and}
  \bibinfo{author}{\bibfnamefont{J.}~\bibnamefont{Hartley}},
  \bibinfo{journal}{Phys. Rev. Lett.} \textbf{\bibinfo{volume}{93}},
  \bibinfo{pages}{190402} (\bibinfo{year}{2004});
%
\bibinfo{author}{\bibfnamefont{A.}~\bibnamefont{Naik}},
  \bibinfo{author}{\bibfnamefont{O.}~\bibnamefont{Buu}},
  \bibinfo{author}{\bibfnamefont{M.~D.} \bibnamefont{LaHaye}},
  \bibinfo{author}{\bibfnamefont{A.~D.} \bibnamefont{Armour}},
  \bibinfo{author}{\bibfnamefont{A.~A.} \bibnamefont{Clerk}},
  \bibinfo{author}{\bibfnamefont{M.~P.} \bibnamefont{Blencowe}},
  \bibnamefont{and} \bibinfo{author}{\bibfnamefont{K.~C.}
  \bibnamefont{Schwab}}, \bibinfo{journal}{Nature}
  \textbf{\bibinfo{volume}{443}}, \bibinfo{pages}{193} (\bibinfo{year}{2006});
%
\bibinfo{author}{\bibfnamefont{L.~F.} \bibnamefont{Wei}},
  \bibinfo{author}{\bibfnamefont{Y.}~\bibnamefont{xi~Liu}},
  \bibinfo{author}{\bibfnamefont{C.~P.} \bibnamefont{Sun}}, \bibnamefont{and}
  \bibinfo{author}{\bibfnamefont{F.}~\bibnamefont{Nori}},
  \bibinfo{journal}{Phys. Rev. Lett.} \textbf{\bibinfo{volume}{97}},
  \bibinfo{pages}{237201} (\bibinfo{year}{2006});
%
\bibinfo{author}{\bibfnamefont{D.}~\bibnamefont{Vitali}},
  \bibinfo{author}{\bibfnamefont{S.}~\bibnamefont{Gigan}},
  \bibinfo{author}{\bibfnamefont{A.}~\bibnamefont{Ferreira}},
  \bibinfo{author}{\bibfnamefont{H.~R.} \bibnamefont{Bohm}},
  \bibinfo{author}{\bibfnamefont{P.}~\bibnamefont{Tombesi}},
  \bibinfo{author}{\bibfnamefont{A.}~\bibnamefont{Guerreiro}},
  \bibinfo{author}{\bibfnamefont{V.}~\bibnamefont{Vedral}},
  \bibinfo{author}{\bibfnamefont{A.}~\bibnamefont{Zeilinger}},
  \bibnamefont{and}
  \bibinfo{author}{\bibfnamefont{M.}~\bibnamefont{Aspelmeyer}},
  \bibinfo{journal}{Phys. Rev. Lett.} \textbf{\bibinfo{volume}{98}},
  \bibinfo{pages}{030405} (\bibinfo{year}{2007}{\natexlab{a}});
%
\bibinfo{author}{\bibfnamefont{D.}~\bibnamefont{Vitali}},
  \bibinfo{author}{\bibfnamefont{P.}~\bibnamefont{Tombesi}},
  \bibinfo{author}{\bibfnamefont{M.~J.} \bibnamefont{Woolley}},
  \bibinfo{author}{\bibfnamefont{A.~C.} \bibnamefont{Doherty}},
  \bibnamefont{and} \bibinfo{author}{\bibfnamefont{G.~J.}
  \bibnamefont{Milburn}}, \bibinfo{journal}{Phys. Rev. A}
  \textbf{\bibinfo{volume}{76}}, \bibinfo{pages}{042336}
  (\bibinfo{year}{2007}{\natexlab{b}}).
%
\bibinfo{author}{\bibfnamefont{K.}~\bibnamefont{Jacobs}},
  \bibinfo{author}{\bibfnamefont{P.}~\bibnamefont{Lougovski}},
  \bibnamefont{and} \bibinfo{author}{\bibfnamefont{M.}~\bibnamefont{Blencowe}},
  \bibinfo{journal}{Phys. Rev. Lett.} \textbf{\bibinfo{volume}{98}},
  \bibinfo{pages}{147201} (\bibinfo{year}{2007});
%
\bibinfo{author}{\bibfnamefont{K.}~\bibnamefont{Jacobs}},
  \bibinfo{journal}{Phys. Rev. Lett.} \textbf{\bibinfo{volume}{99}},
  \bibinfo{pages}{117203} (\bibinfo{year}{2007});
%
\bibinfo{author}{\bibfnamefont{J.~D.} \bibnamefont{Thompson}},
  \bibinfo{author}{\bibfnamefont{B.~M.} \bibnamefont{Zwickl}},
  \bibinfo{author}{\bibfnamefont{A.~M.} \bibnamefont{Jayich}},
  \bibinfo{author}{\bibfnamefont{F.}~\bibnamefont{Marquardt}},
  \bibinfo{author}{\bibfnamefont{S.~M.} \bibnamefont{Girvin}},
  \bibnamefont{and} \bibinfo{author}{\bibfnamefont{J.~G.~E.}
  \bibnamefont{Harris}}, \bibinfo{journal}{Nature}
  \textbf{\bibinfo{volume}{452}}, \bibinfo{pages}{72} (\bibinfo{year}{2008});
%
\bibinfo{author}{\bibfnamefont{A.~A.} \bibnamefont{Clerk}},
  \bibinfo{author}{\bibfnamefont{F.}~\bibnamefont{Marquardt}},
  \bibnamefont{and} \bibinfo{author}{\bibfnamefont{K.}~\bibnamefont{Jacobs}}
  (\bibinfo{year}{2008}), \eprint{arXiv:0802.1842}.

\bibitem[{\citenamefont{Babi\'c et~al.}(2003)\citenamefont{Babi\'c, Furer,
  Sahoo, Farhangfar, and Sch\"onenberger}}]{Babic03}
\bibinfo{author}{\bibfnamefont{B.}~\bibnamefont{Babi\'c}},
  \bibinfo{author}{\bibfnamefont{J.}~\bibnamefont{Furer}},
  \bibinfo{author}{\bibfnamefont{S.}~\bibnamefont{Sahoo}},
  \bibinfo{author}{\bibfnamefont{S.}~\bibnamefont{Farhangfar}},
  \bibnamefont{and}
  \bibinfo{author}{\bibfnamefont{C.}~\bibnamefont{Sch\"onenberger}},
  \bibinfo{journal}{Nano Lett.} \textbf{\bibinfo{volume}{3}},
  \bibinfo{pages}{1577} (\bibinfo{year}{2003}).

\bibitem[{\citenamefont{Sazonova et~al.}(2004)\citenamefont{Sazonova, Yaish,
  \"{U}st\"{u}nel, Roundy, Arias, and McEuen}}]{Sazonova04}
\bibinfo{author}{\bibfnamefont{V.}~\bibnamefont{Sazonova}},
  \bibinfo{author}{\bibfnamefont{Y.}~\bibnamefont{Yaish}},
  \bibinfo{author}{\bibfnamefont{H.}~\bibnamefont{\"{U}st\"{u}nel}},
  \bibinfo{author}{\bibfnamefont{D.}~\bibnamefont{Roundy}},
  \bibinfo{author}{\bibfnamefont{T.~A.} \bibnamefont{Arias}}, \bibnamefont{and}
  \bibinfo{author}{\bibfnamefont{P.~L.} \bibnamefont{McEuen}},
  \bibinfo{journal}{\nat} \textbf{\bibinfo{volume}{431}}, \bibinfo{pages}{284}
  (\bibinfo{year}{2004}).

\bibitem[{\citenamefont{Tian and Zoller}(2004)}]{Tian04}
\bibinfo{author}{\bibfnamefont{L.}~\bibnamefont{Tian}} \bibnamefont{and}
  \bibinfo{author}{\bibfnamefont{P.}~\bibnamefont{Zoller}},
  \bibinfo{journal}{Phys. Rev. Lett.} \textbf{\bibinfo{volume}{93}},
  \bibinfo{pages}{266403} (\bibinfo{year}{2004}).

\bibitem[{\citenamefont{Borgstrom et~al.}(2005)\citenamefont{Borgstrom,
  Zwiller, and Imamoglu}}]{Borgstrom05}
\bibinfo{author}{\bibfnamefont{M.}~\bibnamefont{Borgstrom}},
  \bibinfo{author}{\bibfnamefont{V.}~\bibnamefont{Zwiller}}, \bibnamefont{and}
  \bibinfo{author}{\bibfnamefont{E.~M.~A.} \bibnamefont{Imamoglu}},
  \bibinfo{journal}{Nano Lett.} \textbf{\bibinfo{volume}{5}},
  \bibinfo{pages}{1439} (\bibinfo{year}{2005}).

\bibitem{Olivero05Greentree06}
\bibinfo{author}{\bibfnamefont{P.}~\bibnamefont{Olivero}},
  \bibinfo{author}{\bibfnamefont{S.}~\bibnamefont{Rubanov}},
  \bibinfo{author}{\bibfnamefont{P.}~\bibnamefont{Reichart}},
  \bibinfo{author}{\bibfnamefont{B.~C.} \bibnamefont{Gibson}},
  \bibinfo{author}{\bibfnamefont{S.~T.} \bibnamefont{Huntington}},
  \bibinfo{author}{\bibfnamefont{J.}~\bibnamefont{Rabeau}},
  \bibinfo{author}{\bibfnamefont{A.~D.} \bibnamefont{Greentree}},
  \bibinfo{author}{\bibfnamefont{J.}~\bibnamefont{Salzman}},
  \bibinfo{author}{\bibfnamefont{D.}~\bibnamefont{Moore}},
  \bibinfo{author}{\bibfnamefont{D.~N.} \bibnamefont{Jamieson}},
  \bibnamefont{et~al.}, \bibinfo{journal}{Adv. Mater.}
  \textbf{\bibinfo{volume}{17}}, \bibinfo{pages}{2427} (\bibinfo{year}{2005});
%
\bibinfo{author}{\bibfnamefont{A.~D.} \bibnamefont{Greentree}},
  \bibinfo{author}{\bibfnamefont{P.}~\bibnamefont{Olivero}},
  \bibinfo{author}{\bibfnamefont{M.}~\bibnamefont{Draganski}},
  \bibinfo{author}{\bibfnamefont{E.}~\bibnamefont{Trajkov}},
  \bibinfo{author}{\bibfnamefont{J.~R.} \bibnamefont{Rabeau}},
  \bibinfo{author}{\bibfnamefont{P.}~\bibnamefont{Reichart}},
  \bibinfo{author}{\bibfnamefont{B.~C.} \bibnamefont{Gibson}},
  \bibinfo{author}{\bibfnamefont{S.}~\bibnamefont{Rubanov}},
  \bibinfo{author}{\bibfnamefont{S.~T.} \bibnamefont{Huntington}},
  \bibinfo{author}{\bibfnamefont{D.~N.} \bibnamefont{Jamieson}},
  \bibnamefont{et~al.}, \bibinfo{journal}{J. Phys.: Cond. Mat.}
  \textbf{\bibinfo{volume}{18}}, \bibinfo{pages}{S825} (\bibinfo{year}{2006}).

\bibitem[{\citenamefont{Mancini et~al.}(1998)\citenamefont{Mancini, Vitali, and
  Tombesi}}]{Mancini98}
\bibinfo{author}{\bibfnamefont{S.}~\bibnamefont{Mancini}},
  \bibinfo{author}{\bibfnamefont{D.}~\bibnamefont{Vitali}}, \bibnamefont{and}
  \bibinfo{author}{\bibfnamefont{P.}~\bibnamefont{Tombesi}},
  \bibinfo{journal}{Phys. Rev. Lett.} \textbf{\bibinfo{volume}{80}},
  \bibinfo{pages}{688} (\bibinfo{year}{1998}).

\bibitem[{\citenamefont{Kippenberg and Vahala}(2007)}]{Kippenberg07}
\bibinfo{author}{\bibfnamefont{T.}~\bibnamefont{Kippenberg}} \bibnamefont{and}
  \bibinfo{author}{\bibfnamefont{K.}~\bibnamefont{Vahala}},
  \bibinfo{journal}{Optics Express} \textbf{\bibinfo{volume}{15}},
  \bibinfo{pages}{17172} (\bibinfo{year}{2007}).

\bibitem{Metzger04-Schliesser07}
\bibinfo{author}{\bibfnamefont{C.~H.} \bibnamefont{Metzger}} \bibnamefont{and}
  \bibinfo{author}{\bibfnamefont{K.}~\bibnamefont{Karrai}},
  \bibinfo{journal}{Nature (London)} \textbf{\bibinfo{volume}{432}},
  \bibinfo{pages}{1002} (\bibinfo{year}{2004});
%
\bibinfo{author}{\bibfnamefont{S.}~\bibnamefont{Gigan}},
  \bibinfo{author}{\bibfnamefont{H.~R.} \bibnamefont{B\"{o}hm}},
  \bibinfo{author}{\bibfnamefont{M.}~\bibnamefont{Paternostro}},
  \bibinfo{author}{\bibfnamefont{F.}~\bibnamefont{Blaser}},
  \bibinfo{author}{\bibfnamefont{G.}~\bibnamefont{Langer}},
  \bibinfo{author}{\bibfnamefont{J.~B.} \bibnamefont{Hertzberg}},
  \bibinfo{author}{\bibfnamefont{K.~C.} \bibnamefont{Schwab}},
  \bibinfo{author}{\bibfnamefont{D.}~\bibnamefont{B\"{a}uerle}},
  \bibinfo{author}{\bibfnamefont{M.}~\bibnamefont{Aspelmeyer}},
  \bibnamefont{and}
  \bibinfo{author}{\bibfnamefont{A.}~\bibnamefont{Zeilinger}},
  \bibinfo{journal}{Nature} \textbf{\bibinfo{volume}{444}}, \bibinfo{pages}{67}
  (\bibinfo{year}{2006});
%
\bibinfo{author}{\bibfnamefont{O.}~\bibnamefont{Arcizet}},
  \bibinfo{author}{\bibfnamefont{P.-F.} \bibnamefont{Cohadon}},
  \bibinfo{author}{\bibfnamefont{T.}~\bibnamefont{Briant}},
  \bibinfo{author}{\bibfnamefont{M.}~\bibnamefont{Pinard}}, \bibnamefont{and}
  \bibinfo{author}{\bibfnamefont{A.}~\bibnamefont{Heidmann}},
  \bibinfo{journal}{Nature} \textbf{\bibinfo{volume}{444}}, \bibinfo{pages}{71}
  (\bibinfo{year}{2006});
%
\bibinfo{author}{\bibfnamefont{D.}~\bibnamefont{Kleckner}} \bibnamefont{and}
  \bibinfo{author}{\bibfnamefont{D.}~\bibnamefont{Bouwmeester}},
  \bibinfo{journal}{Nature} \textbf{\bibinfo{volume}{444}}, \bibinfo{pages}{75}
  (\bibinfo{year}{2006});
%
\bibinfo{author}{\bibfnamefont{A.}~\bibnamefont{Schliesser}},
  \bibinfo{author}{\bibfnamefont{P.}~\bibnamefont{DelHaye}},
  \bibinfo{author}{\bibfnamefont{N.}~\bibnamefont{Nooshi}},
  \bibinfo{author}{\bibfnamefont{K.}~\bibnamefont{Vahala}}, \bibnamefont{and}
  \bibinfo{author}{\bibfnamefont{T.}~\bibnamefont{Kippenberg}},
  \bibinfo{journal}{Phys. Rev. Lett.} \textbf{\bibinfo{volume}{97}},
  \bibinfo{pages}{243905} (\bibinfo{year}{2006});
%
\bibinfo{author}{\bibfnamefont{K.~R.} \bibnamefont{Brown}},
  \bibinfo{author}{\bibfnamefont{J.}~\bibnamefont{Britton}},
  \bibinfo{author}{\bibfnamefont{R.~J.} \bibnamefont{Epstein}},
  \bibinfo{author}{\bibfnamefont{J.}~\bibnamefont{Chiaverini}},
  \bibinfo{author}{\bibfnamefont{D.}~\bibnamefont{Leibfried}},
  \bibnamefont{and} \bibinfo{author}{\bibfnamefont{D.~J.}
  \bibnamefont{Wineland}}, \bibinfo{journal}{Phys. Rev. Lett.}
  \textbf{\bibinfo{volume}{99}}, \bibinfo{pages}{137205}
  (\bibinfo{year}{2007});
%
\bibinfo{author}{\bibfnamefont{A.}~\bibnamefont{Schliesser}},
  \bibinfo{author}{\bibfnamefont{R.}~\bibnamefont{Rivi\`ere}},
  \bibinfo{author}{\bibfnamefont{G.}~\bibnamefont{Anetsberger}},
  \bibinfo{author}{\bibfnamefont{O.}~\bibnamefont{Arcizet}}, \bibnamefont{and}
  \bibinfo{author}{\bibfnamefont{T.~J.} \bibnamefont{Kippenberg}},
  \bibinfo{journal}{Nature Physics} \textbf{\bibinfo{volume}{4}},
  \bibinfo{pages}{415} (\bibinfo{year}{2008}). 

\bibitem{WilsonRae07-xue07}
\bibinfo{author}{\bibfnamefont{I.}~\bibnamefont{Wilson-Rae}},
  \bibinfo{author}{\bibfnamefont{N.}~\bibnamefont{Nooshi}},
  \bibinfo{author}{\bibfnamefont{W.}~\bibnamefont{Zwerger}}, \bibnamefont{and}
  \bibinfo{author}{\bibfnamefont{T.~J.} \bibnamefont{Kippenberg}},
  \bibinfo{journal}{Phys. Rev. Lett.} \textbf{\bibinfo{volume}{99}},
  \bibinfo{pages}{093901} (\bibinfo{year}{2007});
%
\bibinfo{author}{\bibfnamefont{F.}~\bibnamefont{Marquardt}},
  \bibinfo{author}{\bibfnamefont{J.~P.} \bibnamefont{Chen}},
  \bibinfo{author}{\bibfnamefont{A.~A.} \bibnamefont{Clerk}}, \bibnamefont{and}
  \bibinfo{author}{\bibfnamefont{S.~M.} \bibnamefont{Girvin}},
  \bibinfo{journal}{Phys. Rev. Lett.} \textbf{\bibinfo{volume}{99}},
  \bibinfo{pages}{093902} (\bibinfo{year}{2007});
%
\bibinfo{author}{\bibfnamefont{F.}~\bibnamefont{Xue}},
  \bibinfo{author}{\bibfnamefont{Y.~D.} \bibnamefont{Wang}},
  \bibinfo{author}{\bibfnamefont{Y.}~\bibnamefont{xi~Liu}}, \bibnamefont{and}
  \bibinfo{author}{\bibfnamefont{F.}~\bibnamefont{Nori}},
  \bibinfo{journal}{Phys. Rev. B} \textbf{\bibinfo{volume}{76}},
  \bibinfo{pages}{205302} (\bibinfo{year}{2007}).

\bibitem{Carr01-savel'ev07}
\bibinfo{author}{\bibfnamefont{S.~M.} \bibnamefont{Carr}},
  \bibinfo{author}{\bibfnamefont{W.~E.} \bibnamefont{Lawrence}},
  \bibnamefont{and} \bibinfo{author}{\bibfnamefont{M.~N.}
  \bibnamefont{Wybourne}}, \bibinfo{journal}{Phys. Rev. B}
  \textbf{\bibinfo{volume}{64}}, \bibinfo{pages}{220101(R)}
  (\bibinfo{year}{2001});
%
\bibinfo{author}{\bibfnamefont{P.}~\bibnamefont{Werner}} \bibnamefont{and}
  \bibinfo{author}{\bibfnamefont{W.}~\bibnamefont{Zwerger}},
  \bibinfo{journal}{Europhys. Lett.} \textbf{\bibinfo{volume}{65}},
  \bibinfo{pages}{158} (\bibinfo{year}{2004});
%
\bibinfo{author}{\bibfnamefont{S.}~\bibnamefont{Savel'ev}} \bibnamefont{and}
  \bibinfo{author}{\bibfnamefont{F.}~\bibnamefont{Nori}},
  \bibinfo{journal}{Phys. Rev. B} \textbf{\bibinfo{volume}{70}},
  \bibinfo{pages}{214415} (\bibinfo{year}{2004});
%
\bibinfo{author}{\bibfnamefont{S.}~\bibnamefont{{Savel'ev}}},
  \bibinfo{author}{\bibfnamefont{X.}~\bibnamefont{{Hu}}}, \bibnamefont{and}
  \bibinfo{author}{\bibfnamefont{F.}~\bibnamefont{{Nori}}},
  \bibinfo{journal}{New Journal of Physics} \textbf{\bibinfo{volume}{8}},
  \bibinfo{pages}{105} (\bibinfo{year}{2006});
%
\bibinfo{author}{\bibfnamefont{S.}~\bibnamefont{Savel'ev}},
  \bibinfo{author}{\bibfnamefont{A.~L.} \bibnamefont{Rakhmanov}},
  \bibinfo{author}{\bibfnamefont{X.}~\bibnamefont{Hu}},
  \bibinfo{author}{\bibfnamefont{A.}~\bibnamefont{Kasumov}}, \bibnamefont{and}
  \bibinfo{author}{\bibfnamefont{F.}~\bibnamefont{Nori}},
  \bibinfo{journal}{Phys. Rev. B} \textbf{\bibinfo{volume}{75}},
  \bibinfo{pages}{165417} (\bibinfo{year}{2007}).

\bibitem[{\citenamefont{Cleland and Roukes}(2002)}]{Cleland02}
\bibinfo{author}{\bibfnamefont{A.~N.} \bibnamefont{Cleland}} \bibnamefont{and}
  \bibinfo{author}{\bibfnamefont{M.~L.} \bibnamefont{Roukes}},
  \bibinfo{journal}{J. Appl. Phys.} \textbf{\bibinfo{volume}{92}},
  \bibinfo{pages}{2758} (\bibinfo{year}{2002}).

\bibitem[{\citenamefont{Cross and Lifshitz}(2001)}]{Cross01}
\bibinfo{author}{\bibfnamefont{M.~C.} \bibnamefont{Cross}} \bibnamefont{and}
  \bibinfo{author}{\bibfnamefont{R.}~\bibnamefont{Lifshitz}},
  \bibinfo{journal}{Phys. Rev. B} \textbf{\bibinfo{volume}{64}},
  \bibinfo{pages}{85324} (\bibinfo{year}{2001}).

\bibitem[{\citenamefont{Wilson-Rae}(2003)}]{phdthesis}
\bibinfo{author}{\bibfnamefont{I.}~\bibnamefont{Wilson-Rae}}, Ph.D. thesis,
  \bibinfo{school}{University of California, Santa Barbara}
  (\bibinfo{year}{2003}).

\bibitem[{\citenamefont{Photiadis and Judge}(2004)}]{Photiadis04}
\bibinfo{author}{\bibfnamefont{D.~M.} \bibnamefont{Photiadis}}
  \bibnamefont{and} \bibinfo{author}{\bibfnamefont{J.~A.} \bibnamefont{Judge}},
  \bibinfo{journal}{Appl. Phys. Lett.} \textbf{\bibinfo{volume}{85}},
  \bibinfo{pages}{482} (\bibinfo{year}{2004}).

\bibitem[{\citenamefont{Chang and Geller}(2005)}]{Chang05}
\bibinfo{author}{\bibfnamefont{C.-M.} \bibnamefont{Chang}} \bibnamefont{and}
  \bibinfo{author}{\bibfnamefont{M.~R.} \bibnamefont{Geller}},
  \bibinfo{journal}{Phys. Rev. B.} \textbf{\bibinfo{volume}{71}},
  \bibinfo{pages}{125304} (\bibinfo{year}{2005}).

\bibitem[{\citenamefont{Caldeira and Leggett}(1983)}]{CaldeiraLeggett83}
\bibinfo{author}{\bibfnamefont{A.~O.} \bibnamefont{Caldeira}} \bibnamefont{and}
  \bibinfo{author}{\bibfnamefont{A.~J.} \bibnamefont{Leggett}},
  \bibinfo{journal}{Ann. Phys.} \textbf{\bibinfo{volume}{149}},
  \bibinfo{pages}{374} (\bibinfo{year}{1983}).

\bibitem[{\citenamefont{Weiss}(1999)}]{Weiss}
\bibinfo{author}{\bibfnamefont{U.}~\bibnamefont{Weiss}},
  \emph{\bibinfo{title}{Quantum Dissipative Systems}}
  (\bibinfo{publisher}{World Scientific}, \bibinfo{year}{1999}).

\bibitem[{\citenamefont{Hu et~al.}(1992)\citenamefont{Hu, Paz, and
  Zhang}}]{Hu92}
\bibinfo{author}{\bibfnamefont{B.~L.} \bibnamefont{Hu}},
  \bibinfo{author}{\bibfnamefont{J.~P.} \bibnamefont{Paz}}, \bibnamefont{and}
  \bibinfo{author}{\bibfnamefont{Y.}~\bibnamefont{Zhang}},
  \bibinfo{journal}{Phys. Rev. D} \textbf{\bibinfo{volume}{45}},
  \bibinfo{pages}{2843} (\bibinfo{year}{1992}).

\bibitem[{\citenamefont{Ford et~al.}(1965)\citenamefont{Ford, Kac, and
  Mazur}}]{Ford65}
\bibinfo{author}{\bibfnamefont{G.~W.} \bibnamefont{Ford}},
  \bibinfo{author}{\bibfnamefont{M.}~\bibnamefont{Kac}}, \bibnamefont{and}
  \bibinfo{author}{\bibfnamefont{P.}~\bibnamefont{Mazur}}, \bibinfo{journal}{J.
  Math. Phys.} \textbf{\bibinfo{volume}{6}}, \bibinfo{pages}{504}
  (\bibinfo{year}{1965}).

\bibitem[{\citenamefont{Rego and Kirczenow}(1998)}]{Rego98}
\bibinfo{author}{\bibfnamefont{L.~G.~C.} \bibnamefont{Rego}} \bibnamefont{and}
  \bibinfo{author}{\bibfnamefont{G.}~\bibnamefont{Kirczenow}},
  \bibinfo{journal}{Phys. Rev. Lett.} \textbf{\bibinfo{volume}{81}},
  \bibinfo{pages}{232} (\bibinfo{year}{1998}).

\bibitem[{\citenamefont{Schwab et~al.}(2000)\citenamefont{Schwab, Henriksen,
  Worlock, and Roukes}}]{Schwab00}
\bibinfo{author}{\bibfnamefont{K.~C.} \bibnamefont{Schwab}},
  \bibinfo{author}{\bibfnamefont{E.~A.} \bibnamefont{Henriksen}},
  \bibinfo{author}{\bibfnamefont{J.~M.} \bibnamefont{Worlock}},
  \bibnamefont{and} \bibinfo{author}{\bibfnamefont{M.~L.}
  \bibnamefont{Roukes}}, \bibinfo{journal}{Nature (London)}
  \textbf{\bibinfo{volume}{404}}, \bibinfo{pages}{974} (\bibinfo{year}{2000}).

\bibitem[{\citenamefont{Lifshitz and Roukes}(2000)}]{Lifshitz00}
\bibinfo{author}{\bibfnamefont{R.}~\bibnamefont{Lifshitz}} \bibnamefont{and}
  \bibinfo{author}{\bibfnamefont{M.~L.} \bibnamefont{Roukes}},
  \bibinfo{journal}{Phys. Rev. B} \textbf{\bibinfo{volume}{61}},
  \bibinfo{pages}{5600} (\bibinfo{year}{2000}).

\bibitem[{\citenamefont{Jiang et~al.}(2004)\citenamefont{Jiang, Yu, Liu, and
  Huang}}]{Jiang04}
\bibinfo{author}{\bibfnamefont{H.}~\bibnamefont{Jiang}},
  \bibinfo{author}{\bibfnamefont{M.-F.} \bibnamefont{Yu}},
  \bibinfo{author}{\bibfnamefont{B.}~\bibnamefont{Liu}}, \bibnamefont{and}
  \bibinfo{author}{\bibfnamefont{Y.}~\bibnamefont{Huang}},
  \bibinfo{journal}{Phys. Rev. Lett.} \textbf{\bibinfo{volume}{93}},
  \bibinfo{pages}{185501} (\bibinfo{year}{2004}).

\bibitem[{\citenamefont{Landau and Lifshitz}(1986)}]{Landau}
\bibinfo{author}{\bibfnamefont{L.~D.} \bibnamefont{Landau}} \bibnamefont{and}
  \bibinfo{author}{\bibfnamefont{E.~M.} \bibnamefont{Lifshitz}},
  \emph{\bibinfo{title}{Theory of Elasticity}}
  (\bibinfo{publisher}{Butterworth-Heinemann, Oxford}, \bibinfo{year}{1986}).

\bibitem[{\citenamefont{Graff}(1991)}]{Graff}
\bibinfo{author}{\bibfnamefont{K.~F.} \bibnamefont{Graff}},
  \emph{\bibinfo{title}{Wave Motion in Elastic Solids}}
  (\bibinfo{publisher}{Dover, New York}, \bibinfo{year}{1991}).

\bibitem[{\citenamefont{Segall et~al.}(2002)\citenamefont{Segall, Ismail-Beigi,
  and Arias}}]{Segall02}
\bibinfo{author}{\bibfnamefont{D.~E.} \bibnamefont{Segall}},
  \bibinfo{author}{\bibfnamefont{S.}~\bibnamefont{Ismail-Beigi}},
  \bibnamefont{and} \bibinfo{author}{\bibfnamefont{T.~A.} \bibnamefont{Arias}},
  \bibinfo{journal}{Phys. Rev. B} \textbf{\bibinfo{volume}{65}},
  \bibinfo{pages}{214109} (\bibinfo{year}{2002}).

\bibitem[{\citenamefont{Ullersma}(1966{\natexlab{a}})}]{Ullersma1}
\bibinfo{author}{\bibfnamefont{P.}~\bibnamefont{Ullersma}},
  \bibinfo{journal}{Physica} \textbf{\bibinfo{volume}{32}}, \bibinfo{pages}{27}
  (\bibinfo{year}{1966}{\natexlab{a}}).

\bibitem[{\citenamefont{Ullersma}(1966{\natexlab{b}})}]{Ullersma2}
\bibinfo{author}{\bibfnamefont{P.}~\bibnamefont{Ullersma}},
  \bibinfo{journal}{Physica} \textbf{\bibinfo{volume}{32}}, \bibinfo{pages}{56}
  (\bibinfo{year}{1966}{\natexlab{b}}).

\bibitem[{\citenamefont{Viviescas and Hackenbroich}(2003)}]{Viviescas03}
\bibinfo{author}{\bibfnamefont{C.}~\bibnamefont{Viviescas}} \bibnamefont{and}
  \bibinfo{author}{\bibfnamefont{G.}~\bibnamefont{Hackenbroich}},
  \bibinfo{journal}{Phys. Rev. A} \textbf{\bibinfo{volume}{67}},
  \bibinfo{pages}{013805} (\bibinfo{year}{2003}), \bibinfo{note}{and references
  therein}.

\bibitem[{\citenamefont{Anetsberger et~al.}(2008)\citenamefont{Anetsberger,
  Rivière, Schliesser, Arcizet, and Kippenberg}}]{Anetsberger08}
\bibinfo{author}{\bibfnamefont{G.}~\bibnamefont{Anetsberger}},
  \bibinfo{author}{\bibfnamefont{R.}~\bibnamefont{Rivi\`ere}},
  \bibinfo{author}{\bibfnamefont{A.}~\bibnamefont{Schliesser}},
  \bibinfo{author}{\bibfnamefont{O.}~\bibnamefont{Arcizet}}, \bibnamefont{and}
  \bibinfo{author}{\bibfnamefont{T.~J.} \bibnamefont{Kippenberg}}
  (\bibinfo{year}{2008}), \eprint{arXiv:0802.4384}.

\bibitem[{\citenamefont{Yakobson et~al.}(1996)\citenamefont{Yakobson, Brabec,
  and Bernholc}}]{Yakobson96}
\bibinfo{author}{\bibfnamefont{B.~I.} \bibnamefont{Yakobson}},
  \bibinfo{author}{\bibfnamefont{C.~J.} \bibnamefont{Brabec}},
  \bibnamefont{and} \bibinfo{author}{\bibfnamefont{J.}~\bibnamefont{Bernholc}},
  \bibinfo{journal}{Phys. Rev. Lett.} \textbf{\bibinfo{volume}{76}},
  \bibinfo{pages}{2511} (\bibinfo{year}{1996}).

\bibitem[{\citenamefont{Robertson et~al.}(1992)\citenamefont{Robertson,
  Brenner, and Mintmire}}]{Robertson92}
\bibinfo{author}{\bibfnamefont{D.~H.} \bibnamefont{Robertson}},
  \bibinfo{author}{\bibfnamefont{D.~W.} \bibnamefont{Brenner}},
  \bibnamefont{and} \bibinfo{author}{\bibfnamefont{J.~W.}
  \bibnamefont{Mintmire}}, \bibinfo{journal}{Phys. Rev. B}
  \textbf{\bibinfo{volume}{45}}, \bibinfo{pages}{12592} (\bibinfo{year}{1992}).

\bibitem[{\citenamefont{Dutta and Horn}(1981)}]{Dutta81}
\bibinfo{author}{\bibfnamefont{P.}~\bibnamefont{Dutta}} \bibnamefont{and}
  \bibinfo{author}{\bibfnamefont{P.~M.} \bibnamefont{Horn}},
  \bibinfo{journal}{Rev. Mod. Phys.} \textbf{\bibinfo{volume}{53}},
  \bibinfo{pages}{497} (\bibinfo{year}{1981}).

\bibitem[{\citenamefont{Paladino et~al.}(2002)\citenamefont{Paladino, Faoro,
  Falci, and Fazio}}]{Paladino02}
\bibinfo{author}{\bibfnamefont{E.}~\bibnamefont{Paladino}},
  \bibinfo{author}{\bibfnamefont{L.}~\bibnamefont{Faoro}},
  \bibinfo{author}{\bibfnamefont{G.}~\bibnamefont{Falci}}, \bibnamefont{and}
  \bibinfo{author}{\bibfnamefont{R.}~\bibnamefont{Fazio}},
  \bibinfo{journal}{Phys. Rev. Lett.} \textbf{\bibinfo{volume}{88}},
  \bibinfo{pages}{228304} (\bibinfo{year}{2002}).

\bibitem[{\citenamefont{Phillips and Lax}(1989)}]{LaxPhyllips}
\bibinfo{author}{\bibfnamefont{R.~S.} \bibnamefont{Phillips}} \bibnamefont{and}
  \bibinfo{author}{\bibfnamefont{P.~D.} \bibnamefont{Lax}},
  \emph{\bibinfo{title}{Scattering theory}} (\bibinfo{publisher}{Academic
  Press}, \bibinfo{year}{1989}).

\bibitem[{\citenamefont{Hakim and Ambegaokar}(1985)}]{Hakim85}
\bibinfo{author}{\bibfnamefont{V.}~\bibnamefont{Hakim}} \bibnamefont{and}
  \bibinfo{author}{\bibfnamefont{V.}~\bibnamefont{Ambegaokar}},
  \bibinfo{journal}{Phys. Rev. A} \textbf{\bibinfo{volume}{32}},
  \bibinfo{pages}{423} (\bibinfo{year}{1985}).

\bibitem[{\citenamefont{Messiah}(1961)}]{Messiah}
\bibinfo{author}{\bibfnamefont{A.}~\bibnamefont{Messiah}},
  \emph{\bibinfo{title}{Quantum Mechanics}} (\bibinfo{publisher}{Dover, New
  York}, \bibinfo{year}{1961}).

\bibitem[{\citenamefont{Wang et~al.}(2000)\citenamefont{Wang, Wong, and
  Nguyen}}]{Wang00}
\bibinfo{author}{\bibfnamefont{K.}~\bibnamefont{Wang}},
  \bibinfo{author}{\bibfnamefont{A.-C.} \bibnamefont{Wong}}, \bibnamefont{and}
  \bibinfo{author}{\bibfnamefont{C.~T.-C.} \bibnamefont{Nguyen}},
  \bibinfo{journal}{J. Microelectromechanical Syst.}
  \textbf{\bibinfo{volume}{9}}, \bibinfo{pages}{347} (\bibinfo{year}{2000}).

\bibitem[{\citenamefont{Wilson-Rae}(unpublished)}]{WilsonRaetobepublished}
\bibinfo{author}{\bibfnamefont{I.}~\bibnamefont{Wilson-Rae}}
  (\bibinfo{year}{unpublished}).

\bibitem[{\citenamefont{Weisstein}()}]{Titchmarsh}
\bibinfo{author}{\bibfnamefont{E.~W.} \bibnamefont{Weisstein}},
  \urlprefix\url{http://mathworld.wolfram.com/TitchmarshTheorem.html}.

\end{thebibliography}

\end{document}